\documentclass[11pt,letterpaper]{article}
\usepackage[T1]{fontenc}
\usepackage{lmodern,mathrsfs}
\usepackage[margin=1.02in]{geometry}
\usepackage{amsmath,amssymb,amsthm,mathtools,booktabs,array,tabularx}
\usepackage{microtype,enumitem,xcolor,hyperref}
\definecolor{ink}{HTML}{263F55}
\hypersetup{colorlinks=true,linkcolor=ink,citecolor=ink,urlcolor=ink,
 pdftitle={Tight Quantum Query Bounds for Collision Search with Prescribed Multiplicities},
 pdfauthor={Tianci Peng, Shiqi Hou, Xiaoyang Dong},
 bookmarksnumbered=true,bookmarksdepth=2}
\allowdisplaybreaks[2]
\newif\ifshowaffiliations
\showaffiliationstrue 
\setlist{topsep=4pt,itemsep=2pt,leftmargin=1.8em}
\newtheorem{theorem}{Theorem}[section]
\newtheorem{lemma}[theorem]{Lemma}
\newtheorem{proposition}[theorem]{Proposition}
\newtheorem{corollary}[theorem]{Corollary}
\theoremstyle{definition}\newtheorem{definition}[theorem]{Definition}
\theoremstyle{remark}\newtheorem{remark}[theorem]{Remark}
\newcommand{\ket}[1]{\lvert #1\rangle}
\newcommand{\bra}[1]{\langle #1\rvert}
\newcommand{\norm}[1]{\left\lVert #1\right\rVert}

\newcommand{\ip}[2]{\langle #1,#2\rangle}
\newcommand{\one}{\mathbf1}
\newcommand{\Id}{I}
\newcommand{\C}{\mathbb C}

\DeclareMathOperator{\dom}{dom}
\DeclareMathOperator{\im}{im}
\DeclareMathOperator{\ran}{ran}
\DeclareMathOperator{\spanop}{span}
\DeclareMathOperator{\Sym}{Sym}
\numberwithin{equation}{section}
\begin{document}
\begin{center}
{\LARGE\bfseries Tight Quantum Query Bounds for Collision Finding\\[3pt]
over Arbitrary Bounded Types\par}
\vspace{1.1em}
\begin{tabular}{ccc}
\ifshowaffiliations
Tianci Peng$^{1}$ & Shiqi Hou$^{1}$ & Xiaoyang Dong$^{1,2,3,*}$\\[3pt]
\else
Tianci Peng & Shiqi Hou & Xiaoyang Dong$^{*}$\\[3pt]
\fi
\small\texttt{ptc@mail.tsinghua.edu.cn}&
\small\texttt{seventeenhsq@gmail.com}&
\small\texttt{xiaoyangdong@tsinghua.edu.cn}
\end{tabular}

{\small
\ifshowaffiliations
$^{1}$Tsinghua University, Beijing, P. R. China\\[2pt]
$^{2}$Zhongguancun Laboratory, Beijing, P. R. China\\[2pt]
$^{3}$State Key Laboratory of Cryptography and Digital Economy Security,\\
Tsinghua University, Beijing, P. R. China\\[3pt]
\fi
$^{*}$Corresponding author\par
}
\end{center}

\begin{abstract}
We determine the bounded-error quantum query complexity of finding
one $r$-collision in an $N$-input function of prescribed type,
which specifies all equality-class sizes. For every fixed maximum
class size and collision order, we give a tight bound as the
maximum of scales determined by the numbers of classes of size
at least $2,\ldots,r$. The lower bound holds for a uniformly
random function of that type, and a matching algorithm succeeds
on every such function. No intermediate class size is required
to occur, and every feasible output alphabet is allowed.

The characterization connects the abundant-collision regime
associated with set equality to the sparse-target regime of
$k$-distinctness. In particular, finding the unique
$(k+1)$-collision in type $(k^n,k+1)$ has complexity
$\Theta_k(N^{3/4-1/[4(2^{k+1}-1)]})$, where $N=kn+k+1$,
resolving an open problem of Belovs.

For the lower bound, we introduce a filtered compressed permutation oracle
that projects onto a harmonic database subspace after each complete
function query. Its $O(T/N)$ comparison error over $T$ queries
overcomes the square-root limitation of the unfiltered process.
Hidden capacity extensions and moment estimates retain the
populations at each intermediate collision order. Matching upper
bounds use sampled collision lists and colored adaptive learning
graphs, with a coherent decision-to-search conversion that avoids
logarithmic overhead. We also give an alternative nested
quantum-walk algorithm for three-collision search, matching the
bound up to polylogarithmic factors.
\end{abstract}

\clearpage
\pdfbookmark[1]{Contents}{contents}
\begingroup
\small
\setlength{\baselineskip}{11pt}
\tableofcontents
\endgroup
\clearpage

\section{Introduction}\label{main:scope}
Quantum query complexity provides a framework for understanding the
power and limitations of quantum algorithms when access to the input
is restricted to oracle queries. In the worst-case setting, an
algorithm must succeed on every input satisfying the promise.
The polynomial method~\cite{BBCMW01} and the adversary
method~\cite{Amb00,LMRSS11} are central tools for proving quantum
query lower bounds. The general adversary bound characterizes
bounded-error query complexity for function evaluation up to constant
factors, but constructing suitable adversary matrices and bounding
their norms can be technically demanding.

In the average-case setting, an algorithm's success probability is
averaged over a prescribed input distribution. A lower bound for
such a distribution also gives a worst-case lower bound whenever
its support consists of valid inputs. The structure of the
distribution offers an additional way to analyze the information
acquired through queries. Zhandry's compressed-oracle
technique~\cite{Zha19} can be viewed as a quantum analogue of lazy
sampling: it represents a random function by a superposition of
small, orthogonal databases. Each query changes the database by a
bounded amount, and successful search can be related to the
amplitude of databases containing the required witnesses. A lower
bound can therefore be obtained by controlling how quickly these
witnesses form.

The original construction exploits the independence of function
values at different inputs. Extending this approach to correlated
distributions is substantially more difficult. For a random
permutation, injectivity couples the values across the domain;
for a random function with prescribed multiplicities, observing one
value constrains the multiplicities remaining elsewhere.
Compressed permutation oracles~\cite{Car25,CM26} provide a way to
represent these dependencies through partial injections. They also
suggest a route to collision bounds under exact-multiplicity
promises, where the independence needed by the original
compressed oracle is unavailable.

Collision search under such promises includes two different
regimes. In the first, many equality classes can supply a target
collision. A basic example is a function of pure type $(k^n)$,
in which every image value has exactly $k$ preimages. For
$2\le r\le k$, every class contains an $r$-collision. This is the
abundant-collision regime familiar from the collision problem and
from the equal-image case of set equality~\cite{BHT98,BR18}.
In the second regime, only a few classes contain a target
collision, while smaller collisions may be plentiful. This is the
situation underlying $r$-distinctness. The type $(k^n,k+1)$
isolates the difficulty: there is one $(k+1)$-collision to find,
surrounded by $n$ classes of size $k$. We study these two
collision-search regimes within a single exact-type formulation.

We consider functions $f:[N]\to\mathbb Z_q$ of exact type
\[
 \lambda=(1^{n_1},\ldots,K^{n_K}),\qquad
 N=\sum_{d=1}^{K}d n_d.
\]
This means that exactly $n_d$ distinct output values have $d$
preimages. Define the tail counts $B_j=\sum_{d=j}^{K}n_d$ and set
$M=B_r\ge1$.
The alphabet is feasible precisely when $q\ge B_1$. An $r$-collision
is a list of $r$ distinct inputs with the same function value.
The algorithm may know the numbers $n_d$, but it receives only the
ordinary function oracle. It is not given the members of any equality
class.
Throughout the main result, $K$ and $r$ are fixed and one collision
is requested.

The type affects which partial collisions are
useful. A repeated pair is progress toward a triple only if its
equality class has another element. With one triple and many
two-element classes, most pairs cannot be extended. When eligible
classes are abundant, on the other hand, tables of partial
collisions can be built and extended by quantum search.
The number $M$ of target classes alone does not capture both
behaviors: the intermediate populations $B_2,\ldots,B_{r-1}$
also matter.

This dependence on the type is the subject of an explicit open
question of Belovs~\cite[Section~7]{Bel26}. After proving a tight
lower bound for unrestricted $r$-distinctness over sufficiently
large alphabets, he asked whether
the same bound holds with one $r$-element class, linearly many
$(r-1)$-element classes, and no classes of other sizes. In our
notation this is the type $(k^n,k+1)$ with $r=k+1$.
He also asked for a tight characterization for an arbitrary
prescribed partition. We resolve both questions for every bounded type, uniformly over all alphabets that realize the type. The resulting formula
connects the abundant-collision and sparse-target regimes through
the full sequence of intermediate populations.

\subsection{Our contributions}\label{intro:contributions}
We determine, up to constant factors, the quantum query complexity
of finding an $r$-collision in a function of arbitrary prescribed
bounded type. Our characterization gives matching lower and upper
bounds throughout the abundant-collision and sparse-target regimes.

\begin{samepage}
For the tail counts $B_j$ defined above, write
\begin{equation}
 \mathcal Q_r(\lambda)=
 \max_{2\le j\le r}
 \left(
 \frac{N^{2^{j-1}}
       \prod_{i=2}^{j-1}B_i^{2^{i-2}}}
      {M^{2^{j-2}}}
 \right)^{1/(2^j-1)},\qquad M=B_r,
 \label{intro:profile-scale}
\end{equation}
where an empty product equals one.
\end{samepage}

\begin{theorem}[Quantum search for a prescribed bounded type]
\label{intro:main}
Fix $K\ge r\ge2$. For every sufficiently large $N$, every exact
type $\lambda$ of maximum class size at most $K$ with $B_r\ge1$,
and every alphabet $q\ge B_1$, the worst-case quantum query
complexity of finding one $r$-collision with probability at least
$2/3$ is
\[
                     \Theta_{K,r}\bigl(\mathcal Q_r(\lambda)\bigr).
\]
The type may be supplied to the algorithm. The lower bound
also holds when success is averaged over the uniform distribution
of functions of that exact type. The constants are independent of
the alphabet and the type.
\end{theorem}

Although the expression in~\eqref{intro:profile-scale} involves
several scales, it is a single characterization valid for every
bounded type. Each scale captures a possible
obstruction at an intermediate collision order. The following
examples illustrate the range of the theorem.

\par\medskip\noindent\textbf{Example 1 (Pure type $(k^n)$):}\enspace
Let $N=kn$ and fix $2\le r\le k$. All relevant tail counts equal
$n=\Theta_k(N)$, so Theorem~\ref{intro:main} gives
\[
 \Theta_{k,r}\!\left(N^{(2^{r-1}-1)/(2^r-1)}\right).
\]
This recovers the single-target specialization of our earlier
multi-collision bounds~\cite{PHD26}. When $r=k$, it also recovers
the tight adaptive bound of Peng and Wu~\cite{PW26} for promised
$k$-to-one finding, with the matching algorithm supplied by Liu
and Zhandry~\cite{LZ19}. Thus both single-target pure-type lower bounds are
contained in the characterization for arbitrary bounded types. In particular,
the pair-finding scale is $\Theta(N^{1/3})$, and higher collision
orders are covered even though the input has no equality classes
of intermediate sizes.

\par\medskip\noindent\textbf{Example 2 (Exceptional type $(k^n,k+1)$):}\enspace
Here $N=kn+k+1$, and the task is to find the unique
$(k+1)$-collision. The tail counts are
$B_2=\cdots=B_k=n+1=\Theta_k(N)$ and $B_{k+1}=1$.
Substitution in Theorem~\ref{intro:main} gives the following
consequence.
\begin{corollary}[Tight search on the exceptional type]
\label{intro:exceptional-tight}
For fixed $k\ge2$, the bounded-error function-query complexity
of finding the unique $(k+1)$-collision in type $(k^n,k+1)$ is
\[
 \Theta_k\!\left(N^{\,3/4-1/[4(2^{k+1}-1)]}\right)
\]
for every feasible alphabet $q\ge n+1$.
\end{corollary}
Indeed, the $j$th scale in~\eqref{intro:profile-scale} is
$\Theta_k(N^{3/4-1/[4(2^j-1)]})$, which increases with $j$.
This resolves the open problem posed by Belovs in
\cite[Section~7]{Bel26}. His analysis of hard input types uses linearly
many classes of every smaller size. Our lower bound attains the
same distinctness exponent when all non-target classes have size
$k$, with no singletons or other intermediate class sizes.
The general theorem likewise imposes no linear-population
assumption on any intermediate class size.
The bound also holds for the smallest feasible alphabet, $q=n+1$,
so no additional large-alphabet assumption is needed.

\par\medskip\noindent\textbf{Example 3 (Type $(1^n,2^b,3)$):}\enspace
Set $N=n+2b+3$. Since $M=1$ and $B_2=b+1$, the tight complexity
of finding the unique triple is
\begin{equation}
 \Theta\!\left(
 \max\left\{N^{2/3},\,N^{4/7}(b+1)^{1/7}\right\}
 \right).
 \label{intro:triple}
\end{equation}
The two terms balance at $b+1\asymp N^{2/3}$. Below this scale,
the element-distinctness term dominates; above it, the population
of non-extendable pairs determines the complexity. In particular,
$b=O(1)$ gives $\Theta(n^{2/3})$, the element-distinctness scale,
whereas $b=\Omega(n)$ gives $b=\Theta(N)$ and complexity
$\Theta(N^{5/7})$, matching the worst-case three-distinctness
exponent. Thus the number of pairs controls a transition between
two familiar query bounds.

Beyond this characterization, our contributions include both
lower-bound techniques and matching algorithms.

\par\medskip\noindent\textbf{A common framework for collision and
distinctness regimes.}\enspace
The complexity formula places the abundant-collision regime associated
with collision and set equality and the sparse-target regime
associated with $k$-distinctness in one quantitative framework.
It identifies how the populations at successive collision orders
govern the transition between them. Our lower-bound argument
tracks partial records within equality classes, so the intermediate
levels remain available even when the input contains no actual
classes of those sizes. This permits a uniform analysis of types
that previously required different treatments.
Our bounds require neither linearly many classes of each
intermediate size nor a large-alphabet assumption: they hold
uniformly for every $q\ge B_1$, including the smallest feasible
alphabet $q=B_1$, with constants independent of $q$.

\par\medskip\noindent\textbf{Filtered CPO beyond the square-root scale.}\enspace
We introduce a \emph{filtered compressed permutation oracle}
(filtered CPO) process for analyzing complete function queries.
After each such query, the process projects the database onto its
harmonic subspace. We prove an $O(T/N)$ comparison with the real
function-query evolution, which remains small for $T=o(N)$
(Corollary~\ref{phase:filtered-comparison}).
This enables lower bounds beyond the square-root horizon of
ordinary CPO soundness~\cite{CM26}, including the distinctness
scales above. We also prove that the unfiltered process has an
intrinsic square-root obstruction even on complete function-query
blocks (Proposition~\ref{phase:unfiltered-barrier}). The filtered
process, together with hidden capacity extensions and
moment estimates in terms of tail counts, supplies the lower bound
in Theorem~\ref{intro:main}.

\par\medskip\noindent\textbf{Matching upper bounds via adaptive
learning graphs.}\enspace
We give function-query algorithms attaining
$O_{K,r}(\mathcal Q_r(\lambda))$ on every input of the prescribed
type, without logarithmic overhead
(Theorem~\ref{search:main}). For scales at most $\sqrt N$,
sampled collision tables and Grover enumeration suffice.
For larger scales, we develop a colored adaptive learning graph
(ALG) whose costs depend on the intermediate populations.
Ordered restriction of the input roles supplies the required
bounds on prefix populations, and a coherent decision-to-search conversion has only
constant overhead. These constructions establish the matching
upper bound across the full range of bounded types.

\par\medskip\noindent\textbf{An alternative nested quantum-walk
algorithm.}\enspace
For three-collision search, we also give a nested quantum random
walk (nested QRW) with query complexity
\[
 \widetilde O_K\!\left(
 \max\left\{N^{2/3}M^{-1/3},\,
           N^{4/7}B_2^{1/7}M^{-2/7}\right\}\right),
\]
matching the type-dependent lower bound up to polylogarithmic
factors (Theorem~\ref{ts:main}). Its main ingredient is a reversible
inner-buffer update that retains all eligible pairs, including
overlapping pairs. The stationary-distribution, spectral-gap,
and initialization analyses are given in Appendix~\ref{app:qrw}.
This provides a separate quantum-walk construction for the
three-collision bound for arbitrary bounded types.

\subsection{Related work}\label{intro:related}
\par\medskip\noindent\textbf{Collision and set equality.}\enspace
Brassard, H{\o}yer, and Tapp~\cite{BHT98} gave a quantum algorithm
that finds a pair in a $k$-to-one function on $N$ inputs using
$O((N/k)^{1/3})$ expected queries. Aaronson and Shi~\cite{AS04}
proved matching collision lower bounds by the polynomial method;
Ambainis~\cite{Amb05} subsequently removed the large-range
restriction. Belovs and Rosmanis~\cite{BR18} gave tight general
adversary bounds for collision and set equality under a sufficiently
large alphabet assumption. In set equality, two injective functions
on $n$-element domains are promised to have equal or disjoint
images. In the equal-image
case their disjoint union has type $(2^n)$, with one preimage in
each domain for every value. This explains the connection with
the abundant-collision regime; the set-equality promise also
specifies how the preimages are distributed between the domains.

\par\medskip\noindent\textbf{Higher-order and multiple collisions.}\enspace
Liu and Zhandry~\cite{LZ19} obtained the exponent
$(2^{r-1}-1)/(2^r-1)$ for finding one $r$-collision. Their
upper-bound construction applies to exact $r$-to-one functions
and, more generally, to sufficiently compressing functions.
Their matching lower bound uses the compressed oracle for
independently sampled random functions. Hamoudi and
Magniez~\cite{HM21} developed strong success-probability bounds
for finding many collision pairs and used them to prove quantum
time-space tradeoffs. These random-function lower bounds do not
directly transfer to exact types: conditioning on the complete
type destroys the independence on which the
compression relies. Our lower bound treats the uniform
distribution on each prescribed type directly. Our earlier
work~\cite{PHD26} established tight pure-type bounds for collisions
in multiple distinct classes. Peng and Wu~\cite{PW26} obtained
tight adaptive and non-adaptive bounds for promised $k$-to-one
finding. The present theorem extends the adaptive single-target
characterization from pure types to arbitrary bounded types.

\par\medskip\noindent\textbf{$k$-distinctness.}\enspace
Ambainis's quantum walk~\cite{Amb07} solves element distinctness
in $O(N^{2/3})$ queries, matching the polynomial lower bound, and
finds $r$ equal elements in $O(N^{r/(r+1)})$ queries.
Belovs~\cite{Bel12} improved the latter bound to
$O(N^{3/4-1/[4(2^r-1)]})$ using adaptive learning graphs.
On the lower-bound side, Bun, Kothari, and Thaler~\cite{BKT18}
and Mande, Thaler, and Zhu~\cite{MTZ20} used dual polynomials to
obtain exponents approaching $3/4$ as $r$ grows.
Belovs~\cite{Bel26} recently proved the matching lower bound for
every fixed $r$, assuming an alphabet of size $\Omega(N^2)$.
His proof uses hard input types with many classes of each smaller
size. Our question fixes the exact type: a bound attained on
those hard input types need not hold on a specified type with missing
class sizes. Theorem~\ref{intro:main} determines this dependence
for bounded types and all feasible alphabets.

\par\medskip\noindent\textbf{Compressed permutation oracles.}\enspace
Rosmanis~\cite{Ros22} introduced a representation-theoretic
decomposition of the permutation-oracle space into knowledge
subspaces, recovering the tight permutation-inversion bound under
forward queries. Majenz, Malavolta, and Walter~\cite{MMW24}
developed a permutation superposition oracle based on strictly
monotone factorizations into transpositions. This gives a nonlocal
database representation of a random permutation, supports both
forward and inverse queries, and yields search bounds for predicates
on a single input-output pair.

Unruh~\cite{Unr23} proposed a compressed permutation oracle in which
inverse queries are implemented by reversing the database;
its soundness was left conjectural.
Carolan~\cite{Car25} modified this construction to preserve
injectivity explicitly and proved its soundness.
Carolan and Majenz~\cite{CM26} improved its soundness horizon
from the $N^{1/12}$ scale to the tight $N^{1/2}$ scale through
a representation-theoretic analysis of the compression isometry.
We use their completion-frame structure to analyze complete
function queries with a harmonic projection after each query.
Keeping the function-query interface is essential for types with
few target classes, where independent permutation access can reveal
additional information. The next subsection explains the filtered
comparison and the estimates in terms of tail counts needed for our lower bounds.

\subsection{Technique overview}\label{intro:technique}
Two issues govern the lower bound. First, a useful database
representation must remain accurate beyond the square-root query
scale. Second, it must distinguish partial collisions that can be
extended to a target from those trapped in smaller classes. We
address the first issue by filtering complete function queries,
and the second by comparing hidden bucket capacities. Moment
estimates connect these two parts and identify the hardest
intermediate collision order. The upper bound balances the same
intermediate populations through sampling and adaptive learning
graphs.

\par\medskip\noindent\textbf{Preserving the function-query interface.}\enspace
We represent a function of the prescribed type as
$f=a\circ\beta\circ\pi$, where $\pi$ is a uniform hidden
permutation of slots, $\beta$ assigns slots to buckets of the
prescribed sizes, and $a$ is an injective table of output labels.
This samples the exact type uniformly for every feasible alphabet.
The algorithm receives only the function oracle; the permutation,
bucket assignment, and label table belong to the representation
used in the analysis. In the phase-query picture, a complete
compressed function query has the form $F=C_xLC_x$: decompress
at the queried input, apply the bucket-label phase, and recompress.
Its temporary response register returns to the blank state.
Analyzing this complete block preserves the restriction on oracle
access that is needed for sparse target classes.

\par\medskip\noindent\textbf{Filtering the database evolution.}\enspace
Completion vectors associate a partial injection with the
superposition of permutations extending it. After removing the
contributions already represented at smaller degrees, their
adjoints identify the harmonic database subspace. Its coefficients
have zero sums along each input- and output-deletion fiber.
The associated Gram operator gives two useful encodings:
$\widetilde V$ preserves norms, while $W$ exactly inverts
completion synthesis. These maps differ by $O(T/N)$ on states
of degree at most $T$ in the range we use. We follow the query
dynamics through $W$ and use $\widetilde V$ at the endpoint to
compare probabilities.

An unfiltered complete query can create a component of norm
$O(N^{-1/2})$ outside the harmonic subspace. This is a substantive
obstruction: Proposition~\ref{phase:unfiltered-barrier} gives a
distinguishing experiment at $T=\Theta(\sqrt N)$ using only
complete function-query blocks. Our analytical process instead
applies the harmonic projector $K$ after each complete query,
without renormalizing:
\[
 \xi_0=\ket\varnothing,\qquad
 \xi_{t+1}=KF_t\mathcal U_t\xi_t,
 \qquad
 \|\xi_T-\widetilde V\psi_T\|=O(T/N).
\]
Here $\mathcal U_t$ is the algorithm's free unitary and $\psi_T$
is its real state; the initial vector includes the same algorithm
registers. The comparison holds for $T\le N/4$ and sufficiently
large $N$ (Corollary~\ref{phase:filtered-comparison}).
Its key estimate concerns the return to the harmonic subspace:
the upward degree component is exact, while conditioning on
fresh and recorded inputs bounds the middle and downward
corrections by $O(1/(N-2t))$. Summing these corrections gives
$O(T/N)$, which remains small at all scales needed here.

\par\medskip\noindent\textbf{Transferring a hidden capacity extension.}\enspace
The type $(k^n,k+1)$ illustrates the second part of the proof.
Its auxiliary database has $n+1$ buckets of capacity $k$;
a uniform hidden mark specifies which receives the extra slot.
The source has $N$ possible inputs and $N-1$ slots and consists
of partial injections, rather than a negative distribution of
complete functions. Following Belovs's transfer
viewpoint~\cite{Bel26}, a map $J$ carries an observed
input-to-bucket assignment from the source to the target.
Different capacities give different numbers of physical
realizations of that assignment. In normalized bucket-orbit
coordinates, this requires the weight
\[
 J_b\ket{\bar d}_{-}
 =\sqrt{\frac{(k+1)_h}{(k)_h}}\ket{\bar d}_{+,b},
 \qquad h=o_b(d),\qquad
 J=\frac1{\sqrt{n+1}}\sum_{b=1}^{n+1}\ket b\otimes J_b.
\]
The map preserves the label lookup and sends the initial source
state to the initial target state. Its image cannot contain all
$k+1$ records in the marked bucket. Consequently, the failure
of $J$ to intertwine the query evolutions measures progress
toward the exceptional collision.

Let $C_j$ count buckets with at least $j$ records, and let
$\mu_j(t)$ be its unnormalized first moment in the filtered
source state. The principal query defect is controlled by
$N^{-1}\sqrt{1+\mu_k(t)}$, so full capacity-$k$ buckets are the
relevant intermediate resource. Filtering requires additional
work: $J$ need not preserve harmonicity, and harmonic projection
need not decrease a collision moment. Deletion identities,
estimates for both directions of leakage, and a separate
counter-projection bound control these effects. The histories
are compared through $S=K^+JK^-$, without assuming that $J$
is an isometry. For $0\le t\le T\le N/4$, the resulting bounds are
\begin{align*}
 \mu_j(t)&=O_k\!\left(
       \frac{t^{\,2-2^{1-j}}}{N^{\,1-2^{1-j}}}\right),
       \qquad 1\le j\le k,\\
 \|P_*\xi_T^+\|&=O_k\!\left(
       \frac TN+
       \frac{T^{\,2-2^{-k}}}{N^{\,3/2-2^{-k}}}\right),
\end{align*}
where $P_*$ tests whether the exceptional bucket is fully
recorded. After the algorithm fixes its output, a constant
number of verification reads converts this amplitude estimate
into its success bound. The terminal conversion contributes
$O_k(T/N+N^{-1/2})$; its $N^{-1/2}$ contribution is incurred
only a constant number of times.

\par\medskip\noindent\textbf{Why arbitrary types give a maximum.}\enspace
For a general type, the useful intermediate populations are
the tail counts $B_j$, rather than just $M=B_r$, the number
of target classes. We extend the comparison to bounded
capacity jumps $c\to d$. Localizing the defect to progress
past occupancy $c$ and averaging the hidden mark gives a
squared-norm estimate that retains both the source candidate
population and the target population.

The moment recursion is proved simultaneously over the target
and its auxiliary capacity multisets. Reducing a capacity to
$j-1$ preserves every lower tail count, allowing the same
induction even when some exact class sizes are absent. We
propagate the additive projection and transfer errors through
every level with their population dependence intact; a coarse
bound that discards this information would lose control beyond
$\sqrt N$.

The proof selects an order maximizing
\eqref{intro:profile-scale}. If target classes are sufficiently
abundant, a moment estimate at this order already bounds
success. Otherwise, the maximizing inequalities identify an
actually present smaller capacity with enough candidate buckets,
and a final hidden extension bounds progress into the target
classes. This yields the maximum over intermediate orders
without requiring linearly many classes of each size. Terminal
verification treats the union of target-bucket events directly,
avoiding an extra factor $\sqrt M$. Every estimate is uniform
for each injective label table; averaging the success
probabilities therefore requires only $q\ge B_1$.

\par\medskip\noindent\textbf{Matching upper bounds.}\enspace
The upper bound chooses raw set sizes so that preparation,
successive prefix extensions, and target completion each cost
$O_{K,r}(\mathcal Q_r)$. The inequalities defining
$\mathcal Q_r$ ensure that these requirements can be balanced.
Realizing the parameters uses different constructions on the
two sides of the square-root scale.

When $\mathcal Q_r\le\sqrt N$, inputs are assigned to disjoint
random roles and sampled at the chosen rates. Querying the
first sample gives a table of values; successive Grover
enumerations extend equal-value prefixes, and a final search
completes an $r$-collision. The parameters give constant
probability that a target survives while controlling the
intermediate populations. Each enumeration has a fixed total
budget, which avoids a logarithmic loss from requiring each
individual discovery to succeed with high probability.

When $\mathcal Q_r\ge\sqrt N$, a colored adaptive learning
graph uses bounds on prefix populations for both positive and
negative inputs. We prove feasibility through the dual-adversary
constraints, including the fault-correction terms. Ordered
restriction of the roles isolates a colorful target with
constant probability while preserving enough space for the
raw sets and controlling every loading cost. To convert
decision into search, parity masks on the first role encode
the isolated target input. A coherent
compute--phase--uncompute operation followed by a Hadamard
transform recovers that input with constant overhead; Grover
search finds the remaining preimages in $O_{K,r}(\sqrt N)$
additional queries (Lemma~\ref{search:parity}). These two
constructions establish the matching query bound without
logarithmic overhead.

\par\medskip\noindent\textbf{A nested-walk alternative for three-collisions.}\enspace
The appendix gives a separate algorithm matching the
three-collision bound up to polylogarithmic factors. Its outer
states are lists of pairs from distinct equality classes.
An inner walk prepares an admissible queried buffer, from
which all eligible pairs are extracted for a reversible batch
exchange. The underlying pair pool may contain overlaps,
while an admissible buffer contains at most two inputs from
each class. Compatible-buffer counts determine a generally
nonuniform stationary distribution. We analyze this
distribution, the spectral gap, and the cost of preparing
the stationary state. Balancing initialization, updates,
and third-input checks gives Theorem~\ref{ts:main}.

\par\medskip\noindent\textbf{Organization.}\enspace
Section~\ref{prelim:section} recalls the CPO, transfer-map,
learning-graph, and quantum-walk frameworks.
Section~\ref{cap:model} defines the oracle representation and
the target and auxiliary databases.
Section~\ref{sec:filtered-comparison} establishes the completion
frames, conditioning estimates, and filtered history comparison.
Section~\ref{sec:capacity-extension} constructs the capacity-extension
map and controls its interaction with harmonic projections and
collision counters. Section~\ref{sec:exceptional-bound} proves
the exceptional-type lower bound, including terminal verification,
and Section~\ref{g:section} extends the analysis to arbitrary
bounded types. Section~\ref{search:overview} gives the matching
search algorithms and completes the proof of
Theorem~\ref{intro:main}. Section~\ref{scope:final} discusses
consequences and further directions. Appendix~\ref{app:qrw}
contains the alternative nested-walk construction for
three-collision search.

\section{Preliminaries}\label{prelim:section}
\par\medskip\noindent\textbf{Conventions.}\enspace We write $[a]=\{1,\ldots,a\}$, $S_a$ for its permutation group, and $(u)_j=u(u-1)\cdots(u-j+1)$ with $(u)_0=1$. Hilbert spaces are finite-dimensional over $\C$; the inner product is linear in its second argument. For an operator $B$, $B^\dagger$ is its adjoint, $\ker B$ its kernel, and $\ran B$ its range. The symbol $\norm{\cdot}$ denotes vector norm or operator norm according to its argument. For self-adjoint operators on the same space, $B\preceq C$ means that $C-B$ is positive semidefinite. A subspace reduces an operator if both it and its orthogonal complement are invariant. We identify a projection with its inclusion into the ambient space when composing maps. All constants implicit in $O_k$ may depend on the fixed integer $k$, but not on the number of buckets, the alphabet or the algorithm's workspace.

\par\medskip\noindent\textbf{Query model.}\enspace
For $f:[N]\to\mathbb Z_q$, a query applies
$O_f\ket{x,z}=\ket{x,z+f(x)}$. Equivalently, using the Fourier
basis in the response register, it applies
\[
 P_f\ket{x,c}=\omega_q^{c f(x)}\ket{x,c},
 \qquad \omega_q=e^{2\pi i/q}.
\]
The two forms are related by free basis changes. A quantum algorithm
alternates queries with input-independent unitaries on its accessible
registers; internal randomness and measurements may be purified.
We count the worst-case number of queries and require success
probability at least $2/3$. Reference registers used to represent
input distributions are inaccessible to the algorithm.

\subsection{Compressed permutation oracles}
\label{prelim:cpo}
We use the compressed permutation oracle of Carolan~\cite{Car25}
in the formulation of Carolan and Majenz~\cite{CM26}.
Its database register has orthonormal basis $\ket D$ indexed by
partial injections $D:[N]\rightharpoonup[N]$.
Write $|D|=|\dom D|$, and let $D[x\mapsto y]$ denote an extension
at an unused input and output. The database starts in $\ket\varnothing$.
For a fixed query input $x$, deleting its record partitions this basis
into the mutually orthogonal spaces
\[
 \mathcal F_{x,H}=\spanop\left(\{\ket H\}\cup
 \{\ket{H[x\mapsto y]}:y\notin\im H\}\right),
 \qquad x\notin\dom H.
\]
On each such space, the compression unitary $C_x$ exchanges
$\ket H$ with
\begin{equation}
 \ket{u_{x,H}}=\frac1{\sqrt{N-|H|}}
       \sum_{y\notin\im H}\ket{H[x\mapsto y]}
 \label{prelim:cpo-extension}
\end{equation}
and fixes the orthogonal complement of their span. Thus $C_x=C_x^\dagger$
and $C_x^2=I$.

Introduce a blank response symbol $\bot$, and let $S_y$ exchange
$\ket\bot$ and $\ket y$ while fixing all other response states.
Let $Q_{x,y}$ project onto databases with $D(x)=y$, and let
$Q_{x,\bot}$ project onto those with $x\notin\dom D$.
The lookup and compressed forward-swap query are
\begin{equation}
 U_x=I\otimes Q_{x,\bot}+\sum_{y\in[N]}S_y\otimes Q_{x,y},
 \qquad \mathsf Q_x=C_xU_xC_x,
 \label{prelim:cpo-query}
\end{equation}
where identities on the response register are implicit in $C_x$.
Coherent queries are controlled by the input register.
An inverse-permutation query is obtained by conjugating the
corresponding forward query by the database flip
$\ket D\mapsto\ket{D^{-1}}$.
Each query can increase $|D|$ by at most one, so after $T$ queries the
database is supported on $|D|\le T$.

For a real permutation $\pi$, the corresponding forward-swap query
applies $S_{\pi(x)}$ to the response. Initializing a reference register
in $(N!)^{-1/2}\sum_{\pi\in S_N}\ket\pi$ purifies a uniform random
permutation. Carolan and Majenz~\cite{CM26} show that, for
$T\le\sqrt N$, the final accessible states in the real and compressed
experiments differ in trace distance by $O(T/\sqrt N)$, even with
forward and inverse queries. In particular, the comparison error
vanishes for $T=o(\sqrt N)$.
Our algorithms have only function-query access. Section~\ref{cap:model}
uses the forward-swap construction inside complete function-query
blocks and extends its database definition to unequal numbers of
inputs and slots. The filtered comparison is established in
Section~\ref{sec:return-section}.

\subsection{Transfer maps and comparison of query histories}
\label{prelim:transfer}
Belovs's transfer operator represents an algorithm under a chosen
input distribution~\cite[Definition~3.4 and Fact~3.5]{Bel26}.
Let $\mathcal F=\mathbb Z_q^{[N]}$, and let $p_{\mu,f}$ be a
probability distribution on pairs consisting of a hidden label
$\mu$ and a truth table $f\in\mathcal F$. The algorithm queries
only $f$. Let $\mathcal X$ and $\mathcal Y$ be the spaces with orthonormal
bases $\{\ket f\}$ and $\{\ket{\mu,f}:p_{\mu,f}>0\}$, respectively.
Define $\Upsilon:\mathcal X\to\mathcal Y$ by
\begin{equation}
 \Upsilon\ket f=q^{N/2}\sum_\mu\sqrt{p_{\mu,f}}\ket{\mu,f}.
 \label{prelim:transfer-map}
\end{equation}
Both input oracles are controlled by the same truth table, so
$O_{\mathcal Y}\Upsilon=\Upsilon O_{\mathcal X}$, with identities
on the algorithm registers understood. The map also commutes with
algorithm-only unitaries. It therefore sends the entire history
with a uniform truth-table prior to the history with prior $p$,
while preserving the function-query interface. The transfer need
not be an isometry on arbitrary vectors.

For our database comparisons, the corresponding intertwining
identity is approximate. Let $A_t^\pm$ be contractions on two
oracle representations, interleaved with the same algorithm-only
unitaries $U_t$, and let $S$ act only on the oracle registers.
Write $\xi_{t+1}^\pm=A_t^\pm U_t\xi_t^\pm$, assume
$S\xi_0^-=\xi_0^+$, and set
$\Delta_t=A_t^+S-SA_t^-$. Subtracting the two updates gives
\[
 \xi_{t+1}^+-S\xi_{t+1}^-
 =A_t^+U_t(\xi_t^+-S\xi_t^-)+\Delta_t U_t\xi_t^-.
\]
Contractivity and induction imply the elementary hybrid estimate
\begin{equation}
 \norm{\xi_T^+-S\xi_T^-}
 \le\sum_{t=0}^{T-1}\norm{\Delta_t U_t\xi_t^-}.
 \label{prelim:transfer-telescope}
\end{equation}
The defect can thus be estimated on the evolving source state,
without a factor $\norm S$ multiplying the accumulated error.

In our application, $S=K^+JK^-$, where $J$ is the weighted
capacity-extension map of Section~\ref{cap:map} and $K^\pm$ are
the harmonic projections. The source is an auxiliary rectangular
space of partial injections, with no associated distribution of
total functions. The map $S$ is constructed directly on these
database spaces; its query defect is bounded in
Lemma~\ref{j:filtered-defect}.

\subsection{Learning graphs and the dual adversary bound}
\label{prelim:learning-graphs}
Let $F:\mathcal X\to\{0,1\}$ be nonconstant, where
$\mathcal X\subseteq\Sigma^m$ is a finite promise domain.
A $1$-certificate for a positive input
$x$ is a restriction $x|_S$ forcing $F=1$ on all consistent promised
inputs. A learning graph is a rooted directed acyclic graph whose
vertices carry sets $S\subseteq[m]$ of loaded inputs; distinct
vertices may carry the same set. Its root carries $\varnothing$,
and each arc $e$ loads one input $j\notin S$, passing from $S$ to
$S\cup\{j\}$. In the basic model each arc has a positive,
input-independent weight $w_e$~\cite{BelLG12}.
For every positive $x$, choose a nonnegative unit flow $p_e(x)$
from the root to vertices containing a $1$-certificate for $x$.
The flow is conserved at all other non-root vertices. Define
\begin{equation}
 C_0=\sum_e w_e,\qquad
 C_1(x)=\sum_e\frac{p_e(x)^2}{w_e},\qquad
 C_1=\max_{x:F(x)=1}C_1(x).
 \label{prelim:lg-complexities}
\end{equation}
These are the negative and positive complexities. The learning-graph
upper bound is $Q(F)=O(\sqrt{C_0C_1})$; equivalently, rescale all
weights to balance $C_0$ and $C_1$ and bound their
maximum~\cite[Section~2.2]{Bel12}.

The underlying dual-adversary formulation also covers the adaptive
construction used below. For each input position $j$, let $X_j$
be a positive semidefinite matrix indexed by $\mathcal X$. If
\begin{equation}
 \sum_{j:x_j\ne y_j}X_j[x,y]=1
 \quad\text{whenever }F(x)=1\text{ and }F(y)=0,
 \label{prelim:lg-dual-feasibility}
\end{equation}
then the general adversary characterization gives
\begin{equation}
 Q(F)=O\!\left(\max_{z\in\mathcal X}\sum_jX_j[z,z]\right).
 \label{prelim:lg-dual-cost}
\end{equation}
The constant is independent of the finite input alphabet~\cite{LMRSS11}.
For an ordinary learning graph, local Gram matrices group inputs
by their common restriction to the source set $S$. For a positive
$x$ and a negative $y$, arcs that first reveal a disagreement form
a cut separating the root from every accepting vertex. The unit
flow across this cut proves~\eqref{prelim:lg-dual-feasibility}.

The same construction permits a weight depending on the source
assignment, provided $w_e(x)=w_e(y)$ whenever $x|_S=y|_S$.
The colored construction in Section~\ref{clg:section} uses
partially revealed assignments, with $\star$ marking a loaded
input whose value is hidden. Inserting a certificate element can
uncover earlier entries, so we use the explicit dual formulation,
as in Belovs's fault-correction construction~\cite[Sections~3 and~6]{Bel12}.
There we write $X_j=\sum_{e\text{ loading }j}\sum_\alpha Y_{e,\alpha}$,
construct each $Y_{e,\alpha}$ as a Gram matrix, and verify
\eqref{prelim:lg-dual-feasibility} directly. Signed off-diagonal
entries are compatible with positive semidefiniteness. The
diagonal estimates then give~\eqref{prelim:lg-dual-cost}.

\subsection{Quantum walk search}\label{prelim:walk}
Let $P$ be a lazy, irreducible, reversible Markov chain on a finite
state space $V$, with stationary distribution $\pi$ and spectral gap
$\delta=1-\lambda_2(P)>0$. Reversibility means
$\pi(v)P(v,w)=\pi(w)P(w,v)$. Let $\mathcal M\subseteq V$ be the
marked set, and suppose that $\pi(\mathcal M)\ge\varepsilon>0$
whenever $\mathcal M$ is nonempty. A quantum walk carries a data
structure $d(v)$ containing the queried information associated with
its current state. Its stationary state, including an edge register,
is
\[
 |\pi_d\rangle=
 \sum_{v,w\in V}\sqrt{\pi(v)P(v,w)}\,
 |v,d(v)\rangle|w,d(w)\rangle.
\]
Write $S$ for its preparation cost, $U$ for the cost of implementing
the coherent update and its inverse, including the associated data
operations, and $C$ for a phase flip on marked states that restores
its ancillary workspace. The quantum walk search theorem of
Magniez, Nayak, Roland, and Santha~\cite[Theorem~3]{MNRS11} gives
a bounded-error algorithm that finds a marked state using
\begin{equation}
 O\!\left(S+\frac1{\sqrt\varepsilon}
                \left(\frac U{\sqrt\delta}+C\right)\right)
 \label{prelim:walk-cost}
\end{equation}
queries. This is a finding guarantee. If a marked state represents
only a partial certificate, extracting and verifying the requested
output must also be included in the cost.

\par\medskip\noindent\textbf{Nested updates.}\enspace
An update may itself require a quantum search. The framework of
Childs, Jeffery, Kothari, and Magniez~\cite[Definition~3.2 and
Theorem~3.3]{CJKM13} permits quantum data depending on the current
vertex and the directed edge. It requires coherent local diffusion
and a reversible swap of the edge and all its data, including any
garbage produced by the inner search. Suppose the inner walks have
uniform marked-mass and gap bounds $\varepsilon',\delta'>0$,
update and checking costs $U',C'$, and the outer local diffusion
with garbage and garbage swap cost at most $T_{\rm loc}$.
Under these implementation
conditions the effective outer update cost is
\[
 U=\widetilde O\!\left(
 T_{\rm loc}+\frac1{\sqrt{\varepsilon'}}
                   \left(\frac{U'}{\sqrt{\delta'}}+C'\right)
 \right).
\]
The setup cost $S$ includes preparation of the inner stationary
states and any search needed to initialize the marked-state data.
Their preparation can be charged once only when
the implementation retains and updates them coherently; any later
preparation or uncomputation is charged where it occurs. Approximate
reflections are implemented with sufficiently small error for the
complete computation, contributing the polylogarithmic factors
suppressed by $\widetilde O$. All costs here count queries to the
original function oracle. Appendix~\ref{app:qrw} verifies these
conditions for the alternative three-collision construction.

\section{The oracle and the two databases}\label{cap:model}
\subsection{Sampling the function and specifying query access}
Let ${N_-}=km$ and $N={N_-}+1$. Partition $[{N_-}]$ into labeled buckets $B_1^-,\ldots,B_m^-$ of size $k$ and let $\beta_-:[{N_-}]\to[m]$ be their index map. Independently sample a uniform mark $b\in[m]$, a uniform permutation $\pi\in S_N$ and a uniform injection $a:[m]\hookrightarrow\mathbb Z_q$. Add the slot $N$ to bucket $b$. Write $B_i^{+,b}$ for the resulting buckets and $\beta_b:[N]\to[m]$ for their index map. Define
\begin{equation}
                 f_{a,b,\pi}(x)=a\bigl(\beta_b(\pi(x))\bigr).
 \label{model:function}
\end{equation}
Buckets contain \emph{slots}; equality classes contain \emph{inputs}. The equality class with label $a(i)$ is $\pi^{-1}(B_i^{+,b})$. We may describe the fixed partitions to the analyst; the algorithm receives no oracle for $\pi$, its inverse, the mark or the label table.

\begin{lemma}[Uniformity]\label{model:uniform}
The function in~\eqref{model:function} is uniform over the stipulated exact type for every $q\ge m$.
\end{lemma}
\begin{proof}
Fix such an $f$. For a specified mark $b$, the value $a(b)$ must be the unique output of multiplicity $k+1$. The other $m-1$ labels may be assigned to the remaining buckets in $(m-1)!$ ways. For each assignment, the restriction of $\pi$ to each equality class can be any bijection onto the corresponding bucket. Hence $m!\,(k+1)!\,(k!)^{m-1}$ triples $(a,b,\pi)$ give $f$, independently of $f$. All triples are equiprobable. This counting requires only that an injection into the alphabet exist.
\end{proof}

We condition on the value of $a$ while proving a uniform estimate. In particular $a_i=a(i)$ is fixed in the phase calculations below. We can then average the resulting success \emph{probabilities} over $a$: if each conditional success amplitude is at most $L$, then the averaged success probability is at most $L^2$.

\begin{lemma}[Phase-query form and final readout]\label{model:bridge}
It suffices to prove the success bound for complete phase queries, arbitrary free unitaries on finite algorithm registers and a final basis readout. A fixed final measurement can be applied to the real and filtered histories without increasing either the number of queries or their comparison error.
\end{lemma}
\begin{proof}
The function oracle is $O_f\ket{x,z}=\ket{x,z+f(x)}$ on $[N]\times\mathbb Z_q$. For $c\in\mathbb Z_q$ put $\ket{\widehat c}=q^{-1/2}\sum_z\exp(-2\pi i cz/q)\ket z$. Translation by $u$ has eigenvalue $\omega_q^{cu}$ on this vector, where $\omega_q=\exp(2\pi i/q)$. Thus one query in the Fourier basis acts diagonally with phase $\omega_q^{c a(\beta_b(\pi(x)))}$; an inverse query changes $c$ to $-c$. The basis changes are free unitaries. More generally, if $u$ indexes a workspace basis, a step may query an input $x_t(u)$ and multiply by any unit phase $z_{t,u}(i)$ for bucket label $i$. Neither $x_t$ nor the table of phases depends on $b$ or $\pi$. This is the phase model used below. In particular ordinary function queries, even conditioned on the now fixed $a$, are covered.

Purify internal coins and keep any intermediate measurement result in an auxiliary register; a later choice can be controlled by that register. For a final measurement with effects $E_j\succeq0$, $\sum_jE_j=I$, the map $v\mapsto\sum_j\ket j\sqrt{E_j}v$ is an isometry, as is seen by taking its squared norm. Extend it to a unitary on a larger workspace, with its ancillas reserved from the start. The resulting basis outcome determines the reported tuple. A failure symbol can be encoded as a fixed tuple with a repeated input. This final unitary commutes with every reference-register encoding and every database-only outcome projector used in the proof. Acting on both histories therefore preserves their distance and the norm on the target event. Branches that stop early can be padded to the query budget with the character $c=0$; the bound is for the worst-case number of queries.

Finally, for fixed $a$ the real reference state starts in $(mN!)^{-1/2}\sum_{b,\pi}\ket{b,\pi}$. Its component in each such sector evolves by precisely the algorithm for \eqref{model:function}. The weights of successful reports in distinct sectors add, since the sectors are orthogonal. Their sum is the average success probability for fixed $a$. Apply Lemma~\ref{model:uniform} after averaging over $a$ to obtain the required distributional interpretation.
\end{proof}

\par\medskip\noindent\textbf{Counters and the two systems.}\enspace
For a partial injection $D$ let $o_i(D)=|\im D\cap B_i|$ be the occupancy of bucket $i$. Put
\[
 C_j(D)=|\{i:o_i(D)\ge j\}|,\qquad
 \widehat C_j\ket D=C_j(D)\ket D.
\]
Unless a superscript indicates otherwise, moments of these counters will refer to the auxiliary system with ${N_-}$ slots. On target branch $b$ let $P_*^b$ project onto $o_b(D)=k+1$, and put $P_*=\bigoplus_bP_*^b$. This is the event that the full exceptional bucket is recorded. The auxiliary system has the same $N$ input positions and ${N_-}=N-1$ slots; it consists of partial injections only. There is no uniform distribution on total injections $[N]\to[{N_-}]$.

\subsection{Partial injections and complete function-query blocks}

Let the query-input set be $[N]$ and let a range set of size $R$ be
partitioned into buckets $B_1,\ldots,B_m$ of capacities
$\kappa_1,\ldots,\kappa_m$.
The database Hilbert space has orthonormal basis indexed by partial
injections $D:[N]\rightharpoonup[R]$.
For fixed $x$, deleting the record at $x$ decomposes the space into
orthogonal subspaces
\[
 \mathcal F_{x,H}
 =\spanop\bigl(\{\ket H\}\cup
       \{\ket{H[x\mapsto y]}:y\notin\im H\}\bigr),
 \qquad x\notin\dom H.
\]
If $s=R-|H|>0$, let $C_x$ exchange $\ket H$ with
\[
 \ket{u_{x,H}}=\frac1{\sqrt s}
    \sum_{y\notin\im H}\ket{H[x\mapsto y]}
\]
and fix the orthogonal complement of their span.  If $s=0$, the local
space is one-dimensional and we define $C_x=I$ there.  This defines
a self-adjoint unitary on the entire database space, even when $R<N$.

For a phase character $c\in\mathbb Z_q$, define the diagonal unitary
\begin{equation}
 L_{x,c}\ket D=
 \begin{cases}
  \ket D,&x\notin\dom D,\\
  \omega_q^{c a_i}\ket D,&D(x)\in B_i,
 \end{cases}
 \qquad \omega_q=e^{2\pi i/q},
 \label{cap:lookup}
\end{equation}
and define the complete compressed function-query block by
\begin{equation}
 F_{x,c}=C_xL_{x,c}C_x,
 \qquad
 F=\sum_{x,c}\ket{x,c}\bra{x,c}\otimes F_{x,c}.
 \label{cap:full-block}
\end{equation}
The label table can be retained coherently; all these definitions are
then controlled by that table.

The forward-swap construction \eqref{prelim:cpo-query} extends
verbatim to this rectangular database, with $y\in[R]$.
Let $D_a$ act on the response register as the identity on $\ket\bot$
and by phase $\omega_q^{ca_i}$ on $\ket y$ when $y\in B_i$.

\begin{lemma}[Exact simplification of a complete block]
\label{cap:complete-simplification}
On an initially blank response register,
\[
 \mathsf Q_xD_a\mathsf Q_x
 \bigl(\ket\bot\otimes v\bigr)
 =\ket\bot\otimes F_{x,c}v.
\]
The response register is exactly blank again at the end of the
compressed block.  One complete block changes at most the record at
$x$; starting from the empty database, the state after $t$ complete
blocks is supported on $|D|\le t$.
\end{lemma}
\begin{proof}
Since $C_x$ and $D_a$ act on different registers,
$\mathsf Q_xD_a\mathsf Q_x=C_xU_xD_aU_xC_x$.
For a fixed database, $U_xD_aU_x$ returns $\ket\bot$ to itself and
applies precisely the phase in \eqref{cap:lookup}.
The local subspace $\mathcal F_{x,H}$ is preserved, proving the
remaining assertions.  Controlled superpositions of $x,c$ obey the
same identities.
\end{proof}

The auxiliary system uses $R={N_-}=km$ and $\kappa_i^-=k$ for all $i$.
It has $N={N_-}+1$ query-input positions.  It is an explicitly defined
unitary database evolution, not a distribution of total injections
$[N]\to[{N_-}]$.  The one-dimensional convention above supplies a
global definition; no terminal subspace is reached when $t\le N/3$.
The target uses $R=N$ and capacities $\kappa^{+,b}$, with an
inaccessible uniform mark register.  Its controlled query is denoted
$F^+=\bigoplus_b F^{+,b}$, while the auxiliary query is $F^-$.

\subsection{Normalized bucket-orbit bases}
\label{cap:orbits}

The following quotient representation makes the comparison explicit.
For a capacity vector $\kappa$, let $\beta$ be the map sending each range slot to its bucket index. A partial bucket assignment is a map $d:[N]\rightharpoonup[m]$ with
$o_i(d)=|d^{-1}(i)|\le\kappa_i$.
Define
\begin{equation}
 \ket{\bar d}_{\kappa}
 =\frac1{\sqrt{\prod_i(\kappa_i)_{o_i(d)}}}
       \sum_{D:\,\beta\circ D=d}\ket D,
 \qquad (a)_h=a(a-1)\cdots(a-h+1).
 \label{cap:orbit-basis}
\end{equation}
These vectors are orthonormal and span the subspace invariant under
permutations of the range slots inside each bucket.  The empty state belongs to this subspace. Complete queries and algorithm-side unitaries preserve it: relabeling slots within a bucket permutes the terms of its uniform extension vector and leaves the lookup phase unchanged. The harmonic projector introduced below also preserves it, since its defining deletion conditions are equivariant under slot relabeling.
We use this basis for the capacity comparison; the canonical-frame
analysis is on the full physical partial-injection space.
We write $H$ for a
partial bucket assignment as well as for its associated orbit basis
vector when no confusion arises.

For a fixed deleted assignment $H$, let
$a_i=\kappa_i-o_i(H)$ and $s=\sum_i a_i$.
The quotient local space has basis $\ket H$ and
$\ket{H[x\mapsto i]}$ for $a_i>0$.
Its uniform extension vector is
\begin{equation}
 u=\sum_{i:a_i>0}\sqrt{a_i/s}\,\ket{H[x\mapsto i]}.
 \label{cap:quotient-extension}
\end{equation}
Indeed, each extension in bucket $i$ has $a_i$ available range
slots, and each extended injection has a unique deletion at $x$.
Applying the physical $C_x$ to \eqref{cap:orbit-basis} gives the
operator exchanging $\ket H$ and \eqref{cap:quotient-extension}.
The lookup phase on the $i$th extension is $\omega_q^{c a(i)}$. In the expression for $u$, $a_i$ denotes the number of unused slots; $a(i)$ is the function label and has no dependence on that count.

\section{A filtered comparison for function queries}\label{sec:filtered-comparison}

\subsection{Completion frames and exact query identities}\label{phase:section}
A completion vector is a superposition of all permutations extending a database. Synthesis with these vectors is exact on a complete function query, but its inverse on the relevant subspace and the isometric encoding of that subspace are different maps. We construct both and bound their difference before analyzing query errors. The completion-frame construction is due to Carolan and Majenz~\cite{CM26}; the incidence and conditioning arguments below prove the properties needed here.

\subsubsection{Completion frames and harmonic databases}
\label{pf:setup}

We construct two encodings of a permutation state: one is an isometry;
the other exactly inverts completion synthesis on its range. We will
need both. We give a counting proof of their comparison, using no
classification of representations of the symmetric group.

Let $S_N$ denote the permutations of $[N]$. Put $\mathcal P=\C[S_N]$,
with orthonormal basis $\{\ket\pi:\pi\in S_N\}$. A size-$h$ database
is a matching of $h$ distinct inputs to $h$ distinct outputs of $[N]$.
Let $\mathcal L_h$ be the Hilbert space with these databases as an
orthonormal basis, and let $L_h$ be the projector onto this summand in
$\mathcal L=\bigoplus_{h=0}^N\mathcal L_h$. The normalized
\emph{completion vector} of $D$ and the synthesis map $E_h$ are
\begin{equation}
 \ket{T_D}=\frac{1}{\sqrt{(N-h)!}}
                   \sum_{\pi\supseteq D}\ket\pi,
 \qquad E_h:\mathcal L_h\longrightarrow\mathcal P,
 \qquad E_h\ket D=\ket{T_D}.
 \label{pf:synthesis}
\end{equation}
Indeed, prescribing $D$ leaves a bijection between the two sets of
$N-h$ unused coordinates undetermined, so there are $(N-h)!$
completions. Let $\mathcal H_{\le h}$ be the span of completion
vectors of sizes at most $h$. For $|H|=h-1$ and $x\notin\dom H$,
\[
  \sum_{y\notin\im H}\ket{T_{H[x\mapsto y]}}
       =\sqrt{N-h+1}\,\ket{T_H};
\]
the analogous identity holds with a fixed unused output and a sum over
inputs. Each identity partitions the completions of $H$ according to
one coordinate. Iteration shows $\ran E_h=\mathcal H_{\le h}$.
Define $\mathcal H_{\le-1}=0$, let $\Pi_h$ project onto the
\emph{exact degree} $\mathcal H_h=\mathcal H_{\le h}\ominus
\mathcal H_{\le h-1}$, and set
\begin{equation}
 A_h=\Pi_h E_h:\mathcal L_h\longrightarrow\mathcal H_h,
 \qquad G_h=A_hA_h^\dagger:\mathcal H_h\longrightarrow\mathcal H_h.
 \label{phase:frames}
\end{equation}
$A_h$ is onto; hence $G_h$ is positive definite on $\mathcal H_h$.
The filtration is exhaustive already at $h=N-1$, since specifying
$N-1$ values specifies the permutation. Thus $\mathcal H_N=0$.

Call a coefficient vector $v\in\mathcal L_h$ \emph{harmonic} if
\begin{equation}
 \sum_{y\notin\im H}v_{H[x\mapsto y]}=0,\qquad
 \sum_{x\notin\dom H}v_{H[x\mapsto y]}=0
 \label{pf:harmonic}
\end{equation}
for every $|H|=h-1$ and each indicated fresh coordinate. The two
conditions are vacuous for $h=0$. Write $\mathscr K_h$ for the subspace
they define. The completion-vector partition above shows immediately
that $A_h^\dagger\psi=E_h^\dagger\psi$ is harmonic for every
$\psi\in\mathcal H_h$: each row sum is a multiple of the inner product
with a lower-degree completion vector. To prove the reverse inclusion
and bound $G_h$, we now evaluate a small, explicit matrix.

\par\medskip\noindent\textbf{Eliminating coordinates outside a fixed set.}\enspace
For a set of outputs $F\subseteq[N]$, suppose $v$ is harmonic and is
invariant under relabeling the outputs by any permutation fixing $F$
pointwise. Fix a database $D$, put $j=|\im D\setminus F|$ and
$d_F=N-|F|$. Let $\mathcal E_F(D)$ be the databases with the same
domain as $D$, all outputs in $F$, and the same edges as $D$ wherever
$D$ takes a value in $F$. With $(u)_j=u(u-1)\cdots(u-j+1)$ and
$(u)_0=1$, the first zero-sum identity gives
\begin{equation}
 v_D=\frac{(-1)^j}{(d_F)_j}\sum_{D'\in\mathcal E_F(D)}v_{D'}.
 \label{pf:eliminate}
\end{equation}
For $j=0$ the sum is $v_D$. Otherwise remove any edge whose output
lies outside $F$. Of the available replacement outputs, $d_F-j+1$
are outside $F$, and every such replacement has coefficient $v_D$:
a transposition of unused outputs carries it back to the original
database. The sum over the available replacements in $F$ is therefore
$-(d_F-j+1)v_D$. Apply the induction hypothesis to each of these
replacements. The resulting completions partition $\mathcal E_F(D)$
according to the output assigned to the chosen input, proving
\eqref{pf:eliminate}. Its denominator is nonzero because $j\le d_F$.
Interchanging inputs and outputs gives the corresponding formula for a
vector invariant under relabelings fixing a set of inputs.

In particular, if $|F|<h$ then $\mathcal E_F(D)$ is always empty,
so the only harmonic vector invariant under this stabilizer is zero.
For a lower-degree database $H$, average any harmonic $v$ over the
output permutations fixing $F=\im H$ pointwise. Equivariance of
completion synthesis means that the averaging does not change
$\langle T_H,E_hv\rangle$, since $T_H$ is fixed. The average remains
harmonic and therefore vanishes. We have proved the purity identity
\begin{equation}
               E_h v=A_h v\in\mathcal H_h
                  \qquad(v\in\mathscr K_h).
 \label{pf:pure}
\end{equation}

\par\medskip\noindent\textbf{The Gram matrix on a fixed pair of coordinate sets.}\enspace
Fix $S,F\subseteq[N]$ with $|S|=|F|=h$ and put $d=N-h$. The base
matchings are the bijections $G:S\to F$. Suppose $v\in\mathscr K_h$
is invariant under separate relabelings of inputs fixing $S$ pointwise
and of outputs fixing $F$ pointwise. Write $w(G)=v_G$ for the base
coefficients. For any $G:S\to F$ we claim that
\begin{align}
 (E_h^\dagger E_h v)_G&=(B w)(G),\label{pf:aligned-gram}\\[-2pt]
 (B w)(G)&=
 \sum_{\substack{R\subseteq S\\ |R|\le d}}
 \frac{1}{(d)_{|R|}}
 \sum_{\substack{G':S\to F\ {\rm bijective}\\
                   G'|_{S\setminus R}=G|_{S\setminus R}}} w(G').
 \label{pf:base-matrix}
\end{align}
To see this, fix a permutation $\pi\supseteq G$. Every size-$h$
restriction $D$ of $\pi$ has domain $(S\setminus R)\cup U$ for
unique sets $R\subseteq S$ and $U\subseteq[N]\setminus S$ of equal
size $j$. Its edges on $S\setminus R$ agree with $G$, and its values
on $U$ belong to $[N]\setminus F$. First use \eqref{pf:eliminate} to
move these $j$ values into $F$. The new edges are indexed by the $j!$
bijections $U\to G(R)$. For each such choice, apply the input version
of \eqref{pf:eliminate} to move the same $j$ edges onto $R$. The
second sum runs exactly once over all $G'$ agreeing with $G$ on
$S\setminus R$. Thus
\[
 v_D=\frac{j!}{(d)_j^2}
     \sum_{G'|_{S\setminus R}=G|_{S\setminus R}}w(G').
\]
For fixed $R$, there are $\binom dj$ choices of $U$, so their total
coefficient is $1/(d)_j$. This proves that $\sum_{D\subseteq\pi,
|D|=h}v_D=(Bw)(G)$, independently of the completion $\pi$. Finally,
$(E_h^\dagger E_hv)_G$ is the average of these sums over the $d!$
completions of $G$, by the two factors $1/\sqrt{d!}$ in
\eqref{pf:synthesis}. This proves \eqref{pf:aligned-gram}.

For a fixed set $R$ the inner sum in \eqref{pf:base-matrix} acts by
an all-ones matrix on each class of bijections with a given
restriction to $S\setminus R$. These blocks are positive semidefinite;
$R=\varnothing$ contributes the identity. Hence $B\succeq\Id$. If
$h\le d$, each row has the same nonnegative sum
\begin{equation}
  \sum_{j=0}^{h}\binom hj\frac{j!}{(d)_j}
   =\frac{d+1}{d-h+1}=1+\frac{h}{N-2h+1}.
 \label{pf:row-sum}
\end{equation}
Indeed, for $u_j=(h)_j/(d)_j$, set $v_j=(d-j+1)u_j$ for $0\le j\le h$
and $v_{h+1}=0$. Then $v_j-v_{j+1}=(d-h+1)u_j$, including $j=h$.
Since the matrix is symmetric and nonnegative entrywise, its operator
norm is at most this row sum (use a largest coordinate of an
eigenvector). The positivity statement itself holds for every $h$.

\begin{lemma}[Harmonic range and Gram bounds]
\label{pf:gram}
For every $h$, $\mathscr K_h=\ran A_h^\dagger$ and
$G_h\succeq\Id$ on $\mathcal H_h$. For $2h\le N$ one has
\begin{equation}
 \Id\preceq G_h\preceq
 \left(1+\frac{h}{N-2h+1}\right)\Id.
 \label{pf:gram-interval}
\end{equation}
In particular $G_h\preceq 2\Id\preceq e\Id$ if $3h+2\le N$.
\end{lemma}
\begin{proof}
First suppose that a harmonic vector $v$ satisfies $E_hv=0$. If it
is nonzero, choose a matching $D_0$ with $v_{D_0}\ne0$. Take
$S=\dom D_0$, $F=\im D_0$ and average $v$ over the two pointwise
stabilizers used above. Their actions preserve harmonicity and commute
with synthesis in the sense of relabeling. The resulting vector
$\bar v$ is still annihilated by $E_h$ and has unchanged coefficient at
$D_0$. Equation~\eqref{pf:aligned-gram} gives $Bw=0$ for its nonzero
base vector $w$, contradicting $B\succeq\Id$. Thus $E_h$, hence also
$A_h$ by \eqref{pf:pure}, is injective on $\mathscr K_h$.
We already proved $\ran A_h^\dagger\subseteq\mathscr K_h$. The
orthogonal complement of that range is $\ker A_h$; no nonzero
harmonic vector belongs to this complement. This proves equality.

Now let $G_h\psi=g\psi$ with $\psi\ne0$. Because $A_h$ is onto,
$v=A_h^\dagger\psi\ne0$. By purity $E_h v=A_hv=g\psi$, and
$E_h^\dagger\psi=A_h^\dagger\psi$, so $E_h^\dagger E_hv=gv$.
Choose a nonzero coefficient and average as before. The operator
$E_h^\dagger E_h$ commutes with these relabelings, hence $Bw=gw$ for
a nonzero $w$. The lower bound for $B$ and, when $2h\le N$, its row
sum~\eqref{pf:row-sum} are therefore bounds for every eigenvalue
of $G_h$. At $h=0$ the same argument gives $G_0=\Id$.
\end{proof}

\par\medskip\noindent\textbf{The two encodings.}\enspace
Inverses in the following formulas act on $\mathcal H_h$. Define
\begin{equation}
 W_h=A_h^\dagger G_h^{-1},\qquad
 \widetilde V_h=A_h^\dagger G_h^{-1/2},\qquad
 K_h=W_hA_h.
 \label{phase:encodings}
\end{equation}
$K_h$ is self-adjoint and idempotent with range $\ran A_h^\dagger$;
by Lemma~\ref{pf:gram} it is the orthogonal projector onto the harmonic
databases. The polar factor $\widetilde V_h$ is an isometry, because
$\widetilde V_h^\dagger\widetilde V_h=\Id$, whereas
$W_h=\widetilde V_hG_h^{-1/2}$ is a contraction. Define
$E=\sum_h E_hL_h$, and take the block direct sums $A,W,\widetilde V,K$
of the corresponding degree maps; put $Z=\Id-K$. At size $N$ take
$K_N=A_N=0$. Operators on the database or permutation register act as
the identity on the algorithm registers. We have
\begin{equation}
 EW=\Id,\qquad WA=K,\qquad AZ=0,\qquad (E-A)K=0.
 \label{phase:frame-identities}
\end{equation}
Indeed, purity gives $E_hW_h=A_hW_h=\Id$, and the other relations
follow from the range of $K_h$ and purity again. Consequently
\begin{equation}
 \mathcal T=WE=K+\mathcal R,\qquad
 \mathcal R=W(E-A)Z,\qquad \mathcal T^2=\mathcal T.
 \label{pf:canonicalization}
\end{equation}
The identity $EW=\Id$ implies idempotence, but does not imply
orthogonality of $\mathcal T$.

For $\Pi_{\le s}=\sum_{h=0}^s\Pi_h$ and $2s\le N$ the same spectrum
gives the endpoint estimate
\begin{equation}
 \norm{(\widetilde V-W)\Pi_{\le s}}
       \le \frac{s}{2(N-2s+1)}.
 \label{pf:endpoint}
\end{equation}
In fact the norm in degree $h$ is the maximum of $1-g^{-1/2}$ over
its Gram eigenvalues. For $g\ge1$ this is at most $(g-1)/2$;
use \eqref{pf:gram-interval} and maximize over $h\le s$.
Finally, if $a$ is harmonic and supported on sizes $h\le s$ with
$3s+2\le N$, then $a=W\psi$ for $\psi=Aa$, and
$\norm\psi\le\sqrt{2}\norm a\le\sqrt{e}\norm a$ by the same bound.

\begin{table}[ht]
\centering\small
\begin{tabularx}{0.96\linewidth}{@{}lX@{}}\toprule
Operator & Space and role \\ \midrule
$L_h$, $K_h$ & $L_h$ selects physical size $h$. Within that size, $K_h$ selects the joint deletion kernel. \\
$\Pi_h$ & Selects exact degree $h$ on the real permutation space. \\
$E_h$, $A_h$ & Send size-$h$ database amplitudes to permutation amplitudes; $A_h=\Pi_h E_h$ keeps only exact degree $h$. \\
$G_h$ & The Gram operator $A_hA_h^\dagger$ on the real exact-degree space. \\
$W_h$, $\widetilde V_h$ & Return a real degree-$h$ state to the harmonic database. $E_hW_h=I$; $\widetilde V_h$ preserves norms. \\
$J_{h,x}$, $J$ & $J_{h,x}$ is the frame for a recorded input in degree $h$ (Section~\ref{pk:section}); $J$ transfers between capacities (Section~\ref{cap:map}). \\
\bottomrule\end{tabularx}
\caption{Physical database size and real permutation degree index different spaces.}\label{tab:operators}
\end{table}

\subsubsection{Exact query identities}\label{phase:exact-section}
Fix an input $x$ and phases $z_y$ indexed by slots $y\in[N]$, with $|z_y|=1$. On the permutation space let $Q_y\ket\pi=\one_{\{\pi(x)=y\}}\ket\pi$ and $U_z=\sum_y z_yQ_y$. The symbols $Q_y$ here act on permutations, whereas $Q^{\rm db}_{x,y}$ below tests a database record. Write $M_y=C_xQ^{\rm db}_{x,y}C_x$, and let $F_z=C_xL_zC_x$ be the complete database block in~\eqref{cap:full-block}, with slot phases $z_y$. All formulas are uniform in these phases.

\begin{lemma}[Synthesis of one query]\label{exact:synthesis}
On the harmonic subspace,
\begin{equation}
 EF_zK=U_zEK,\qquad WU_z\psi=KF_zW\psi+\mathcal RF_zW\psi.
 \label{phase:exact-dynamics}
\end{equation}
The real query and the compressed block have no matrix element between degrees, respectively sizes, differing by more than one.
\end{lemma}
\begin{proof}
In a deleted fiber $H$ put $s=N-|H|$, denote its absent coordinate by $0$, and write a local vector as $(v_0,(v_y)_{y\notin\im H})$. Let $u=s^{-1/2}\sum_y\ket y$ be its uniform recorded component. Put $w=\ket0-u$. The compression is the reflection $C_x=I-ww^\dagger$. For $\pi\supseteq H$ the fiber contributes to the coefficient of $Ev$ at $\pi$ the quantity
\[
   ((N-|H|)!)^{-1/2}\bigl(v_0+\sqrt{s}\,v_{\pi(x)}\bigr).
\]
For $\pi\not\supseteq H$ the contribution vanishes. The functional in parentheses vanishes on $w$ and hence is unchanged by $C_x$. If $v$ is harmonic, the recorded components in each fiber sum to zero by~\eqref{pf:harmonic}; this also holds for a direct sum of homogeneous harmonic vectors. Thus $(C_xv)_0=0$. Applying $L_z$ after the first $C_x$ multiplies the functional by exactly $z_{\pi(x)}$. The final $C_x$ leaves it unchanged. Summing over the fibers proves the first identity. The second follows by applying $W$ to the first, since $EW=I$ and $WE=K+\mathcal R$.

A database fiber consists of just the two sizes $|H|,|H|+1$. On the real side, $Q_y\ket{T_D}$ is either zero, $\ket{T_D}$, or a scalar multiple of $\ket{T_{D[x\mapsto y]}}$. Thus $U_z$ and its adjoint both raise the filtration by at most one. Orthogonality to lower degrees then proves the claimed two-sided band bound.
\end{proof}

For $d=N-h$ and fixed $x,y$, write $R^{00}_{x,y;h}$ for the diagonal projector on size-$h$ databases missing both $x$ and $y$, and $R^{11}_{x,y;h}$ for the projector onto those containing the edge $(x,y)$. Define operators on $\mathcal H_h$ by
\begin{equation}
 A_{y,h}=\Pi_hQ_y\Pi_h,\quad H^{00}_{y,h}=d^{-1}A_hR^{00}_{x,y;h}A_h^\dagger,
 \quad H^{11}_{y,h}=A_hR^{11}_{x,y;h}A_h^\dagger .
 \label{phase:assertion-frames}
\end{equation}
A subscript $z$ in place of $y$ denotes the sum over $y$ with coefficients $z_y$; in particular $A_{z,h}=\Pi_hU_z\Pi_h$. Put $Q^-_{z,h}=\Pi_{h-1}U_z\Pi_h$ and define
\begin{align}
 D^0_{z,h}&=K_hL_hF_zW_h-W_h\Pi_hU_z\Pi_h,\notag\\
 D^-_{z,h}&=L_{h-1}F_zW_h-W_{h-1}Q^-_{z,h}.
 \label{phase:two-corrections}
\end{align}
The second expression is zero at $h=0$.

\begin{lemma}[The three degree components]\label{phase:component-identities}
For $3h+2\le N$, the middle and downward corrections are
\begin{align}
 D^0_{z,h}&=W_h\bigl(H^{00}_{z,h}G_h^{-1}+H^{11}_{z,h}G_h^{-1}-A_{z,h}\bigr),
 \label{phase:middle-identity}\\
 D^-_{z,h}&=W_{h-1}\bigl(G_{h-1}Q^-_{z,h}G_h^{-1}-Q^-_{z,h}\bigr).
 \label{phase:down-identity}
\end{align}
The upward component has no correction. In particular,
\begin{equation}
                  \mathcal RF_zW_h=-D^0_{z,h}-D^-_{z,h}.
 \label{phase:return-components}
\end{equation}
\end{lemma}
\begin{proof}
Take homogeneous harmonic vectors $a,c$ of size $h$. The absent coefficient of $C_xa$ is zero, so on this input $F_z=\sum_y z_y M_y$. In a fiber with $|H|=h$, $a$ is a multiple of $\ket H$ and $C_x$ sends it to its uniform extensions; measuring slot $y$ contributes $d^{-1}\overline{c_H}a_H$ when that slot is available. In a fiber with $|H|=h-1$, both vectors have only recorded components, with sum zero. The compression fixes these vectors, and measuring $y$ keeps precisely the term containing $(x,y)$. No other fiber contributes. It follows that
\[
 \langle c,L_hM_yL_h a\rangle=\langle c,(d^{-1}R^{00}_{x,y;h}+R^{11}_{x,y;h})a\rangle.
\]
Test this formula against $c=A_h^\dagger\phi$, put $a=W_h\psi=A_h^\dagger G_h^{-1}\psi$, sum with weights $z_y$ and use $K_h=W_hA_h$. This gives~\eqref{phase:middle-identity}.

For the downward component let $h\ge1$, $\varphi\in\mathcal H_h$, and start with $a=A_h^\dagger\varphi$, whose coefficients are $\langle T_D,\varphi\rangle$. At a size-$(h-1)$ matching $H$ missing $x$ and $y$, the coefficient of $M_ya$ is $a_{H[x\mapsto y]}/\sqrt{N-h+1}$. Indeed the first reflection fixes the recorded row of $a$, and the absent coefficient after the last reflection is its uniform average after measuring $y$. This equals $\langle Q_y T_H,\varphi\rangle$ by completion normalization. If $y$ is used at another input, both coefficients vanish. If $H$ already contains $x$, the physical coefficient is zero because its fiber has no size-$h$ member. The real coefficient is zero as well: $Q_yT_H$ is either zero or a vector of degree at most $h-1$. Thus
\[
 L_{h-1}M_y A_h^\dagger\varphi=E_{h-1}^\dagger Q_y\varphi
   =A_{h-1}^\dagger\Pi_{h-1}Q_y\varphi.
\]
For the last equality, $Q_y\varphi$ has no components below degree $h-1$, and a size-$(h-1)$ completion column has none above it. Substitute $\varphi=G_h^{-1}\psi$, sum the phases, and use $A_{h-1}^\dagger=W_{h-1}G_{h-1}$. This yields~\eqref{phase:down-identity} and also proves that the entire physical down block is harmonic.

Finally project the synthesis identity onto degree $h+1$. The sizes below $h+1$ have no such component under $E$, and only sizes at most $h+1$ occur. Therefore $A_{h+1}L_{h+1}F_zW_h=\Pi_{h+1}U_z\Pi_h$. Application of $W_{h+1}$ gives equality of the upward canonical blocks. Subtracting the two remaining blocks from~\eqref{phase:exact-dynamics} proves~\eqref{phase:return-components}.
\end{proof}

We will repeatedly use the following elementary inequality: for positive operators $B_y$ and scalars $|z_y|\le1$,
\begin{equation}
       \left\|\sum_y z_yB_y\right\|\le\left\|\sum_y B_y\right\|.
       \label{phase:positive-family}
\end{equation}
Indeed, set $S((v_y)_y)=\sum_y B_y^{1/2}v_y$. The two sums are respectively $S\operatorname{diag}(z_y)S^\dagger$ and $SS^\dagger$.

\subsection{Conditioning on a permutation entry}\label{sec:literal-section}
We need two estimates for the assertion $\pi(x)=y$. On states whose harmonic records know the input $x$, a normalized completion frame is close to the identity. On the complementary states we can delete $x$ and $y$ and compare the resulting smaller completion frame with the original one. The first estimate uses the energy of a masked vector under deletion; the second uses insertion into a permutation.

\subsubsection{A fixed input in a harmonic database}
\label{pk:section}

Fix an input $x$. A completion vector can have some database coefficients
containing $x$ and others missing $x$. Simply keeping the former need not
preserve harmonicity. We first quantify this loss in the database basis, and
then transfer the estimate to the real permutation space.

\par\medskip\noindent\textbf{The deletion gap.}\enspace
For positive integers $a,b$ and $h\le\min(a,b)$, let
$\mathcal L_{a,b,h}$ have as orthonormal basis the size-$h$ partial
injections from $[a]$ to $[b]$. For $h>0$ define the normalized row maps
$B_{\rm d}$ and $B_{\rm i}$ by
\[
 (B_{\rm d}v)_{H,u}=(b-h+1)^{-1/2}
     \sum_{y\notin\im H} v_{H[u\mapsto y]},\qquad
 (B_{\rm i}v)_{H,y}=(a-h+1)^{-1/2}
     \sum_{u\notin\dom H} v_{H[u\mapsto y]}.
\]
Here $|H|=h-1$, and the coordinate named in the row is absent from $H$.
Write $\partial=(B_{\rm d},B_{\rm i})$ and let $K_{a,b,h}$ project onto
its kernel. At $h=0$ take $\partial=0$ and $K_{a,b,0}=I$. In the square
space $a=b=N$ this harmonic projector is $K_h$ from the completion
construction.

\begin{lemma}[Deletion gap]\label{pk:dgap}
Let $1\le h\le\min(a,b)$ and $2h\le\min(a,b)+1$. Then
\begin{equation}\label{pk:gapineq}
 \|\partial v\|^2\ \ge\ \min\!\left\{
  \frac{a-2h+2}{a-h+1},\frac{b-2h+2}{b-h+1}\right\}
  \|(I-K_{a,b,h})v\|^2 .
\end{equation}
\end{lemma}
\begin{proof}
Put $s=b-h+1$, $t=a-h+1$, $D=B_{\rm d}^{\dagger}B_{\rm d}$ and
$F=B_{\rm i}^{\dagger}B_{\rm i}$. For an input $u$, the domain-deletion
rows with missing input $u$ have disjoint supports and unit norm: each
has exactly $s$ extensions, and deleting the record at $u$ recovers its
unique row. Their Gram on the database space is consequently an orthogonal
projector $D_u$, and $D=\sum_u D_u$. Let $M_u$ be the diagonal mask for
$u\in\dom D'$ on a basis database $D'$. Thus $M_u D_u=D_u$.

To bound $D^2$, consider two distinct inputs $u,v$. Both $D_u w$ and
$D_v w$ can contribute to their inner product only on matchings containing
both inputs. Remove their two records. In each resulting fiber the two
remaining outputs range over distinct pairs $i,j$ of an $(s+1)$-element
set. Write $w_{ij}$ for the corresponding coefficients. On the fiber,
$D_u w$ has the constant value $c_j=s^{-1}\sum_{i\ne j}w_{ij}$ in
column $j$, while $D_v w$ has the constant value
$r_i=s^{-1}\sum_{j\ne i}w_{ij}$ in row $i$.
Since $\sum_i r_i=\sum_j c_j$, their inner product on the fiber is
\[
 \sum_{i\ne j}\overline{c_j}r_i
 =\Big|\sum_i r_i\Big|^2-\sum_i\overline{c_i}r_i.
\]
Its real part is at least $-\tfrac12\sum_i(|c_i|^2+|r_i|^2)$.
The squared norms on the fiber are $s\sum_i|c_i|^2$ and
$s\sum_i|r_i|^2$. Summing the fiber inequalities gives
\[
 2\operatorname{Re}\langle D_u w,D_v w\rangle
 \ge -s^{-1}\bigl(\|M_vD_u w\|^2+\|M_uD_v w\|^2\bigr).
\]
For $h=1$ both sides vanish because the common support is empty. Each
matching in the support of $D_u w$ has exactly $h-1$ other occupied inputs.
Expanding $\|\sum_u D_u w\|^2$ and summing over unordered pairs therefore
yields
\[
 \|Dw\|^2\ge\left(1-\frac{h-1}{s}\right)
          \sum_u\|D_u w\|^2
 =\left(1-\frac{h-1}{s}\right)\langle w,Dw\rangle.
\]
Thus every positive eigenvalue of $D$ is at least $1-(h-1)/s$.
Transposing matchings exchanges inputs and outputs and gives the analogous
lower bound $1-(h-1)/t$ for $F$.

It remains to check that these two gaps apply to their joint kernel.
Let $T_{\rm out}$ and $T_{\rm in}$ sum, respectively, the actions on
matchings of all unordered output and input transpositions. On a matching,
a transposition of two unused coordinates fixes it; one between a used and
an unused coordinate replaces one image or one input. Swapping two used
outputs gives the same matching as swapping their two preimages. The
used--used terms thus cancel in the difference. The operator $sD$ has
diagonal $h$ and off-diagonal entry one precisely for one replacement of
an output by an unused one; this follows directly by counting shared
deletion rows. Likewise $tF$ describes input replacements. Consequently
\[
 T_{\rm out}-T_{\rm in}
 =sD-tF+\left[\binom{b-h}{2}-\binom{a-h}{2}\right]I.
\]
Relabeling either coordinates permutes the rows of $B_{\rm d}$ and the
columns by the induced relabeling of the database, so $D$ commutes with
both transposition sums. The identity gives $[D,F]=0$. In a common
orthonormal eigenbasis the nonzero eigenvalues of $D+F$ are sums with at
least one positive summand. Its kernel is exactly $\ker\partial$, since
$\langle v,(D+F)v\rangle=\|\partial v\|^2$. This proves
\eqref{pk:gapineq}.
\end{proof}

\par\medskip\noindent\textbf{Masking one input.}\enspace
From now on $a=b=N$. On $\mathcal L_h=\mathcal L_{N,N,h}$ use $M_x$ for
the input mask above and $M^y$ for the mask of $y\in\im D$. Their harmonic
compressions are self-adjoint positive contractions. Only the input
compression needs the following stronger bound.

\begin{lemma}[Harmonic occupancy mask]\label{pk:mask}
If $2h\le N$, the spectrum of $\mathsf M_x=K_h M_x|_{\ran K_h}$ lies in
\[
 \{0\}\ \cup\ \left[1-\frac1{N-2h+2},1\right].
\]
Its kernel on $\ran K_h$ consists precisely of the harmonic vectors
supported on databases missing $x$.
\end{lemma}
\begin{proof}
At $h=0$ the mask vanishes. Let $h\ge1$, $s=N-h+1$ and $K_h v=v$.
On any domain-deletion row the occupancy of $x$ is constant across the
row, so $B_{\rm d}M_x v=0$. On an image-deletion row $(H,y)$ with
$x\in\dom H$ the row sum likewise vanishes by harmonicity. If
$x\notin\dom H$, exactly one term remains, $v_{H[x\mapsto y]}/\sqrt{s}$.
Deleting the record at $x$ counts each matching containing $x$ exactly
once. Thus
\begin{equation}\label{pk:energy}
 \|\partial M_xv\|^2=s^{-1}\|M_xv\|^2,\qquad
 \|(I-K_h)M_xv\|^2\le\frac{\|M_xv\|^2}{N-2h+2},
\end{equation}
where the second inequality is Lemma~\ref{pk:dgap}. The two quadratic
forms $\|M_xv\|^2$ and $\|K_hM_xv\|^2$ equal
$\langle v,\mathsf M_x v\rangle$ and $\langle v,\mathsf M_x^2 v\rangle$,
respectively. Subtracting the loss in \eqref{pk:energy} gives
$\mathsf M_x^2\succeq(1-1/(N-2h+2))\mathsf M_x$. Its eigenvalues are
in $[0,1]$, proving the assertion. The first quadratic form also proves
the kernel statement.
\end{proof}

\par\medskip\noindent\textbf{Frames on the permutation space.}\enspace
A masked completion frame has two indices that must be kept separate:
the size $s$ of the databases used to form it, and the degree $j$ of the
vectors to which it is applied. Define the following positive operators
on $\mathcal P=\mathbb C[S_N]$:
\[
 \mathcal J_{s,x}=E_s M_x E_s^\dagger,\qquad
 \mathcal O_{s,y}=E_s M^y E_s^\dagger,\qquad
 \mathcal G_s=E_s E_s^\dagger.
\]
Write $Q_y\ket\pi=\mathbf1_{\{\pi(x)=y\}}\ket\pi$ for a fixed $x$.

\begin{lemma}[Fixed-coordinate frames reduce the degrees]\label{pk:convolution}
Each of these frames commutes with every degree projector $\Pi_j$.
The full frame $\mathcal G_s$ commutes with both masked frames, and
$\mathcal J_{s,x}$ commutes with $\mathcal O_{s',y}$ for any sizes $s,s'$.
Moreover $[\mathcal J_{s,x},Q_y]=0$ for every $y$.
\end{lemma}
\begin{proof}
The $(\pi,\sigma)$ matrix entry of a frame is $(N-s)!^{-1}$ times the
number of selected $s$-matchings contained in both permutations. For
$\mathcal J_{s,x}$ this count is unchanged if both permutations are
multiplied on the left by the same permutation: this relabels outputs
and does not change which matchings contain $x$. Its kernel therefore
has the form $j(\pi^{-1}\sigma)$, and the action on a function $f$ is
$\sum_\tau j(\tau) f(\pi\tau)$. For $\mathcal O_{s,y}$ the kernel is
invariant under common right multiplication instead; its action is a
linear combination of left translations. For $\mathcal G_s$ both
invariances hold. Since left and right translations commute, this proves
all the asserted frame commutations. Each translation sends a completion
column of degree $\le j$ to a completion column of the same size and is
unitary, so it preserves both $\mathcal H_{\le j}$ and
$\mathcal H_{\le j-1}^\perp$, and hence $\mathcal H_j$. Finally, two
permutations sharing a matching containing $x$ must agree on $x$.
Thus the kernel of $\mathcal J_{s,x}$ has no entries between distinct
values of $\pi(x)$, which proves its commutation with every $Q_y$.
\end{proof}

Restrict the size-$h$ frames to $\mathcal H_h$, writing
\[
 J_{h,x}=A_hM_xA_h^\dagger,\qquad O_{h,y}=A_hM^yA_h^\dagger,
 \qquad G_h=A_h A_h^\dagger .
\]
By Lemma~\ref{pk:convolution} these are actual restrictions, not just
compressions. Set $\mathcal F_{h,x}=\ker J_{h,x}$ (the \emph{fresh}
space) and call $\mathcal F_{h,x}^{\perp}$ the \emph{known} space. Both
are subspaces of $\mathcal H_h$. Indeed,
$\langle\psi,J_{h,x}\psi\rangle=\|M_xA_h^\dagger\psi\|^2$: freshness
means exactly that no coefficient of the harmonic analysis $A_h^\dagger
\psi$ records $x$. The word ``known'' refers to its orthogonal complement,
not to the unprojected mask $M_x$.

\begin{lemma}[Normalized frame on the known space]\label{pk:transfer}
For $2h\le N$, put $B_{h,x}=J_{h,x}G_h^{-1}$. It vanishes on
$\mathcal F_{h,x}$ and is a positive contraction, with
\begin{equation}\label{pk:ratio}
 \left\|(I-B_{h,x})|_{\mathcal F_{h,x}^{\perp}}\right\|
 \le (N-2h+2)^{-1}.
\end{equation}
The degree-space operators $G_h^{\pm1}$, $O_{h,y}G_h^{-1}$ and
$A_{y,h}=\Pi_h Q_y|_{\mathcal H_h}$ preserve both summands. On the
harmonic database space $A_h^\dagger A_h$ commutes with $\mathsf M_x$ and
with $\mathsf M^y=K_h M^y|_{\ran K_h}$; the latter two operators also
commute. For a harmonic $v$, one has
\begin{equation}\label{pk:masked-synthesis}
 E_h K_h M_x v=E_h M_x v.
\end{equation}
\end{lemma}
\begin{proof}
By Lemma~\ref{pk:convolution}, $J_{h,x}$ commutes with $G_h$, $O_{h,y}$,
and $A_{y,h}$. For the last assertion use the commutation of the full
$\mathcal J_{h,x}$ with both $Q_y$ and $\Pi_h$. Since these frames are
self-adjoint, their commuting operators preserve both $\ker J_{h,x}$ and
its orthogonal complement; so do $G_h^{-1}$ and $O_{h,y}G_h^{-1}$.
In particular the positive operator $J_{h,x}G_h^{-1}$ has the same kernel
as $J_{h,x}$. Using $K_h=A_h^\dagger G_h^{-1}A_h$ gives the intertwining
identity
\begin{equation}\label{pk:mask-intertwining}
 \mathsf M_x A_h^\dagger
 = A_h^\dagger G_h^{-1}J_{h,x}=A_h^\dagger B_{h,x}.
\end{equation}
The map $A_h^\dagger$ is injective from $\mathcal H_h$ onto
$\ran K_h$. Thus $B_{h,x}$ has the spectrum in Lemma~\ref{pk:mask},
which gives \eqref{pk:ratio}. With outputs in place of inputs the same
calculation intertwines $\mathsf M^y$ with $O_{h,y}G_h^{-1}$; in
particular the latter is also a positive contraction. Moreover
$(A_h^\dagger A_h)A_h^\dagger=A_h^\dagger G_h$. Transfer the commutations
of $G_h$, $B_{h,x}$ and $O_{h,y}G_h^{-1}$ across the same invertible map
to obtain all three database-space commutations. Finally write
$v=A_h^\dagger\psi=E_h^\dagger\psi$ with $\psi\in\mathcal H_h$.
Then $E_h M_x v=\mathcal J_{h,x}\psi\in\mathcal H_h$. On the other hand
$E_h K_h=A_h$ (synthesis on the harmonic space is pure of degree $h$,
and $A_h K_h=A_h$). Thus $E_h K_h M_xv=\Pi_h E_h M_xv=E_h M_xv$.
This last identity does not assert that $M_xv$ itself is harmonic.
\end{proof}

\subsubsection{Conditioning on a fresh input and output}
\label{px:section}

The estimate on the known subspace concerns a recorded input.  On its
orthogonal complement we need to understand what a literal assertion
$\pi(x)=y$ sees.  The complement of this one input and one output is a
smaller permutation problem.  The completion vectors in the two problems
have different normalizations; keeping those normalizations will give both
a relative bound and the exact commutation needed for the downward query.

Fix $x,y\in[N]$, put $n=N-1$ and $d=N-h$, and assume $2h\le n$.
The small input and output sets are $X=[N]\setminus\{x\}$ and
$Y=[N]\setminus\{y\}$.  Identifying them separately with $[n]$ identifies
their bijections with $S_n$.  Superscripts $(n)$ and $(N)$ indicate which
universe is meant.  Write $\mathscr K^{(s)}_h=\ran K^{(s)}_h$ for the
harmonic coefficient space and let $\iota:\mathcal L^{(n)}_h\to
\mathcal L^{(N)}_h$ extend coefficients by zero on matchings that use $x$
or $y$.  Thus $\iota^\dagger$ simply discards those coordinates.
On $\mathscr K^{(n)}_h$ define
\begin{equation}
 B=(A_h^{(n)})^\dagger A_h^{(n)}\big|_{\mathscr K_h^{(n)}},\qquad
 S=E_h^{(N)}\iota\big|_{\mathscr K_h^{(n)}},\qquad
 \widehat B=S^\dagger S .
 \label{px:two-grams}
\end{equation}
In particular, $B$ is a Gram operator on coefficients; $G_h^{(N)}$ acts
on permutation amplitudes.  Put $M_u=\mathbf1_{\{u\in\dom D\}}$ on the
small database, and let $\mathsf M_u=K_h^{(n)}M_u|_{\mathscr K_h^{(n)}}$.

\begin{lemma}[Insertion identity]
\label{px:insertion}
For every $v\in\mathcal L^{(n)}_h$,
\begin{equation}
 d\norm{E_h^{(N)}\iota v}^2=
 \norm{E_h^{(n)}v}^2+\sum_{u\in X}\norm{E_h^{(n)}(I-M_u)v}^2 .
 \label{px:insert-all}
\end{equation}
On the small harmonic space this gives the operator identity
\begin{equation}
 \widehat B=B-\frac1d\sum_{u\in X}B(\mathsf M_u-\mathsf M_u^2).
 \label{px:insert-operator}
\end{equation}
Consequently $B\widehat B=\widehat B B$ and
\begin{equation}
 (1-\rho_h)B\preceq\widehat B\preceq B,
 \qquad \rho_h=\frac{h}{d(n-2h+2)} .
 \label{px:relative}
\end{equation}
\end{lemma}
\begin{proof}
A large permutation is uniquely specified by a pair $(u,\sigma)$, where
$u\in X\cup\{x\}$ and $\sigma:X\to Y$ is a bijection.  If $u=x$, extend
$\sigma$ by $x\mapsto y$. Otherwise replace the edge
$u\mapsto\sigma(u)$ by $u\mapsto y$ and $x\mapsto\sigma(u)$.  The
inverse reads $u=\pi^{-1}(y)$, deletes $x,y$, and restores $u\mapsto
\pi(x)$ if $u\ne x$.  A small matching $D$ is completed by this $\pi$
precisely when $\sigma\supseteq D$ and $u\notin\dom D$.  Also
$(N-h)!=d(n-h)!$.  Thus the coefficient of $E_h^{(N)}\iota v$ at
$(u,\sigma)$ is $d^{-1/2}$ times the coefficient of
$E_h^{(n)}(I-M_u)v$ at $\sigma$, with $M_x=0$ for the omitted input.
Summing the squares proves \eqref{px:insert-all}.

For harmonic $v$, Lemma~\ref{pk:transfer} gives
$E_h^{(n)}M_uv=E_h^{(n)}\mathsf M_u v$ and
$B\mathsf M_u=\mathsf M_u B$.  This is where purity of the
\emph{synthesized} masked vector matters; $M_uv$ itself need not be
harmonic.  Every size-$h$ matching records exactly $h$ inputs, so
$\sum_u\mathsf M_u=hI$.  Expanding the right side of
\eqref{px:insert-all} on a harmonic vector gives the quadratic form of
\[
 (n+1)B-2hB+\sum_u B\mathsf M_u^2
 =dB-\sum_u B(\mathsf M_u-\mathsf M_u^2).
\]
Both sides are self-adjoint; complex polarization yields
\eqref{px:insert-operator}.  Each summand commutes with $B$, proving the
commutation with $\widehat B$.  Finally, Lemma~\ref{pk:mask} gives
\[
 0\preceq \mathsf M_u-\mathsf M_u^2
 \preceq (n-2h+2)^{-1}\mathsf M_u.
\]
Because these operators commute with the positive $B$, the same
inequalities hold after multiplication by $B$.  Sum over $u$ and use
\eqref{px:insert-operator} to obtain \eqref{px:relative}. For $h=0$
all masks vanish and $\widehat B=B$.  Notice that it is the polarized
\emph{identity}, not the two inequalities in \eqref{px:relative}, that
supplies commutation.
\end{proof}

We next identify the space on which this comparison applies. In the
large universe, let $M_x$ and $M^y$ be the masks for a recorded input
$x$ and output $y$.  On $\mathcal H_h^{(N)}$ abbreviate $J_x=J_{h,x}$, $J^y=O_{h,y}$ and
$\mathcal F_x=\mathcal F_{h,x}$; thus put
\[
 J_x=A_h^{(N)}M_x(A_h^{(N)})^\dagger,\qquad
 J^y=A_h^{(N)}M^y(A_h^{(N)})^\dagger,\qquad
 \mathcal F_x=\ker J_x,\quad
 \mathcal F_{xy}=\mathcal F_x\cap\ker J^y.
\]
Let $P_{xy,h}$ denote the orthogonal projection onto
$\mathcal F_{xy}$. The symbol $A_{y,h}=\Pi_h Q_y\Pi_h$ refers to the
degree-$h$ assertion for the fixed input $x$; $Q_y$ is the literal
projector $\mathbf1_{\{\pi(x)=y\}}$ on permutation amplitudes.

\begin{lemma}[The support obtained by deleting two coordinates]
\label{px:both-fresh}
$S$ maps $\mathscr K_h^{(n)}$ bijectively onto $\mathcal F_{xy}$, and
\begin{equation}
 G_h^{(N)}S=S\widehat B.
 \label{px:gram-transport}
\end{equation}
On $\mathcal F_x$, the assertion has support in $\mathcal F_{xy}$:
$A_{y,h}=P_{xy,h}A_{y,h}P_{xy,h}$ there.
\end{lemma}
\begin{proof}
The image $\iota\mathscr K_h^{(n)}$ is exactly the large harmonic
space supported on matchings avoiding $x,y$. Indeed, a deletion row
with a nonzero summand in the extension by zero must have its retained
matching and its fixed coordinate in the smaller rectangle $X\times Y$.
For such a row the large sum equals the small sum: all added terms are
zero.  All other rows are zero throughout. This applies separately to
the input and output deletion rows. Conversely, restricting a harmonic
vector supported in $X\times Y$ therefore satisfies all small deletion
rows. This also proves the assertion for $h=0$, when there are no rows.
In particular $S=A_h^{(N)}\iota|_{\mathscr K_h^{(n)}}$.

For brevity write $A=A_h^{(N)}$, $G=AA^\dagger$.
For $\psi\in\mathcal H_h^{(N)}$ one has
$\ip{\psi}{J_x\psi}=\norm{M_x A^\dagger\psi}^2$, and likewise for
$J^y$. Thus $\psi\in\mathcal F_{xy}$ precisely when
$A^\dagger\psi\in\iota\mathscr K_h^{(n)}$. On the large harmonic
space $A^\dagger A$ commutes with both compressed masks by
Lemma~\ref{pk:transfer} (the output version follows by inversion).
The kernel of a compressed mask on that space is the kernel of the
original mask there, since $\ip{v}{K_h^{(N)}M_x v}=\norm{M_xv}^2$.
Hence $A^\dagger A$ preserves $\iota\mathscr K_h^{(n)}$, so $Sv$
belongs to $\mathcal F_{xy}$ for every small harmonic $v$.
Conversely $G$ and $G^{-1}$ preserve $\mathcal F_{xy}$ by their
commutation with $J_x,J^y$ (Lemma~\ref{pk:convolution}). For
$\psi\in\mathcal F_{xy}$, we may therefore write
$A^\dagger G^{-1}\psi=\iota v$, whence $Sv=AA^\dagger G^{-1}\psi=\psi$.
Injectivity follows from the injectivity of $A$ on its harmonic space.
On $\mathcal F_{xy}$ we have $SS^\dagger=G$: inserting the orthogonal
projection onto $\iota\mathscr K_h^{(n)}$ between $A$ and $A^\dagger$
does not change $AA^\dagger\psi$ there. Applying this identity to $S$
gives \eqref{px:gram-transport}.

It remains to check support of the assertion; this does not follow
merely from its energy on $\mathcal F_{xy}$. On the full permutation
space let the recorded-input frame be $E_h^{(N)}M_x(E_h^{(N)})^\dagger$
and define the recorded-output frame analogously. They preserve degree
$h$ by Lemma~\ref{pk:convolution}. On a completion column recording
$x$, $Q_y$ acts as the identity if the column contains $x\mapsto y$
and as zero otherwise. The same holds on a column recording $y$, since
a permutation has only one preimage of $y$. Multiplying the rank-one
summands of either full frame by $Q_y$, on either side, consequently
selects the same fixed-edge frame. Projecting to degree $h$ gives
\[
 J_x A_{y,h}=A_{y,h}J_x=J^y A_{y,h}
 = A_h^{(N)}\mathbf1_{\{D(x)=y\}}(A_h^{(N)})^\dagger.
\]
All three products vanish on $\mathcal F_x$. Hence
$A_{y,h}\mathcal F_x\subseteq\mathcal F_{xy}$; self-adjointness on
$\mathcal F_x$ gives the claimed support on both sides.
\end{proof}

\begin{lemma}[Literal conditioning]
\label{px:restriction}
Let $R$ restrict a large permutation amplitude to the fibre
$\pi(x)=y$, identified with the small bijections $X\to Y$, with
counting measure on both sets. Then $Q_y=R^\dagger R$,
$R\mathcal F_x\subseteq\mathcal H_h^{(n)}$, and
\begin{equation}
 RSv=d^{-1/2}E_h^{(n)}v
 \qquad (v\in\mathscr K_h^{(n)}).
 \label{px:literal-conditioning}
\end{equation}
\end{lemma}
\begin{proof}
The adjoint $R^\dagger$ is extension by zero outside the fibre, by the
definition of the counting inner products. Restricting any completion
column either kills it or removes the fixed edge $x\mapsto y$ if
present; the remaining column has at most the original degree. Hence
$R\psi\in\mathcal H_{\le h}^{(n)}$ for $\psi\in\mathcal F_x$. For
$h\ge1$, zero-extending a small column of size at most $h-1$ gives a
scalar multiple of a large column of size at most $h$ containing
$x\mapsto y$. If its size is less than $h$, it is orthogonal to $\psi$
by degree; at size $h$, its coefficient against $\psi$ vanishes since
$M_x (A_h^{(N)})^\dagger\psi=0$. By adjointness $R\psi$ is orthogonal
to $\mathcal H_{\le h-1}^{(n)}$. At $h=0$ this space is zero.
Finally, on the fibre, every embedded matching has exactly the same
satisfaction indicator as its small matching. Its completion coefficient
changes by the factorial factor $d^{-1/2}$, giving
\eqref{px:literal-conditioning}. More generally, for a large harmonic
$v'$ supported on matchings missing $x$, columns using $y$ cannot
contribute on this fibre. Thus
$R E_h^{(N)}v'=d^{-1/2}E_h^{(n)}\iota^\dagger v'$.
The left side has pure degree $h$: $E_h^{(N)}v'\in\mathcal F_x$
by the same mask-kernel argument used in Lemma~\ref{px:both-fresh}.
One can therefore insert $K_h^{(n)}$ before synthesis on the right;
one cannot simply assume $\iota^\dagger v'$ itself is harmonic.
\end{proof}

\begin{proposition}[Fresh assertion and its Gram commutation]
\label{px:fresh-conclusion}
On $\mathcal F_x$, with $d=N-h$,
\begin{equation}
 P_{xy,h}\preceq d A_{y,h}\preceq(1+\varepsilon_h) P_{xy,h},
 \qquad [G_h^{(N)},A_{y,h}]=0,
 \qquad \varepsilon_h=\frac{h}{(N-2h)(N-h+1)}.
 \label{px:fresh-bounds}
\end{equation}
In particular the nonzero eigenvalues of $d A_{y,h}$ there belong to
$[1,1+\varepsilon_h]$.
\end{proposition}
\begin{proof}
For small harmonic $u,v$, literal conditioning gives
\begin{equation}
 \ip{Su}{A_{y,h}Sv}=\ip{RSu}{RSv}=d^{-1}\ip{u}{Bv}.
 \label{px:conditional-pair}
\end{equation}
Put $\psi=Sv$. By \eqref{px:relative} the quantities
$\norm{\psi}^2=\ip{v}{\widehat B v}$ and
$d\ip{\psi}{A_{y,h}\psi}=\ip{v}{Bv}$ satisfy
\[
 \norm{\psi}^2\le d\ip{\psi}{A_{y,h}\psi}
 \le(1-\rho_h)^{-1}\norm{\psi}^2.
\]
Indeed $(1-\rho_h)(1+\varepsilon_h)=1$, as one sees from
$d(n-2h+2)-h=(N-2h)(N-h+1)>0$. This proves the sandwich on
$\mathcal F_{xy}$, and the support identity of Lemma~\ref{px:both-fresh}
proves it on $\mathcal F_x$.

For commutation retain the full sesquilinear identity
\eqref{px:conditional-pair}. Since $S$ is a bijection onto
$\mathcal F_{xy}$ and $\widehat B=S^\dagger S$ is invertible,
\[
 A_{y,h}S=d^{-1}S\widehat B^{-1}B.
\]
Testing both sides against $Su$ proves this equality; both ranges lie
in $\mathcal F_{xy}$. In the same coordinates $G_h^{(N)}S=S\widehat B$
by \eqref{px:gram-transport}. The two operators commute there because
$\widehat B B=B\widehat B$. On $\mathcal F_x\ominus\mathcal F_{xy}$
the assertion vanishes, while the Gram preserves this subspace by
self-adjointness and invariance of $\mathcal F_x,\mathcal F_{xy}$.
This completes the commutator proof on all of $\mathcal F_x$.

The deleted input and output were chosen arbitrarily. More explicitly,
independent relabelings $a,b$ act by $D\mapsto bDa^{-1}$ and
$\pi\mapsto b\pi a^{-1}$. These are coordinate unitaries on both
spaces; they preserve $D\subseteq\pi$ and the factor $(N-h)!$.
They therefore intertwine synthesis, its adjoint, both harmonic
projections, the masks and the assertions. No comparison between
irreducible labels for different outcomes is required.
\end{proof}

\subsection{Comparing a complete function query}\label{sec:return-section}
The preceding estimates describe a fixed degree. We now put them together across adjacent degrees, first for a whole history of filtered function queries and then for a bounded number of unfiltered terminal reads.

\subsubsection{The middle and downward return}

We now estimate the two terms in \eqref{phase:return-components}.
Fix the query input \(x\) and write \(\mathcal F_h\)
for the subspace \(\ker J_{h,x}\) of \(\mathcal H_h\), and \(\mathcal F_h^\perp\)
for its orthogonal complement \emph{in \(\mathcal H_h\)}.
These are the fresh and known subspaces, respectively.

\begin{lemma}[Middle return]\label{phase:middle-bound}
If \(3h+2\le N\), then
\[
 \|D^0_{z,h}\|\le \frac{e+1}{N-2h}.
\]
\end{lemma}
\begin{proof}
Set \(d=N-h\), \(G=G_h\), \(J=J_{h,x}\) and \(A_y=A_{y,h}\)
as in \eqref{phase:assertion-frames}.
Define the output-occupied frame
\(O_y=A_h\mathbf1_{\{y\in\im D\}}A_h^\dagger\).
Lemma~\ref{pk:convolution} gives \([J,A_y]=[J,O_y]=[G,J]=[G,O_y]=0\).

Write \(A_z=\sum_y z_yA_y\). Since it is the compression of \(U_z\),
\(\|A_z\|\le1\). We use the identities
\[
 H^{11}_{y,h}=A_yJ,\qquad
 \sum_y H^{00}_{y,h}=G-J,\qquad \sum_y O_y=hG.
\]
For the first, multiply the fixed-input frame by the literal \(Q_y\):
among its completion vectors only those recording \(x\mapsto y\)
survive. Projection onto degree \(h\) gives the assertion.
For the other two, each matching of size \(h\) missing \(x\) has
\(d\) unused outputs, and each matching has \(h\) occupied outputs.
All of these operators reduce \(\mathcal F_h\) and \(\mathcal F_h^\perp\).
Indeed \(\psi\in \mathcal F_h\) means \(A_h^\dagger\psi\) vanishes on
matchings recording \(x\). Hence
\(d H^{00}_{y,h}\psi=(G-O_y)\psi\) on \(\mathcal F_h\).
This expression preserves \(\mathcal F_h\); self-adjointness gives the
invariance of \(\mathcal F_h^\perp\). The other invariances follow from
the commutations just proved.

On \(\mathcal F_h^\perp\), the fixed-input ratio estimate \eqref{pk:ratio} gives
\(\|I-JG^{-1}\|\le\eta_h=(N-2h+2)^{-1}\).
By the positive-family inequality~\eqref{phase:positive-family}, \(\|H^{00}_{z,h}\|\le\|G-J\|\le e\eta_h\) on this
subspace. Also
\(H^{11}_{z,h}G^{-1}-A_z=A_z(JG^{-1}-I)\).
As \(\|G^{-1}\|,\|W_h\|\le1\), \eqref{phase:middle-identity}
bounds the known part by \((e+1)\eta_h\).

On \(\mathcal F_h\), put \(B_y=(O_yG^{-1})|_{\mathcal F_h}\) and let \(P_y\)
project onto \(\mathcal F_h\cap\ker O_y\). We have \(0\le B_y\le I\),
since \(0\le O_y\le G\) and the two commute. Thus
\(I-B_y\ge P_y\); the cross terms vanish since \(B_yP_y=0\).
The conditioning estimate of Proposition~\ref{px:fresh-conclusion} gives
\[
 P_y\ \preceq\ d A_y\ \preceq\ (1+\varepsilon_h)P_y,\qquad
 \varepsilon_h=\frac{h}{(N-2h)(N-h+1)}.
\]
On this subspace \(H^{11}_{y,h}=0\), and
\(H^{00}_{y,h}G^{-1}=d^{-1}(I-B_y)\).
Thus the real operator in parentheses in \eqref{phase:middle-identity}
is the signed sum of the differences between two positive families,
\(d^{-1}(I-B_y)\) and \(A_y\). Both sum to \(I\), since
\(\sum_yB_y=hI\) and \(\sum_y A_y=I\). Both dominate \(P_y/d\).
Subtract this common part before applying the positive-family
inequality. The norm is at most
\[
 2\left\|I-\frac1d\sum_yP_y\right\|
 \le 2\frac{\varepsilon_h}{1+\varepsilon_h}\le2\varepsilon_h.
\]
This cancellation controls all outcomes at once. Since
\(\varepsilon_h\le1/(N-2h)\), the fresh estimate is sufficient.
Finally the two restrictions of the real operator have orthogonal
ranges; its norm is their maximum. Applying \(W_h\) cannot increase
that norm.
\end{proof}

To handle neighboring degrees we must use \emph{one and the same
frame size} on both sides of the query. For integers \(s,j\), let
\(J_x^{[s;j]}\) be the restriction to \(\mathcal H_j\) of the
full-space frame \(E_s\mathbf1_{\{x\in\dom D\}}E_s^\dagger\).
In particular \(J_x^{[s;s]}=J_{s,x}\).
Because the full-space operator preserves all degrees and commutes
with every literal at \(x\), it satisfies
\begin{equation}
 J_x^{[s;i]}\Pi_iU_z\Pi_j
 =\Pi_iU_z\Pi_j J_x^{[s;j]}.
 \label{pr:intertwining}
\end{equation}

\begin{lemma}[Lower frame on the fresh subspace]\label{pr:lower-frame}
Let \(j<N\), \(d=N-j\) and \(A_y=A_{y,j}\). On \(\mathcal F_j\) one has
\[
 J_x^{[j+1;j]}=d\sum_y A_yG_j A_y.
\]
Moreover \(Q=\Pi_jU_z\Pi_{j+1}\) has image in \(\mathcal F_j\) and vanishes
on \(\mathcal F_{j+1}\).
\end{lemma}
\begin{proof}
For \(\phi\in\mathcal H_{j+1}\) and a matching \(D\) of size \(j\),
the coefficient \(\langle T_D,Q_y\phi\rangle\) is zero if \(D\)
records \(x\): the literal acts on \(T_D\) as either the identity or
zero, and \(\phi\perp\mathcal H_{\le j}\).
If \(D\) misses \(x\) but uses \(y\), the coefficient is again zero.
In the remaining case it is
\(d^{-1/2}\langle T_{D\cup\{(x,y)\}},\phi\rangle\),
by the normalization of the two completion vectors.
Projection onto degree \(j\) does not change these coefficients,
since \(Q_y\phi\) has no component below \(j\). Thus the analysis
of \(\Pi_jQ_y\phi\) vanishes on matchings recording \(x\); it
vanishes identically when \(\phi\in \mathcal F_{j+1}\).
The size-\(j\) completion vectors span \(\mathcal H_{\le j}\), so these
statements imply the last two claims, after summing over \(y\).

For \(u\in \mathcal F_j\), this same vanishing applied to
\(\Pi_{j-1}Q_yu\) shows that \(Q_yu\) has no component below \(j\)
(at \(j=0\) there is nothing to show). Its size-\(j\) analysis
therefore equals \(A_j^\dagger A_yu\). This analysis is supported on
matchings missing both \(x\) and \(y\): if \(D\) records \(x\), use
\(\langle T_D,u\rangle=0\), and if it misses \(x\) but uses \(y\),
\(Q_yT_D=0\). Now enumerate size-\((j+1)\) matchings containing the
edge \((x,y)\) by deleting that edge. For \(u,v\in \mathcal F_j\) the
resulting frame identity in quadratic-form notation is
\[
 \langle u,J_x^{[j+1;j]}v\rangle
  =d\sum_y \langle A_j^\dagger A_yu,A_j^\dagger A_yv\rangle
  =d\sum_y \langle u,A_yG_j A_yv\rangle .
\]
Here the factor \(d\) follows from
\(T_{D\cup\{(x,y)\}}=\sqrt d\,Q_yT_D\).
This proves equality of the compressions to \(\mathcal F_j\).
They are also restrictions: the right side preserves \(\mathcal F_j\), and
the size-\((j+1)\) fixed-input frame commutes with the size-\(j\)
fixed-input frame. To see the latter commutation, write their full-space
kernels as convolutions supported on the subgroup fixing \(x\).
The coefficients depend only on the number of other fixed inputs of
the relative permutation and are class functions on that subgroup.
Convolution by two such functions commutes. Restricting to degree \(j\)
preserves this relation.
\end{proof}

\begin{lemma}[Downward return]\label{phase:down-bound}
For \(h=j+1\) satisfying \(3h+2\le N\),
\begin{equation}
 \|G_j Q G_h^{-1}-Q\|\le\frac4{N-h},
 \qquad \|D^-_{z,h}\|\le\frac4{N-h}.
 \label{phase:down-bound-eq}
\end{equation}
\end{lemma}
\begin{proof}
Set \(R=J_{h,x}G_h^{-1}\), \(S=J_x^{[h;j]}G_j^{-1}\), and
\(\Delta=G_j QG_h^{-1}-Q\).
Lemma~\ref{pr:lower-frame} and degree-wise invariance under the Gram
operators show that \(\Delta\) vanishes on \(\mathcal F_h\) and takes values in
\(\mathcal F_j\). By \eqref{pr:intertwining} with \(s=h\),
\begin{equation}
 S\Delta=Q(R-I)-(S-I)Q.
 \label{pr:ratio}
\end{equation}
The point is to estimate \(S\) on the actual range of \(\Delta\),
rather than comparing unrelated eigenvalue lists across degrees.

On \(\mathcal F_j\), Proposition~\ref{px:fresh-conclusion} gives
\([A_y,G_j]=0\). Hence the preceding frame identity reduces to
\(S=d\sum_y A_y^2\), \(d=N-j\). Each \(A_y\) is positive,
\(\sum_y A_y=I\), and every nonzero eigenvalue \(a\) satisfies
\(1\le da\le1+\varepsilon_j\). Functional calculus gives
\(
 A_y\preceq dA_y^2\preceq(1+\varepsilon_j)A_y
\);
summing and taking norms yields
\begin{equation}
 I\preceq S|_{\mathcal F_j}\preceq(1+\varepsilon_j)I,\qquad
 \|w\|\le\|Sw\|\quad(w\in \mathcal F_j).
 \label{pr:coercive}
\end{equation}
For \(v\in \mathcal F_h^\perp\), use \(\|Q\|\le1\),
\(\|(R-I)v\|\le\eta_h\|v\|\), and \(Qv,\Delta v\in \mathcal F_j\) in
\eqref{pr:ratio}--\eqref{pr:coercive} to get
\(\|\Delta v\|\le(\eta_h+\varepsilon_j)\|v\|\).
Since \(\Delta\) annihilates \(\mathcal F_h\), this is its full norm bound.
Here \(\eta_h=1/(N-2j)\), \(\varepsilon_j\le1/(N-2j)\), and
\(2/(N-2j)\le4/(N-j-1)\) by \(3h+2\le N\).
Finally use \eqref{phase:down-identity} and \(\|W_j\|\le1\).
\end{proof}

\begin{theorem}[Complete-phase canonical return]\label{phase:return}
If \(3D+2\le N\), then
\begin{equation}
 \|\mathcal RF_zW\Pi_{\le D}\|
 =\|(KF_zW-WU_z)\Pi_{\le D}\|
 \le\frac{e+5}{N-2D}.
 \label{phase:return-bound}
\end{equation}
\end{theorem}
\begin{proof}
The canonical upward correction vanishes by the exact query identities.
Among the middle blocks, different source degrees have orthogonal
target degrees; the same holds among the downward blocks. Thus the
norm of either family is its largest block norm. By
\eqref{phase:return-components} the two families together cost at most
\((e+1)/(N-2D)+4/(N-2D)\). Coherent controls by the queried input or
by the phase vector give an orthogonal direct sum of such operators.
\end{proof}

\subsubsection{An intrinsic square-root barrier for the unfiltered process}
\label{phase:unfiltered-barrier-section}

The following obstruction concerns the unfiltered compressed evolution,
not merely the accumulation of a particular one-query error estimate.
It is witnessed by an algorithm that measures only its own registers.

\begin{proposition}[Unfiltered CPO has a square-root soundness barrier]
\label{phase:unfiltered-barrier}
Let $N$ be even. Fix signs $z_y\in\{-1,1\}$ satisfying
$\sum_{y=1}^N z_y=0$, and let $F_x=C_xL_zC_x$ be the corresponding
complete compressed phase-query block on square partial-injection
databases. For every odd integer $s\ge1$ with $s^2\le N/24$,
there is an algorithm using $s$ such blocks whose acceptance
probability is zero for every real permutation, but is at least
\begin{equation}
 p_{\rm CPO}\ge \frac{s^2}{8N}
 \label{phase:unfiltered-barrier-probability}
\end{equation}
for the unfiltered CPO initialized at the empty database.
Each block can be implemented with two forward-swap queries and
a known diagonal response operation.
\end{proposition}

\begin{proof}
Let $\ket u=N^{-1/2}\sum_x\ket x$ be the uniform query-input state,
and put $P=\ket u\bra u\otimes I$ and $R=2P-I$.
Both the reflection $R$ and the final measurement $\{P,I-P\}$
act only on the algorithm's input register. Write
\[
 F=\sum_x\ket x\bra x\otimes F_x,
 \qquad B=\bra uF\ket u=\frac1N\sum_xF_x.
\]
The operator $F$ is a self-adjoint unitary, and $B$ is a
self-adjoint contraction. Starting from $\ket u\ket\varnothing$,
the algorithm applies $(RF)^s$ and accepts upon outcome $P$.

For a real permutation $\pi$, the phase query is
$U_\pi\ket x=z_{\pi(x)}\ket x$. Hence
$\bra uU_\pi\ket u=N^{-1}\sum_yz_y=0$.
The identity below therefore shows that the real acceptance
probability is zero for every odd $s$.

To calculate the compressed acceptance probability, let
$T_j$ be the polynomials defined by
$T_0(b)=1$, $T_1(b)=b$ and
$T_{j+1}(b)=2bT_j(b)-T_{j-1}(b)$.
For $V=RF$ we have
$P(V+V^\dagger)=2PFP=2BP$, with $B$ identified with its action
on $\ran P$. Since $V^\dagger=V^{-1}$, multiplication by $V^j$
gives the recurrence
\[
 PV^{j+1}P+PV^{j-1}P=2B\,PV^jP.
\]
The initial values imply $PV^jP=T_j(B)$.
If $\nu$ is the spectral measure of $B$ in $\ket\varnothing$,
then, for odd $s$,
\begin{equation}
 p_{\rm CPO}
 =\norm{T_s(B)\ket\varnothing}^2
 =\int_{-1}^1\sin^2(s\arcsin b)\,d\nu(b).
 \label{phase:unfiltered-spectral-probability}
\end{equation}
Here $T_j(b)=\cos(j\arccos b)$ follows from the same recurrence.

We next compute two moments of $\nu$. Write $\ket{x,y}$ for
the database containing just $x\mapsto y$. The local reflection
and the balanced-sign condition give
\[
 B\ket\varnothing=N^{-3/2}\sum_{x,y}z_y\ket{x,y},
 \qquad \int b^2\,d\nu(b)=\frac1N.
\]
For a size-two matching $D=\{(x_1,y_1),(x_2,y_2)\}$, put
$z(D)=z_{y_1}z_{y_2}$. A second application gives
\begin{align}
 B^2\ket\varnothing
 ={}&\frac1N\ket\varnothing
     -\frac1{N^{5/2}}\sum_{x,y}\ket{x,y}\notag\\
 &+\frac{2}{N^{5/2}\sqrt{N-1}}
       \sum_{|D|=2}\left(z(D)+\frac1{N-1}\right)\ket D.
 \label{phase:unfiltered-second-image}
\end{align}
Indeed, querying the input already present in the signed
one-record row gives
$\sum_yz_y F_x\ket{x,y}=\sqrt N\ket\varnothing$.
For $w\ne x$, the mean of the signs available after recording
$x\mapsto y$ is $-z_y/(N-1)$, so
\[
 F_w\ket{x,y}
 =-\frac{z_y}{N-1}\ket{x,y}
 +\frac1{\sqrt{N-1}}\sum_{v\ne y}
    \left(z_v+\frac{z_y}{N-1}\right)
    \ket{\{(x,y),(w,v)\}}.
\]
The two possible insertion orders give the size-two coefficient
in~\eqref{phase:unfiltered-second-image}. Since
$\sum_{y_1\ne y_2}z_{y_1}z_{y_2}=-N$, counting the
$\binom N2$ input pairs yields
\[
 \sum_{|D|=2}\left(z(D)+\frac1{N-1}\right)^2
 =\frac{N^3(N-2)}2.
\]
The three sizes in~\eqref{phase:unfiltered-second-image} are
orthogonal. Consequently
\begin{equation}
 \int b^4\,d\nu(b)
 =\frac1{N^2}+\frac1{N^3}
      +\frac{2(N-2)}{N^2(N-1)}
 \le\frac3{N^2}.
 \label{phase:unfiltered-fourth-moment}
\end{equation}

For $|b|\le1/(2s)$, the inequalities
$|b|\le|\arcsin b|\le2|b|$ give $|s\arcsin b|\le1$.
Using $|\sin x|\ge|x|/2$ for $|x|\le1$, we obtain
$\sin^2(s\arcsin b)\ge s^2b^2/4$.
The two moment estimates imply
\[
 \int_{|b|\le1/(2s)}b^2\,d\nu(b)
 \ge\frac1N-4s^2\int b^4\,d\nu(b)
 \ge\frac1N\left(1-\frac{12s^2}N\right)
 \ge\frac1{2N}.
\]
Substitution into~\eqref{phase:unfiltered-spectral-probability}
proves~\eqref{phase:unfiltered-barrier-probability}.
\end{proof}

\begin{corollary}[Optimal scale of uniform state comparison]
\label{phase:unfiltered-barrier-norm}
For the algorithm above, let $\zeta_s$ be its unfiltered CPO state
and $\psi_s$ its real purified state. For every isometric encoding
$V_{\rm db}$ acting only on the real reference register,
\[
 \norm{\zeta_s-(I\otimes V_{\rm db})\psi_s}
 \ge\frac{s}{\sqrt{8N}}.
\]
In particular, a soundness comparison uniform over these algorithms
cannot have vanishing error throughout the scale $s=\Theta(\sqrt N)$.
\end{corollary}
\begin{proof}
The real final state has $P\psi_s=0$, and a reference-only
encoding commutes with $P$. Projecting the difference onto $P$
and applying Proposition~\ref{phase:unfiltered-barrier} gives
the norm bound. Choose odd $s$ equal to a sufficiently small
constant multiple of $\sqrt N$ to obtain a constant lower bound.
\end{proof}

\begin{remark}[The obstruction persists for complete function queries]
\label{phase:unfiltered-function-barrier}
The construction does not require revealing permutation images to
the algorithm. For pure type $(k^n)$ with even $n$ and alphabet
$\mathbb Z_n$, the character $c=n/2$ gives the phase
$(-1)^{f(x)}$. Exactly half the $N=kn$ slots have each sign,
for every injective label table. The same algorithm and proof
therefore apply to unfiltered complete function-query blocks.
The filtered comparison avoids this obstruction by replacing each
block with its harmonic compression. It does not assert a smaller
error for the original unfiltered evolution.
In this example, the initial averaged component is
$B\ket\varnothing=N^{-3/2}\sum_{x,y}z_y\ket{x,y}$.
For each fixed $y$ its coefficients are constant in $x$, so it
belongs to the range of the adjoint image-deletion map. Thus
$K_1B\ket\varnothing=0$: the harmonic projection removes the
component that initiates this distinguishing experiment.
\end{remark}

\subsubsection{Comparison along a filtered history}

Let \(\psi_t\) denote the real state after \(t\) queries and let
\(\mathcal U_t\) be the oracle-independent unitary before query \(t+1\).
With \(F_t\) the complete compressed query at this step, define
\begin{equation}
 \xi_0=\ket\varnothing,\qquad
 \xi_{t+1}=KF_t\mathcal U_t\xi_t .
 \label{phase:filtered-process}
\end{equation}
The initial vector includes the same algorithm and reference registers
as the real initial state. The maps \(K,W,\widetilde V\) act only on
the oracle register; they commute with the intervening free unitaries.
Thus \(\|\xi_t\|\le1\), even though the states are not renormalized.

\begin{corollary}[Filtered history comparison]\label{phase:filtered-comparison}
For \(3T+2\le N\),
\begin{align}
 \|\xi_T-W\psi_T\|&\le(e+5)T/(N-2T),\label{phase:filtered-W}\\
 \|\xi_T-\widetilde V\psi_T\|
 &\le(e+5)T/(N-2T)+T/[2(N-2T+1)].\label{phase:filtered-isometry}
\end{align}
In particular the error is \(O(T/N)\) for \(T\le N/4,\ N\ge8\).
\end{corollary}
\begin{proof}
The initial real permutation state has degree zero and is sent to the
empty database by both encodings. Just before query \(t+1\) the real
state has degree at most \(t\); the free unitary preserves this cutoff.
Theorem~\ref{phase:return} therefore bounds the new discrepancy against
\(W\psi_{t+1}\) by \((e+5)/(N-2t)\), in addition to the preceding
discrepancy propagated by the contraction \(KF_t\mathcal U_t\).
Sum these bounds for \(0\le t<T\) and use the endpoint estimate
\eqref{pf:endpoint} once at the final degree. This proves
both inequalities, including \(T=0\).
\end{proof}

\subsubsection{The unprojected phase and the coherent response}

A terminal read is performed without projecting back to the canonical
subspace. To handle it we need a one-step estimate for the part discarded
by \(K\). There is a short way to obtain this estimate from the blocks
already analyzed: \(W^\dagger W=G^{-1}\), so a change in the norm of
an encoded vector is governed by the change in the inverse Gram operator.
Here \(G^{-1}\) denotes \(\bigoplus_jG_j^{-1}\) on the real degree
decomposition.

\begin{lemma}[Inverse Gram under one query]\label{pr:inverse-gram}
For \(3D+5\le N\),
\begin{equation}
 \|(G^{-1}U_z-U_zG^{-1})\Pi_{\le D}\|
 \le\frac{10}{N-2D-2}.
 \label{pr:gram-comm}
\end{equation}
\end{lemma}
\begin{proof}
Only the middle, downward and upward blocks occur. On degree \(h\)
the middle block is \([G_h^{-1},A_z]\). It vanishes on \(\mathcal F_h\),
by the fresh Gram commutation in Proposition~\ref{px:fresh-conclusion}.
On \(\mathcal F_h^\perp\), use an inverse of \(J=J_{h,x}\) \emph{only on this
subspace}. Since \(G_h\succeq I\) and \(JG_h^{-1}\succeq(1-\eta_h)I\)
there, commutation implies \(J\succeq(1-\eta_h)I\). Also \([J,A_z]=0\).
Consequently
\[
 \|G_h^{-1}-J^{-1}\|
 =\|J^{-1}(JG_h^{-1}-I)\|
 \le\frac{\eta_h}{1-\eta_h},\qquad
 \|[G_h^{-1},A_z]\|\le\frac{2}{N-2h+1}.
\]
For the block \(Q=\Pi_{h-1}U_z\Pi_h\) multiply
\eqref{phase:down-bound-eq} by \(G_{h-1}^{-1}\) to obtain
\(\|G_{h-1}^{-1}Q-QG_h^{-1}\|\le4/(N-h)\).
The upward block from \(h\) to \(h+1\) is the adjoint, up to a sign,
of this commutator bound for the downward query with conjugated
phases, at source degree \(h+1\). Its norm is at most \(4/(N-h-1)\).
This is why we require stability also at \(D+1\).
Within each of the three families different sources give orthogonal
targets. Summing their three suprema for \(h\le D\) proves
\eqref{pr:gram-comm}.
\end{proof}

\begin{proposition}[Outgoing square leakage]\label{pr:leak}
If \(3D+5\le N\), then
\[
 \|(I-K)F_zW\Pi_{\le D}\|\le\frac8{\sqrt{N-D}},\qquad
 \|(I-K)F_zK L_{\le D}\|\le\frac{8\sqrt e}{\sqrt{N-D}} .
\]
Also \(\|(F_zW-WU_z)\Pi_{\le D}\|\le 28/\sqrt{N-D}\).
\end{proposition}
\begin{proof}
Fix a unit vector \(\psi\in\mathcal H_{\le D}\), and set
\(r=KF_zW\psi-WU_z\psi\). By Theorem~\ref{phase:return},
\(\|r\|\le(e+5)/(N-2D)\).
Pythagoras, unitarity of \(F_z\), and \(\|WU_z\psi\|\le1\) give
\begin{align*}
 \|(I-K)F_zW\psi\|^2
 &= \|W\psi\|^2-\|WU_z\psi+r\|^2\\
 &\le
 \langle\psi,(G^{-1}-U_z^\dagger G^{-1}U_z)\psi\rangle
       +2\|r\|
 \le\frac{26}{N-2D-2}.
\end{align*}
For the final inequality apply Lemma~\ref{pr:inverse-gram} and
\(e+5<8\). Since \(N-D\le2(N-2D-2)\), the first claimed bound follows.
For the second, write a canonical vector \(a= W\psi\);
\(\|\psi\|\le\sqrt e\|a\|\) on the specified degrees.
Finally add \(\|r\|\) to the first bound. Here
\(\sqrt{N-D}\le N-2D\), since \(N-D\le2(N-2D-2)\)
and \(N-2D\ge 2\). This gives even \(16/\sqrt{N-D}\);
the stated constant leaves room in the following estimates.
\end{proof}

Introduce an answer register with orthonormal basis indexed by
\(\{\perp\}\cup[N]\); \(S_y\) exchanges \(\perp\) and \(y\) and
fixes the other entries. The real forward read is
\(\mathsf P=\sum_yS_y\otimes Q_y\).
The compressed forward read \(\mathsf Q\) decompresses by \(C_x\),
applies \(S_y\) if the decompressed record at \(x\) is \(y\)
(and the identity if there is no record), and recompresses.
We need a bound for an arbitrary initial response: the verifier can
have retained earlier responses and be entangled with them.

\begin{proposition}[Comparison for a coherent forward read]\label{pr:swap}
For \(3D+5\le N\),
\[
 \|(\mathsf Q\widetilde V-\widetilde V\mathsf P)
                     (I\otimes\Pi_{\le D})\|
 \le\frac{60}{\sqrt{N-D}}+\frac{D+1}{N-2D-1}.
\]
The bound is uniform in the input \(x\), including coherent controls.
\end{proposition}
\begin{proof}
First use \(W\) in both encodings and define outcome defects
\[
 B_y=(M_y W-WQ_y)\big|_{\mathcal H_{\le D}} .
\]
The decompressed unused-answer projector annihilates \(W\), by
the domain harmonicity relations. Hence
\(\sum_yB_y=0\) and, for signs \(\sigma_y\in\{-1,1\}\),
\(\sum_y\sigma_y B_y=(F_\sigma W-WU_\sigma)|_{\mathcal H_{\le D}}\).
Proposition~\ref{pr:leak} gives this signed sum norm at most
\(\delta=28/\sqrt{N-D}\), for every sign choice.

The full response defect with \(W\) is \(\sum_y S_y\otimes B_y\).
On a vector \((v_\perp,(v_y)_y)\) its components are
\[
 \left(\sum_yB_yv_y,\ (B_yv_\perp-B_yv_y)_y\right),
\]
where the minus sign uses \(\sum_yB_y=0\).
For independent uniform signs, expanding the squared norm gives
\[
 \sum_y\|B_yv\|^2
 =\mathbb E_\sigma\left\|\sum_y\sigma_y B_yv\right\|^2
 \le\delta^2\|v\|^2 .
\]
The same holds for \(B_y^\dagger\) by taking adjoints of the signed
sums. Thus both the column \(v\mapsto(B_yv)_y\) and the row
\((v_y)_y\mapsto\sum_yB_yv_y\) are contractions up to the factor
\(\delta\). The off-diagonal part of the displayed response defect
has norm at most \(\delta\), since its row and column have disjoint
input and output summands. The diagonal part has norm at most
\(\max_y\|B_y\|\le\delta\). This proves the comparison with \(W\)
with error \(2\delta\), independently of the number of outcomes.

Replace \(W\) by \(\widetilde V\) once on each side. The starting
real degree is at most \(D\) and the final one at most \(D+1\).
The two reads are unitary, and the endpoint bound
\eqref{pf:endpoint} contributes at most
\[
 \frac D{2(N-2D+1)}+
 \frac{D+1}{2(N-2D-1)}\le\frac{D+1}{N-2D-1}.
\]
Use \(2\delta\le60/\sqrt{N-D}\). Direct sums over the input and
the phase controls preserve each bound.
\end{proof}

The two comparisons play different roles. We sum the canonical
return, of order \(1/(N-2D)\), throughout the algorithm. The
\(1/\sqrt{N-D}\) comparison is used only for a bounded number of
terminal reads. Summing the latter error throughout the history
would lose the range needed for the lower bound.

\section{Capacity extension and harmonic projections}\label{sec:capacity-extension}

\subsection{The weighted capacity-extension map}
\label{cap:map}

The target has one more slot than the auxiliary database. Carrying an
assignment across unchanged preserves all observed function labels, but
changes the number of physical databases realizing the assignment.
The square root of the ratio of these orbit sizes determines the transfer weight.
Throughout this subsection we work on the normalized bucket-orbit spaces of
Section~\ref{cap:orbits}.  All identities extend by the identity on
the algorithm registers. Let $P_*$ be the diagonal projector which, in
mark branch $b$, tests whether all $k+1$ slots of the marked bucket are
recorded. Write $C_j(d)=\#\{i:o_i(d)\ge j\}$ for a feasible
bucket assignment, and use the same notation on physical databases
via their induced assignment. Write $\widehat C_j$ for the diagonal
operator with these eigenvalues; the underlying space (source, or
one marked target branch) will be clear from its argument.

\begin{definition}[Capacity extension]
For a source orbit assignment $d$ and a mark $b$, let
\begin{equation}
 J_b\ket{\bar d}_{\kappa^-}
 =r_{o_b(d)}\ket{\bar d}_{\kappa^{+,b}},
 \qquad
 r_h=\sqrt{\frac{(k+1)_h}{(k)_h}}
     =\sqrt{\frac{k+1}{k+1-h}},
 \quad 0\le h\le k.
 \label{cap:global-Jb}
\end{equation}
Define
\begin{equation}
 Jv=\frac1{\sqrt m}\sum_{b=1}^m\ket b\otimes J_bv.
 \label{cap:global-J}
\end{equation}
On the physical space this means first extracting the source orbit
coefficients and then lifting the weighted coefficients to the target
orbits. Explicitly, if $E_-$ and $E_{+,b}$ send feasible assignment
coordinates to their normalized physical orbit vectors, the physical
map in branch $b$ is $E_{+,b} J_b E_-^\dagger$, with $J_b$ on the
right understood in assignment coordinates. It vanishes on the
orthogonal complement of the source orbit space. It leaves a retained
injective label table unchanged.
\end{definition}

\begin{lemma}[Global properties]
\label{cap:global-properties}
The map $J$ sends the auxiliary initial state to the target initial
state. On the source orbit subspace,
\begin{equation}
 I\le J^\dagger J\le(k+1)I,
 \qquad P_*J=0,
 \qquad L^+J=JL^-.
 \label{cap:global-properties-equations}
\end{equation}
These statements hold with arbitrary auxiliary registers and with a
coherently retained injective function-label table.
\end{lemma}
\begin{proof}
Different source bucket assignments have orthogonal images, and
\[
 J^\dagger J\ket{\bar d}
 =\left(\frac1m\sum_b\frac{k+1}{k+1-o_b(d)}\right)\ket{\bar d}.
\]
Each summand lies between $1$ and $k+1$.  For the empty assignment,
all weights are $1$, which gives the uniform target mark and empty
database.  A source assignment has $o_b(d)\le k$, so its image never
contains $k+1$ records in the marked bucket.  Finally, $L^-$ and $L^+$ denote the lookup part of a complete query
in the source and the marked target, respectively. Its phase depends
only on the algorithm control and, when $x$ is recorded, on its
bucket label.  These are unchanged by every $J_b$, proving the
intertwining identity.
\end{proof}

\par\medskip\noindent\textbf{Relation to the hidden transposition.}\enspace
Fix a mark $b$ and relabel its target slots so the unmarked buckets
agree with those of the source. Write the marked target bucket as
$B_b\cup\{s_0\}$, with $|B_b|=k$.
Embed a source orbit with $h$ marked records into the subspace with
no record occupying the added slot.  Let $\Sym_b$ average permutations of these
$k+1$ slots.  The fraction of the target orbit that avoids $s_0$ is
\begin{equation}
 p_h=\frac{(k)_h}{(k+1)_h}=\frac{k+1-h}{k+1}.
 \label{cap:avoid-singleton}
\end{equation}
Hence $\Sym_b$ sends the embedded source vector to $\sqrt{p_h}$
times the normalized target orbit, whereas $J_b$ sends it to
$p_h^{-1/2}$ times that orbit.  Equivalently, on this source orbit,
$J_b=p_h^{-1}\Sym_b$ after the natural embedding.
This is the normalization supplied by the hidden-transposition
argument; it is not an unweighted averaging map.

\subsubsection{The local capacity-comparison estimate}
\label{cap:local-comparison}

We first estimate the failure of the capacity-extension map to
intertwine the two local decompression operators.  This estimate is
stable under tensoring with arbitrary auxiliary registers.

Fix integers $k\ge2$ and $m\ge1$, and put ${N_-}=km$ and $N={N_-}+1$.
Let $H$ be a source database after deletion of the queried input.
Write
\[
 h_i=o_i(H),\qquad a_i=k-h_i\quad(1\le i\le m),\qquad
 s={N_-}-|H|=\sum_{i=1}^{m}a_i,\qquad d=s+1=N-|H|.
\]
Here $a_i$ denotes an available capacity, not a function label.
For the local comparison, assume $|H|\le N/3$.  In particular,
$s\ge1$, $d\ge2$, and $m\le d$.  Let
\[
 I_H=\{i\in[m]:a_i>0\},\qquad
 F_H=\#\{i\in[m]:a_i=0\}=C_k(H).
\]

The source local space has orthonormal basis
$\ket v$ and $\ket i$ for $i\in I_H$, where $\ket v$ denotes the
database with the queried input unrecorded.  For a target mark $b$,
the target local space has orthonormal basis $\ket v$ and $\ket i$
for $i\in I_H\cup\{b\}$.  The corresponding uniform extension
vectors in the normalized orbit bases are
\begin{equation}
 u^- =\sum_{i\in I_H}\sqrt{\frac{a_i}{s}}\,\ket i,
 \qquad
 u_b^+ =\sum_{i\in I_H\cup\{b\}}
          \sqrt{\frac{a_i+\one_{\{i=b\}}}{d}}\,\ket i.
 \label{cap:local-extension-vectors}
\end{equation}
Both vectors are unit vectors orthogonal to $\ket v$. All
coordinates excluded by capacity have zero amplitude. A deleted
assignment that is itself infeasible in the source likewise has no
source component. These facts also hold just after a source query:
each local query preserves the span of feasible extensions.
For such a vector $u$, define
\[
 C(u)=I-\ket v\bra v-\ket u\bra u
             +\ket v\bra u+\ket u\bra v.
\]
Thus $C(u)$ exchanges $\ket v$ and $u$ and fixes their orthogonal
complement.  We use $C^-=C(u^-)$ and $C_b^+=C(u_b^+)$ on their
respective local spaces.  Identical basis labels in different local
spaces are understood through their canonical coordinate embeddings.

On this local space the capacity-extension map is
\begin{equation}
 \begin{aligned}
  J_b\ket v&=r_b\ket v,
       &r_b&=\sqrt{\frac{k+1}{a_b+1}},\\
  J_b\ket i&=r_b\ket i &&(i\in I_H,\ i\ne b),\\
  J_b\ket b&=r_b\sqrt{\frac{a_b+1}{a_b}}\,\ket b
       &&(a_b>0).
 \end{aligned}
 \label{cap:local-J}
\end{equation}
When $a_b=0$, the source has no $\ket b$ coordinate, so the last
line is absent.  In particular, no division by $a_b$ is used in that
case.  Set
\[
 D_b=C_b^+J_b-J_bC^-.
\]

\begin{lemma}[Local decompression defect]
\label{cap:local-decompression-defect}
For every auxiliary Hilbert space $\mathcal A$ and every vector
$\varphi$ in the source local space tensored with $\mathcal A$,
\begin{equation}
 \frac1m\sum_{b=1}^{m}
       \norm{(D_b\otimes I_{\mathcal A})\varphi}^{2}
 \le
 \frac{32(k+1)(1+C_k(H))}{md}\,\norm{\varphi}^{2}.
 \label{cap:local-defect-bound}
\end{equation}
\end{lemma}

\begin{proof}
We suppress identity operators on $\mathcal A$.  Put
\[
 \rho=\sqrt{\frac d{s}},\qquad
 \alpha=1-\rho^{-1}.
\]
Since $s=d-1$ and $d\ge2$,
\begin{equation}
 0\le\alpha\le\frac1d,
 \qquad
 0\le\rho-1\le\frac2d,
 \qquad
 r_b^2\le k+1.
 \label{cap:local-elementary-bounds}
\end{equation}
Define a vector in the source local space and a vector in the target
local space, respectively, by
\[
 p_b=J_b^\dagger u_b^+-r_bu^- ,\qquad
 q_b=J_bu^- -r_bu_b^+ .
\]
Expanding the two decompression operators, and using
$J_b\ket v=r_b\ket v$ and $J_b^\dagger\ket v=r_b\ket v$, gives
the exact identity
\begin{equation}
 D_b=(\ket v-\ket{u_b^+})\bra{p_b}
            +\ket{q_b}(\bra{u^-}-\bra v).
 \label{cap:local-rank-two}
\end{equation}

For $a_b>0$, direct substitution in
\eqref{cap:local-extension-vectors}--\eqref{cap:local-J} yields
\begin{equation}
 \begin{aligned}
  J_bu^-&=r_b\rho\,u_b^+,\\
  p_b&=-r_b\alpha\,u^-
             +\frac{r_b}{\sqrt{a_bd}}\ket b,\\
  q_b&=r_b(\rho-1)u_b^+.
 \end{aligned}
 \label{cap:local-nonfull-case}
\end{equation}
If $a_b=0$, the target has one additional coordinate $\ket b$,
orthogonal to the embedded source space, and instead
\begin{equation}
 \begin{aligned}
  u_b^+&=\rho^{-1}u^-+d^{-1/2}\ket b,\\
  p_b&=-r_b\alpha\,u^-,\\
  q_b&=r_b\bigl(\alpha u^- -d^{-1/2}\ket b\bigr),
       \qquad r_b^2=k+1.
 \end{aligned}
 \label{cap:local-full-case}
\end{equation}
The orthogonality of $u^-$ and the new coordinate $\ket b$ in the
last case is essential for the estimate below.

Write the auxiliary-register-valued coordinates of the input as
\[
 \varphi=\ket v\otimes\varphi_v+
              \sum_{i\in I_H}\ket i\otimes\varphi_i,
 \qquad
 \xi=(\bra{u^-}\otimes I_{\mathcal A})\varphi.
\]
Then $\norm\xi\le\norm\varphi$ and
$\sum_{i\in I_H}\norm{\varphi_i}^2\le\norm\varphi^2$.
Equations~\eqref{cap:local-nonfull-case} and
\eqref{cap:local-full-case} imply
\[
 (\bra{p_b}\otimes I_{\mathcal A})\varphi
 =-r_b\alpha\,\xi+
   \begin{cases}
    \displaystyle\frac{r_b}{\sqrt{a_bd}}\varphi_b,&a_b>0,\\[4pt]
    0,&a_b=0.
   \end{cases}
\]
Using $\norm{x+y}^2\le2\norm x^2+2\norm y^2$, we obtain
\begin{equation}
 \begin{aligned}
 \frac1m\sum_b
   \norm{(\bra{p_b}\otimes I_{\mathcal A})\varphi}^2
 &\le 2(k+1)\alpha^2\norm\xi^2
       +\frac{2(k+1)}{md}
           \sum_{b\in I_H}\frac{\norm{\varphi_b}^2}{a_b}\\
 &\le 2(k+1)\left(\frac1{d^2}+\frac1{md}\right)
           \norm\varphi^2.
 \end{aligned}
 \label{cap:local-p-average}
\end{equation}
In particular, the second term is summed over orthogonal source
coordinates.  It is not estimated by $m$ copies of the full input norm.

For $a_b>0$, \eqref{cap:local-nonfull-case} gives
$\norm{q_b}^2\le4(k+1)/d^2$.  For $a_b=0$,
\eqref{cap:local-full-case} gives
\[
 \norm{q_b}^2=(k+1)\left(\alpha^2+\frac1d\right)
       \le(k+1)\left(\frac1{d^2}+\frac1d\right).
\]
Consequently,
\begin{equation}
 \frac1m\sum_b\norm{q_b}^2
 \le\frac{4(k+1)}{d^2}
       +\frac{(k+1)C_k(H)}{md}.
 \label{cap:local-q-average}
\end{equation}

Finally, $\norm{\ket v-\ket{u_b^+}}=\sqrt2$ and
\[
 \norm{((\bra{u^-}-\bra v)\otimes I_{\mathcal A})\varphi}
 \le\sqrt2\,\norm\varphi.
\]
Applying \eqref{cap:local-rank-two}, followed by
\eqref{cap:local-p-average} and \eqref{cap:local-q-average}, yields
\begin{equation}
 \begin{aligned}
 \frac1m\sum_b\norm{D_b\varphi}^2
 &\le
 \frac4m\sum_b
   \norm{(\bra{p_b}\otimes I_{\mathcal A})\varphi}^2
       +\frac4m\sum_b\norm{q_b}^2\norm\varphi^2\\
 &\le
 (k+1)\left(
       \frac{24}{d^2}+
       \frac{8+4C_k(H)}{md}\right)\norm\varphi^2\\
 &\le
 \frac{32(k+1)(1+C_k(H))}{md}\norm\varphi^2.
 \end{aligned}
 \label{cap:local-defect-final-calculation}
\end{equation}
The last step uses $m\le d$ and
$32+4C_k(H)\le32(1+C_k(H))$.
\end{proof}

We next pass from decompression to a complete function-query block.
The phase operation causes no additional comparison error: a source
record assigned to bucket $i$ and its image under $J_b$ have the same
function label, including when $i=b$.

\begin{corollary}[Local complete-query defect]
\label{cap:local-complete-query-defect}
Let $L^-$ and $L_b^+$ be unitary lookup operators on the source and
target local spaces, with common auxiliary registers, such that
\[
 L_b^+J_b=J_bL^-.
\]
Suppose that $L^-$ is the same source operator for every $b$.
Define
\[
 F^-=C^-L^-C^-,\qquad F_b^+=C_b^+L_b^+C_b^+.
\]
Then, for every source vector $\varphi$, including arbitrary
auxiliary registers,
\begin{equation}
 \frac1m\sum_b
   \norm{(F_b^+J_b-J_bF^-)\varphi}^2
 \le
 \frac{128(k+1)(1+C_k(H))}{md}\norm\varphi^2.
 \label{cap:local-function-defect-bound}
\end{equation}
\end{corollary}

\begin{proof}
The intertwining assumption and the definition of $D_b$ give
\begin{equation}
 F_b^+J_b-J_bF^-
     =C_b^+L_b^+D_b+D_bL^-C^-.
 \label{cap:local-full-query-identity}
\end{equation}
Take the direct sum over $b$, normalized by $m^{-1/2}$.
The first term has the same direct-sum norm as
$m^{-1/2}\bigoplus_bD_b\varphi$, because each $C_b^+L_b^+$ is
unitary.  Apply Lemma~\ref{cap:local-decompression-defect} to this
term.  Apply the same lemma to the second term with the common
source vector $L^-C^-\varphi$, whose norm equals $\norm\varphi$.
The triangle inequality gives twice the square root of the bound
in \eqref{cap:local-defect-bound}.  Squaring proves
\eqref{cap:local-function-defect-bound}.
\end{proof}

\begin{remark}
\label{cap:local-query-registers}
For a function phase query, the lookup operators act by the phase
associated with bucket $i$ on coordinate $\ket i$, and by the
identity on $\ket v$.  Thus the intertwining hypothesis holds even
when the function-label table and the phase character are retained
as quantum registers.  All preceding estimates are uniform over
these registers.  They also apply to superpositions of queried
inputs: different queried inputs and deletion databases $H$ define
orthogonal local components, and the mark register makes the
different target branches $b$ orthogonal.
\end{remark}

\subsubsection{A state-dependent complete-query defect}
\label{cap:hybrid}

The local estimate yields a state-dependent query comparison in
terms of the auxiliary collision counter.

\begin{lemma}[Global one-query comparison]
\label{cap:global-defect}
For any bucket-invariant source vector $v$ supported on
$|D|\le t\le N/3$,
\begin{equation}
 \norm{(F^+J-JF^-)v}
 \le\frac{K_k^{(0)}}N
      \sqrt{\norm v^2+\ip v{\widehat C_kv}},
 \label{cap:global-defect-bound}
\end{equation}
where $K_k^{(0)}$ depends only on $k$.
\end{lemma}
\begin{proof}
Decompose $v$ into orthogonal queried-input, phase-character, and
deleted-assignment components $v_{x,c,H}$.
Their images under the query defect remain orthogonal, including
the target mark register.  Corollary~\ref{cap:local-complete-query-defect}
therefore gives
\[
 \norm{(F^+J-JF^-)v}^2
 \le\frac{128(k+1)}{m(N-t)}
       \sum_{x,c,H}(1+C_k(H))\norm{v_{x,c,H}}^2.
\]
Within each component, $C_k(H)\le C_k(D)$.  Since
$m=(N-1)/k\ge N/(2k)$ and $N-t\ge2N/3$, the right-hand side is
at most
\[
 \frac{384k(k+1)}{N^2}
       \bigl(\norm v^2+\ip v{\widehat C_kv}\bigr).
\]
Thus, for example, $K_k^{(0)}=\sqrt{384k(k+1)}$ is sufficient.
\end{proof}

\subsection{Harmonic projections and the capacity-extension map}
\label{proj:section}

We next show that inserting canonical projections preserves the estimates
needed for the capacity comparison.  The source projection will be defined
directly on rectangular partial matchings.  In particular, this construction
does not posit a total injection from $N$ inputs to ${N_-}=N-1$ outputs.

\subsubsection{Deletion maps and square harmonicity}
\label{proj:deletion-maps}

Use the normalized deletion maps of Section~\ref{pk:section}, now with their dimensions displayed: $B^{\mathrm{dom}}_{A,B,h}=B_{\rm d}$ and $B^{\mathrm{im}}_{A,B,h}=B_{\rm i}$. As before, $\partial_{A,B,h}$ is their direct sum, $K_{A,B,h}$ projects onto the joint kernel, and $Z_{A,B,h}=I-K_{A,B,h}$. In degree zero $\partial=0$ and $K=I$.
We use superscripts $+$ and $-$ for $(A,B)=(N,N)$ and $(N,{N_-})$,
respectively, and write $K^\pm=\bigoplus_hK_h^\pm$ and
$Z^\pm=I-K^\pm$.  A subscript $\le D$ denotes the sum of the size
blocks up to $D$; $L_h^\pm$ and $L_{\le D}^\pm$ denote the
projectors onto the entire size-$h$ block and the sum of these
physical size blocks, respectively.

For square spaces the harmonic subspace is the canonical image in
each size block, as shown in the completion-frame construction.
Lemma~\ref{pk:dgap}, applied with $A=B=N$, yields the following
slightly weakened version of its gap:
\begin{equation}
 (\partial_h^+)^\dagger\partial_h^+
 \succeq\frac{N-2h+1}{N-h+1}Z_h^+
 \qquad(1\le h\le N/2).
 \label{proj:square-gap}
\end{equation}
We also need the following consequence: testing whether the output
contains a given set almost preserves harmonicity. This is specific
to a fixed set of slots; below we prove a separate estimate for a
collision counter.
 
\begin{lemma}[A fixed target and the canonical projection]
\label{proj:target-validity}
Let $R\subseteq[N]$ have size $r\ge1$, and let
$P_R=\one_{\{R\subseteq\im D\}}$ on the square database space.
For $a\in\operatorname{ran}K_h^+$ and $h\le N/2$,
\begin{equation}
 \|Z_h^+P_Ra\|^2
 \le\frac r{N-2h+1}\|P_Ra\|^2.
 \label{proj:target-validity-bound}
\end{equation}
Consequently, if $v$ is supported on sizes at most $D\le N/2$, then
\begin{equation}
 \|P_RK^+v\|
 \le\|P_Rv\|+
       \sqrt{\frac r{N-2D+1}}\|Z^+v\|.
 \label{proj:target-one-sided}
\end{equation}
\end{lemma}

\begin{proof}
The assertion is immediate when $h<r$.  For an image-deletion row
$(H,y)$, whether $H[x\mapsto y]$ covers $R$ is independent of $x$.
The image-harmonic zero sum therefore gives
$B_h^{+,\mathrm{im}}P_Ra=0$.
For a domain-deletion row $(H,x)$, the row is zero if $H$ already
covers $R$, by harmonicity, and also if $H$ misses at least two elements
of $R$.  If it misses exactly one, say $y_*$, its value is
$a_{H[x\mapsto y_*]}/\sqrt{N-h+1}$.  Each degree-$h$ database covering
$R$ occurs in exactly $r$ such rows, obtained by deleting one of its
records whose image lies in $R$.  Hence
\[
 \|\partial_h^+P_Ra\|^2
 =\frac r{N-h+1}\|P_Ra\|^2.
\]
Equation~\eqref{proj:square-gap} proves
\eqref{proj:target-validity-bound}.

Orthogonal summation over the size blocks gives
$\|Z^+P_Ra\|\le\varepsilon\|P_Ra\|$ for canonical $a$ of degree
at most $D$, where $\varepsilon=\sqrt{r/(N-2D+1)}$.
Write $a=K^+v$ and $b=Z^+v$.  Since $b$ is orthogonal to the canonical
space,
\[
 \|P_Ra\|^2
 =\operatorname{Re}\langle P_Ra,P_Rv\rangle
   -\operatorname{Re}\langle Z^+P_Ra,b\rangle
 \le\|P_Ra\|\bigl(\|P_Rv\|+\varepsilon\|b\|\bigr).
\]
Division by $\|P_Ra\|$ proves the result when this norm is nonzero;
the zero case is immediate.
\end{proof}

\subsubsection{The rectangular projection from a square embedding}
\label{proj:rectangular}

Add a distinguished slot $s_*$ to the source range, and let
$i_h:\mathcal L_{N,{N_-},h}\to\mathcal L_{N,N,h}$ be the basis embedding
whose images never use $s_*$.  In this construction only, write
$P_{\mathrm{dum}}=\one_{\{s_*\in\im D\}}$ and
$P_0=I-P_{\mathrm{dum}}$.  This one-slot event is different from the
full marked-bucket event $P_*$ used in the capacity comparison.

\begin{lemma}[Intersection and angle]
\label{proj:angle}
For $0\le h<N/2$,
\begin{equation}
 i_h\operatorname{ran}K_h^-
 =\operatorname{ran}K_h^+\cap\operatorname{ran}P_0.
 \label{proj:kernel-intersection}
\end{equation}
If $1\le h<N/2$ and $v\in\mathcal L_{N,{N_-},h}$, then
\begin{equation}
 \|Z_h^-v\|
 \le
 \left(1-\frac1{N-2h+1}\right)^{-1/2}
 \|Z_h^+i_hv\|.
 \label{proj:angle-bound}
\end{equation}
At $h=0$ both invalid subspaces are zero.
\end{lemma}

\begin{proof}
Embedding a rectangular vector adds zero coefficients for all databases
using $s_*$.  Its square zero-sum equations are exactly the rectangular
equations, together with equations whose coefficients are all zero.
The different positive normalizing factors in the domain equations do
not change their kernels.  This proves
\eqref{proj:kernel-intersection} in both directions.

On $\operatorname{ran}K_h^+$, let
$B=K_h^+P_{\mathrm{dum}}K_h^+$ and
$\varepsilon_h^2=1/(N-2h+1)$.  Lemma~\ref{proj:target-validity} gives
\[
 B-B^2\preceq\varepsilon_h^2B.
\]
Every nonzero eigenvalue of $B$ therefore lies in
$[1-\varepsilon_h^2,1]$.  It follows that
$K_h^+P_0K_h^+$ has eigenvalue $1$ precisely on the intersection in
\eqref{proj:kernel-intersection}, and its remaining eigenvalues are
at most $\varepsilon_h^2$.  The nonzero singular-value squares of
$K_h^+P_0$ and its adjoint coincide.  Consequently, for a vector $w$
in $\operatorname{ran}P_0$ orthogonal to the intersection,
\[
 \|K_h^+w\|^2\le\varepsilon_h^2\|w\|^2,
 \qquad
 \|Z_h^+w\|^2\ge(1-\varepsilon_h^2)\|w\|^2.
\]
Apply this to $w=i_hZ_h^-v$.  The vector $i_hK_h^-v$ belongs to the
square canonical space, so $Z_h^+w=Z_h^+i_hv$.  This proves
\eqref{proj:angle-bound}.
\end{proof}

\begin{corollary}[A sufficient rectangular incidence gap]
\label{proj:rect-gap}
For $1\le h<N/2$,
\begin{equation}
 (\partial_h^-)^\dagger\partial_h^-
 \succeq\frac{N-2h}{N-h+1}Z_h^-.
 \label{proj:rect-gap-bound}
\end{equation}
\end{corollary}

\begin{proof}
Apply Lemma~\ref{pk:dgap} with $A=N$ and $B=N-1$.
Its coefficient is $(N-2h+1)/(N-h)$, which is at least the
coefficient in~\eqref{proj:rect-gap-bound}.
\end{proof}

\subsubsection{Outgoing and incoming leakage}
\label{proj:leakage}

The square leakage estimate was proved for a complete phase query
on a permutation. Here we transfer it to the rectangular auxiliary
space and record its adjoint form. The estimates hold for the inverse
query, coherent control of the queried input and phase, and retained
label tables.

\begin{lemma}[Leakage of a complete query]
\label{proj:outgoing}
Suppose $3D+5\le N$ and $N-2D-1>1$.  Define
\begin{align}
 b_D^+&=\frac{8\sqrt e}{\sqrt{N-D}},
 \label{proj:square-leakage-constant}\\
 b_D^-&=\left(1-\frac1{N-2D-1}\right)^{-1/2}
 \left(b_D^++2\sqrt{\frac2{{N_-}-D}}\right).
 \label{proj:rect-leakage-constant}
\end{align}
Then
\begin{equation}
 \|Z^\pm F^\pm K^\pm_{\le D}\|\le b_D^\pm.
 \label{proj:outgoing-bound}
\end{equation}
Taking adjoints and using one-step size bandedness also gives
\begin{equation}
 \|K^\pm F^\pm Z^\pm L^\pm_{\le D}\|\le b_{D+1}^\pm
 \label{proj:incoming-bound}
\end{equation}
provided the hypotheses for the outgoing estimate also hold at $D+1$.  In particular, all these
bounds are $O(N^{-1/2})$ for $D\le N/4$ and sufficiently large $N$.
\end{lemma}

\begin{proof}
In the square case this is Proposition~\ref{pr:leak}, applied to a
unit-modulus phase, with the coherent controls retained. Its proof uses
the canonical return estimate before any capacity transfer.
 
For the rectangular bound, fix a query address and a deleted database
$H$, and put $s={N_-}-|H|$.  Give the added square slot any phase of
modulus one, for example $1$.  The square and rectangular lookup phases
then intertwine under the embedding.  Their local normalized extension
vectors satisfy
\[
 u^+=a\,iu^-+b\ket{s_*},\qquad
 a=\sqrt{\frac s{s+1}},\quad b=\frac1{\sqrt{s+1}}.
\]
Let $i=\bigoplus_h i_h$ and work in one deleted-database component.
The compression operators agree under $i$ on the source subspace
orthogonal to the vacuum $\ket v$ and $u^-$. On their span,
$(C^+i-iC^-)\ket v=u^+-iu^-$ and
$(C^+i-iC^-)u^-=iu^--a u^+-(1-a)\ket v$. Thus the two columns
have Gram matrix
\[
 \begin{pmatrix}2(1-a)&-b^2\\-b^2&2(1-a)\end{pmatrix}.
\]
Its largest eigenvalue is $2(1-a)+b^2=b^2(1+2/(1+a))$.
Since $2/(1+a)=1+(1-a)/(1+a)\le1+1/(s+1)$, this eigenvalue
is at most $2/(s+1)+1/(s+1)^2\le2/s$.  Expanding the two compression
factors, using exact intertwining of the lookup phases, therefore gives
\begin{equation}
 \|(F^+i-iF^-)L^-_{\le D}\|
 \le2\sqrt{\frac2{{N_-}-D}}.
 \label{proj:embedded-query-defect}
\end{equation}
The second compression factor stays in the same deleted-database
component, so the same lower bound on $s$ applies to both terms.

For $a\in\operatorname{ran}K^-_{\le D}$, equation
\eqref{proj:kernel-intersection} makes $ia$ square canonical.  The square
bound and \eqref{proj:embedded-query-defect} imply
\[
 \|Z^+iF^-a\|
 \le\left(b_D^++2\sqrt{\frac2{{N_-}-D}}\right)\|a\|.
\]
The rectangular output has size at most $D+1$.  Apply
Lemma~\ref{proj:angle} at that cutoff to obtain
\eqref{proj:rect-leakage-constant}.
Finally, the adjoint of an incoming block is an outgoing block for
$F^\dagger$.  A query changes size by at most one, so its canonical
source can be restricted to size at most $D+1$.  This proves
\eqref{proj:incoming-bound}.  Coherent controls and auxiliary registers
are handled by orthogonal direct sums throughout.
\end{proof}

\subsubsection{Collision counters under projection}
\label{proj:counters}

Let $\kappa$ bound the bucket capacities. We use $C_j$ and
$\widehat C_j$ as in Section~\ref{cap:map}, for $1\le j\le\kappa$.
The common harmonic kernel is invariant under every permutation of
input positions and every permutation of range slots.  Its orthogonal
projector therefore commutes with permutations within each bucket.
In particular, applying $K^\pm$ preserves the normalized bucket-orbit
subspace of Section~\ref{cap:orbits}.

\begin{lemma}[Counter boundary estimate]
\label{proj:counter-boundary}
Let $a$ be harmonic at size $h$, with range size $B$, and let
$2\le j\le\kappa$.  Then
\begin{equation}
 \|\partial_h \widehat C_ja\|^2
 \le\frac{j(\kappa-j+1)}{B-h+1}
       \langle a,\widehat C_j\widehat C_{j-1}a\rangle.
 \label{proj:counter-incidence}
\end{equation}
In the square case this implies
\begin{equation}
 \|Z_h^+\widehat C_ja\|^2
 \le\frac{j(\kappa-j+1)}{N-2h+1}
       \langle a,\widehat C_j\widehat C_{j-1}a\rangle.
 \label{proj:square-counter-boundary}
\end{equation}
In the rectangular source case it implies
\begin{equation}
 \|Z_h^-\widehat C_ja\|^2
 \le\frac{j(\kappa-j+1)(N-h+1)}{(N-h)(N-2h)}
       \langle a,\widehat C_j\widehat C_{j-1}a\rangle.
 \label{proj:rect-counter-boundary}
\end{equation}
\end{lemma}

\begin{proof}
The image-deletion component vanishes because $C_j(H[x\mapsto y])$
is independent of $x$.  In a domain-deletion row, subtract $C_j(H)$
times the harmonic row sum.  Only outputs in buckets having occupancy
exactly $j-1$ in $H$ remain.  The number of such available slots is at
most $(\kappa-j+1)C_{j-1}(H)$, so Cauchy--Schwarz bounds this row's
squared norm by that number times the sum of the corresponding squared
coefficients, divided by $B-h+1$.

For a final database $D$, these deletions remove one of the $j$ records
in a bucket of occupancy exactly $j$.  Thus $D$ is counted
$jB_{=j}(D)$ times, where $B_{=j}(D)\le C_j(D)$; also
$C_{j-1}(H)\le C_{j-1}(D)$.  Summing the rows proves
\eqref{proj:counter-incidence}.  Apply the square gap
\eqref{proj:square-gap} or the rectangular gap
\eqref{proj:rect-gap-bound} to obtain the remaining assertions.
\end{proof}

\begin{corollary}[One-sided counter comparison]
\label{proj:counter-projection}
In the rectangular source, let $v$ be supported on sizes at most
$D<N/2$, and put
\[
 a=K^-v,\quad b=Z^-v,\quad
 q_j=\langle a,\widehat C_ja\rangle,\quad
 p_j=\langle v,\widehat C_jv\rangle,\quad \delta=\|b\|.
\]
For $2\le j\le\kappa$,
\begin{equation}
 q_j\le p_j+
 2\delta\sqrt{\frac{2(\kappa-j+1)D}{N-2D}\,q_{j-1}}.
 \label{proj:counter-projection-bound}
\end{equation}
For the square projection the same statement holds with the coefficient
inside the square root replaced by
$(\kappa-j+1)D/(N-2D+1)$.
\end{corollary}

\begin{proof}
Orthogonal summation of \eqref{proj:rect-counter-boundary} over sizes
up to $D$, and $(N-h+1)/(N-h)\le2$, give
\[
 \|Z^-\widehat C_ja\|^2
 \le\frac{2j(\kappa-j+1)}{N-2D}
       \langle a,\widehat C_j\widehat C_{j-1}a\rangle
 \le\frac{2(\kappa-j+1)D}{N-2D}q_{j-1},
\]
since $C_j\le D/j$.  The exact expansion
\[
 q_j-p_j
 =-2\operatorname{Re}\langle Z^-\widehat C_ja,b\rangle
   -\langle b,\widehat C_jb\rangle
\]
and positivity of $C_j$ prove \eqref{proj:counter-projection-bound}.
Use \eqref{proj:square-counter-boundary} for the square version.
\end{proof}

\subsubsection{Compatibility of the weighted extension}
\label{j:compatibility}

We return to the capacity-extension map $J$ from
\eqref{cap:global-Jb}--\eqref{cap:global-J}.  It preserves database
size and satisfies $P_*J=0$, where $P_*$ now denotes the event that
the marked bucket contains all $k+1$ records.
We will estimate both off-diagonal blocks of $J$ relative to the two
harmonic projections.  The second estimate is necessary because $J$
is not an isometry; the first estimate alone does not imply it.

Here are the incidence equations in the normalized bucket-orbit basis.
If $H$ has size $h-1$ and available bucket capacities $a_i$, the
unnormalized domain row is
\begin{equation}
 \sum_{i:a_i>0}\sqrt{a_i}\,u_{H[x\mapsto i]},
 \label{j:orbit-dom-row}
\end{equation}
and its contribution to the squared incidence norm is divided by the
range-side normalization $B-h+1$.  The image zero sums are
\begin{equation}
 \sum_{x\notin\dom H}u_{H[x\mapsto i]}=0.
 \label{j:orbit-im-row}
\end{equation}
To verify the normalization, let
$w(H)=\prod_i(\kappa_i)_{o_i(H)}$.  A physical extension in bucket $i$
has coefficient $u_{Hi}/\sqrt{w(H)a_i}$.  Summing its $a_i$ available
slots gives $\sqrt{a_i}u_{Hi}/\sqrt{w(H)}$.  Summing squared physical
rows over the $w(H)$ realizations of $H$ cancels this last denominator.
For image rows, summing also over the $a_i$ choices of the fixed output
cancels the factor $a_i$ in the squared coefficient.  This proves
\eqref{j:orbit-dom-row} and \eqref{j:orbit-im-row}, including their
norm interpretation.

\begin{lemma}[Forward compatibility]
\label{j:forward}
For $u\in\operatorname{ran}K_h^-$ in the bucket-orbit subspace and
$h\le N/2$,
\begin{equation}
 \|Z_h^+Ju\|^2
 \le f_h^2\|u\|^2,
 \qquad
 f_h^2:=\frac{(k+1)h}{2m(N-2h+1)}.
 \label{j:forward-bound}
\end{equation}
Set $f_0=0$.
\end{lemma}

\begin{proof}
Fix a source deleted assignment $H$, a query input $x$, and a mark $b$.
Put $a_i=k-o_i(H)$ and $r=r_{o_b(H)}$.  If $a_b>0$, the target domain
row of $J_bu$ is
\[
 r\left(\sum_{i\ne b}\sqrt{a_i}\,u_{Hi}
       +\frac{a_b+1}{\sqrt{a_b}}u_{Hb}\right)
 =\frac r{\sqrt{a_b}}u_{Hb},
\]
where source harmonicity was used in the equality.  If $a_b=0$, the
row is zero: the source has no $b$-extension, and $J_b$ has zero
coefficient on the target extension that would complete the marked
bucket.  Target deleted assignments that are not source-feasible
likewise give zero rows.  All image-harmonic rows vanish because the
extension weight in $J_b$ is independent of the newly added input.

For a source database $d$ with marked occupancy $s=o_b(d)\ge1$,
each deletion of one of its $s$ marked records contributes coefficient
squared
\[
 \frac{r_{s-1}^2}{k+1-s}
 =\frac{k+1}{(k+2-s)(k+1-s)}\le\frac{k+1}{2}.
\]
There are $h$ records in all.  Summing the domain-row norms over the
orthogonal mark branches, with the factor $1/m$ from $J$, gives
\[
 \|\partial_h^+Ju\|^2
 \le\frac{(k+1)h}{2m(N-h+1)}\|u\|^2.
\]
The square incidence gap proves \eqref{j:forward-bound}.
\end{proof}

\begin{lemma}[Reverse compatibility]
\label{j:reverse}
Let $v=\bigoplus_bv^b\in\operatorname{ran}K_h^+$ belong to the
controlled target bucket-orbit subspace, and suppose $h<N/2$.
Let $C_k$ count buckets with occupancy at least $k$ in each branch.
Then
\begin{equation}
 \|Z_h^-J^\dagger v\|^2
 \le\frac{(k+1)^2(N-h+1)}{m(N-h)(N-2h)}
       \langle P_*v,\widehat C_kP_*v\rangle.
 \label{j:reverse-weighted}
\end{equation}
In particular,
\begin{equation}
 \|K_h^+JZ_h^-\|\le g_h,
 \qquad
 g_h^2:=\frac{(k+1)^2h(N-h+1)}{km(N-h)(N-2h)}.
 \label{j:reverse-bound}
\end{equation}
Here the operator is between the corresponding bucket-orbit subspaces,
and $g_0=0$.  If $P_*v=0$, then $J^\dagger v$ is exactly source harmonic.
\end{lemma}

\begin{proof}
Fix a source deleted assignment $H$ and input $x$, and again put
$a_i=k-o_i(H)$ and $r=r_{o_b(H)}$.  If $a_b>0$, the source domain row
of $J_b^\dagger v^b$ is
\[
 r\left(\sum_{i\ne b}\sqrt{a_i}\,v^b_{Hi}
               +\sqrt{a_b+1}\,v^b_{Hb}\right)=0,
\]
by target harmonicity.  If $a_b=0$, the source omits the last target
extension.  The row is consequently
$-\sqrt{k+1}\,v^b_{Hb}$.  Image harmonicity is preserved in both cases,
since the weights are independent of $x$.

Unlike the forward direction, different marks can interfere after
$J^\dagger$.  The normalized source domain row is exactly
\[
 -\sqrt{\frac{k+1}{m(N-h)}}
       \sum_{b:o_b(H)=k}v^b_{H[x\mapsto b]}.
\]
Cauchy--Schwarz bounds the squared sum by
$C_k(H)\sum_{b:o_b(H)=k}\|v^b_{H[x\mapsto b]}\|^2$.
Each target database with a full marked bucket is counted exactly
$k+1$ times, once for each deletion of a marked record.  Deleting such
a record leaves its number of buckets of occupancy at least $k$
unchanged.  Therefore
\[
 \|\partial_h^-J^\dagger v\|^2
 \le\frac{(k+1)^2}{m(N-h)}
       \langle P_*v,\widehat C_kP_*v\rangle.
\]
The rectangular gap \eqref{proj:rect-gap-bound} proves
\eqref{j:reverse-weighted}.  Since $C_k\le h/k$, taking adjoints gives
\eqref{j:reverse-bound}.  When $P_*v=0$, both deletion components
of $J^\dagger v$ vanish, proving the last assertion.
\end{proof}

\begin{corollary}[Endpoint leakage of the projected extension]
\label{j:endpoint}
Let $S=K^+JK^-$.  On source size $h<N/2$,
\begin{equation}
 \|P_*S L_h^-\|
 \le e_h,
 \qquad
 e_h^2:=\frac{(k+1)^2h}{2m(N-2h+1)^2}.
 \label{j:endpoint-bound}
\end{equation}
In particular, $e_h=O_k(\sqrt h/N^{3/2})$ for $h\le N/4$.
\end{corollary}

\begin{proof}
For $u\in\operatorname{ran}K_h^-$, apply
\eqref{proj:target-one-sided} to $Ju$, separately in each mark branch,
with $R$ its $k+1$ marked slots.  The term $P_*Ju$ vanishes, and
Lemma~\ref{j:forward} bounds the remaining term.  Thus
\[
 \|P_*K_h^+Ju\|
 \le\sqrt{\frac{k+1}{N-2h+1}}\,f_h\|u\|=e_h\|u\|.
\]
\end{proof}

\begin{lemma}[Defect between the filtered queries]
\label{j:filtered-defect}
Let $A^\pm=K^\pm F^\pm K^\pm$ and $S=K^+JK^-$.  Then
\begin{equation}
 A^+S-SA^-
 =K^+(F^+J-JF^-)K^-
  -K^+F^+Z^+JK^-
  +K^+JZ^-F^-K^-.
 \label{j:filtered-defect-identity}
\end{equation}
For $u\in\operatorname{ran}K^-_{\le D}$ in the bucket-orbit subspace,
the sum of the norms of the last two terms applied to $u$ is at most
\begin{equation}
 \bigl(b_{D+1}^+f_D+g_{D+1}b_D^-\bigr)\|u\|.
 \label{j:filtered-new-terms}
\end{equation}
Consequently, for $D\le N/4$ and sufficiently large $N$,
\begin{equation}
 \|(A^+S-SA^-)u\|
 \le\|(F^+J-JF^-)u\|
       +\frac{C_k\sqrt{D+1}}{N^{3/2}}\|u\|,
 \label{j:filtered-defect-bound}
\end{equation}
with $C_k$ depending only on $k$.
\end{lemma}

\begin{proof}
Insert $K^+=I-Z^+$ immediately before $J$ in the first product, and
$K^-=I-Z^-$ immediately after $J$ in the second.  Adding and subtracting
$K^+JF^-K^-$ gives \eqref{j:filtered-defect-identity} exactly.

The vector $Z^+Ju$ has size at most $D$.  The incoming bound
\eqref{proj:incoming-bound} and Lemma~\ref{j:forward} bound the second
term by $b_{D+1}^+f_D\|u\|$.  The vector $Z^-F^-u$ has size at most
$D+1$.  Lemma~\ref{j:reverse} and the rectangular outgoing bound give
$g_{D+1}b_D^-\|u\|$ for the third term.  Size blocks are orthogonal,
and $J,K^\pm$ preserve size, so the uniform bounds up to the indicated
cutoffs are the maxima of the displayed constants.  These constants
are increasing in their cutoffs in the range used here.

For $D\le N/4$, $f_D,g_{D+1}=O_k(\sqrt{D+1}/N)$ and
$b_D^\pm,b_{D+1}^\pm=O(N^{-1/2})$ for sufficiently large $N$.
The first term in \eqref{j:filtered-defect-identity} is bounded by the
unprojected capacity defect because $K^+$ is a contraction.
This proves \eqref{j:filtered-defect-bound}.  All steps allow arbitrary
auxiliary registers and controlled queries.  The reverse estimate
already accounts for coherent summation over the hidden mark.
\end{proof}

\section{Lower bound for the exceptional type}\label{sec:exceptional-bound}

\subsection{Filtered moments and the exceptional-bucket amplitude}
\label{hybrid:section}

We now combine the local capacity estimate with the projection
estimates.  This requires checking the moment recurrence for the
filtered source itself; a bound for an unfiltered process would not
suffice.

\subsubsection{A one-query moment estimate before projection}

We use $\widehat C_j$ for the diagonal operator with eigenvalues
$C_j(d)$ defined above. The value on a deleted assignment $H$ does
not include a record at the queried input.

\begin{lemma}[Local counter gain]
\label{hybrid:local-moment}
Let $v$ be a source vector of norm at most one, supported on database
sizes at most $D<{N_-}$, in the bucket-invariant subspace, with arbitrary
auxiliary registers.  For a complete source phase query $F^-$ and
$2\le j\le k$,
\begin{equation}
 \left|\ip{F^-v}{\widehat C_jF^-v}
                -\ip v{\widehat C_jv}\right|
 \le 2\sqrt{2(k-j+1)}
       \sqrt{\frac{\ip v{\widehat C_{j-1}v}}{{N_-}-D}}.
 \label{hybrid:local-gain}
\end{equation}
\end{lemma}
\begin{proof}
Fix a queried input, phase character, label table, and deleted
assignment $H$.  Let $b_j(H)$ count buckets with occupancy exactly
$j-1$.  On this local space,
$\widehat C_j=C_j(H)I+B$, where $B$ projects onto extensions in
those buckets.  For the normalized extension vector $u$,
\[
 p_H:=\norm{Bu}^2=
 \frac{(k-j+1)b_j(H)}{{N_-}-|H|}.
\]
Writing $v_0=\ket H$ and $w=v_0-u$, the compression reflection is
$C_x=I-\ket w\bra w$.  Direct expansion gives
\[
 [B,C_x]=\ket{Bu}\bra w-\ket w\bra{Bu},\qquad
 \norm{[B,C_x]}^2=2p_H-p_H^2\le2p_H.
\]
Indeed, the rank-two skew-adjoint operator has squared singular value
$\norm w^2\norm{Bu}^2-|\ip w{Bu}|^2=2p_H-p_H^2$.
The lookup commutes with $B$, so
$\norm{[B,F^-]}\le2\sqrt{2p_H}$.

Decompose $v$ into the orthogonal controlled-input, character, table,
and deletion components $v_H$, suppressing the other indices.
The identity
$(F^-)^\dagger\widehat C_jF^--\widehat C_j
=(F^-)^\dagger[\widehat C_j,F^-]$
and the local norm bound show that the absolute gain is at most
\[
 \frac{2\sqrt{2(k-j+1)}}{\sqrt{{N_-}-D}}
 \sum_H\sqrt{b_j(H)}\,\norm{v_H}^2.
\]
Every database in the $H$-component satisfies
$b_j(H)\le C_{j-1}(d)$.  Cauchy--Schwarz and
$\sum_H\norm{v_H}^2\le1$ give \eqref{hybrid:local-gain}.
\end{proof}

\subsubsection{The filtered recurrence}

Let the source start in the empty database and evolve as
\[
 \xi_{t+1}^-=K^-F_t^-\mathcal U_t\xi_t^-,
 \qquad \xi_0^-=\ket\varnothing,
\]
where $\mathcal U_t$ is an arbitrary algorithm-only unitary; a
tensor factor equal to the algorithm's initial vector (of norm at
most one) is implicit in $\xi_0^-$. These states have norm at most one and database size at most $t$.
We do not renormalize after projecting. The compression commutes
with every permutation within a bucket, the lookup is constant on
each bucket, and $K^-$ preserves invariance. Starting from the empty
database, the whole trajectory thus stays in the orbit space on which
the transfer estimates hold. Define the unnormalized first moment
\[
 \mu_j(t)=\ip{\xi_t^-}{\widehat C_j\xi_t^-}.
\]

\begin{proposition}[Filtered first moments]
\label{hybrid:moments}
For sufficiently large $N$ and $0\le t\le T\le N/4$,
\begin{equation}
 \mu_j(t)\le C_{k,j}\frac{t^{\alpha_j}}{N_-^{\beta_j}},
 \qquad
 \alpha_j=2-2^{1-j},\quad\beta_j=1-2^{1-j},
 \quad 1\le j\le k,
 \label{hybrid:moment-bound}
\end{equation}
where $C_{k,j}$ depends only on $k,j$.
\end{proposition}
\begin{proof}
The first layer satisfies $\mu_1(t)\le t$, since the process is
contractive and has degree at most $t$.
For $j\ge2$, put $v=F_t^-\mathcal U_t\xi_t^-$ and $D=t+1$.
The outgoing estimate of Lemma~\ref{proj:outgoing} gives
$\delta:=\norm{Z^-v}\le C/\sqrt N$.
The counter projection estimate \eqref{proj:counter-projection-bound}
and Lemma~\ref{hybrid:local-moment} imply
\begin{align}
 \mu_j(t+1)\le\mu_j(t)
 &+\frac{C_k}{\sqrt {N_-}}\sqrt{\mu_{j-1}(t)}\notag\\
 &+\frac{C_k\sqrt{t+1}}{N}\sqrt{\mu_{j-1}(t+1)}.
 \label{hybrid:refined-moment}
\end{align}
Here the projection estimate uses the already projected vector
$K^-v=\xi_{t+1}^-$.  Its same-time lower-layer term is handled by
induction on the layer, not by an assumption that this term is bounded
by the preceding-time moment.

Because $t+1\le N/4$ and ${N_-}=N-1$, increasing $C_k$ gives
\[
 \mu_j(t+1)\le\mu_j(t)+\frac{C_k}{\sqrt {N_-}}
 \left(\sqrt{\mu_{j-1}(t)}+
       \sqrt{\mu_{j-1}(t+1)}\right).
\]
Assume \eqref{hybrid:moment-bound} for layer $j-1$ throughout
$0\le t\le T$.  Since $\mu_j(0)=0$, summation and
$\sum_{s=0}^{t-1}(s+1)^{\alpha_{j-1}/2}
\le t^{1+\alpha_{j-1}/2}$ give the claim with
$C_{k,j}=2C_k\sqrt{C_{k,j-1}}$.  The identities
$\alpha_j=1+\alpha_{j-1}/2$ and
$\beta_j=(1+\beta_{j-1})/2$ complete the induction.
\end{proof}

\subsubsection{Comparison through the projected capacity map}

Run the same algorithm against the filtered target process,
$\xi_{t+1}^+=K^+F_t^+\mathcal U_t\xi_t^+$, starting with the
uniform hidden mark and empty database.  Define
\[
 S=K^+JK^-,\qquad \mathcal A_t^\pm=K^\pm F_t^\pm K^\pm.
\]
Both $\mathcal A_t^\pm$ are contractions, and
$S\xi_0^-=\xi_0^+$.

\begin{theorem}[Filtered exceptional-bucket amplitude]
\label{hybrid:theorem}
For sufficiently large $N$ and $0\le T\le N/4$,
\begin{equation}
 \norm{P_*\xi_T^+}\le C_k\left(
 \frac TN+
 \frac{T^{\,2-2^{-k}}}{N^{\,3/2-2^{-k}}}\right).
 \label{hybrid:amplitude}
\end{equation}
The estimate is uniform in the alphabet and the injective label table.
\end{theorem}
\begin{proof}
Apply Lemma~\ref{j:filtered-defect} and the raw comparison
\eqref{cap:global-defect-bound} to the canonical prequery vector
$\mathcal U_t\xi_t^-$. The unitary acts on algorithm registers
alone and therefore commutes with $S$, $K^\pm$ and the counters.
Consequently,
\[
 \norm{(\mathcal A_t^+S-S\mathcal A_t^-)\mathcal U_t\xi_t^-}
 \le\frac{C_k}{N}\sqrt{1+\mu_k(t)}
       +\frac{C_k\sqrt{t+1}}{N^{3/2}}.
\]
The equality
\[
 \xi_{t+1}^+-S\xi_{t+1}^-
 =\mathcal A_t^+\mathcal U_t(\xi_t^+-S\xi_t^-)
  +(\mathcal A_t^+S-S\mathcal A_t^-)
          \mathcal U_t\xi_t^-
\]
now telescopes without a multiplicative factor $\norm J$.
Using Proposition~\ref{hybrid:moments} yields
\begin{align*}
 \norm{\xi_T^+-S\xi_T^-}
 &\le\frac{C_k}{N}\sum_{t<T}\sqrt{1+\mu_k(t)}
       +\frac{C_k}{N^{3/2}}\sum_{t<T}\sqrt{t+1}\\
 &\le C_k\left(
 \frac TN+
 \frac{T^{\,2-2^{-k}}}{N^{\,3/2-2^{-k}}}
 +\left(\frac TN\right)^{3/2}\right).
\end{align*}
Here $1+\alpha_k/2=2-2^{-k}$ and
$1+\beta_k/2=3/2-2^{-k}$. The endpoint estimate of
Corollary~\ref{j:endpoint} gives
$\norm{P_*S\xi_T^-}=O_k(\sqrt T/N^{3/2})$.
For $1\le T\le N/4$ this term and $(T/N)^{3/2}$ are bounded by
$O_k(T/N)$.  At $T=0$ the target amplitude is zero.
The triangle inequality proves \eqref{hybrid:amplitude}.
\end{proof}

\subsection{From database amplitude to collision probability}
\label{terminal:section}

Write $r=k+1$. The database projector $P_*$ tests whether all $r$
slots of the exceptional bucket have been recorded. A successful
algorithm, however, reports inputs rather than entries of the database.
We connect the two events by reading the reported inputs twice. The
first pass stores their permutation images; the second tests whether
those stored pairs are present. Both passes are part of the proof,
and are appended only after the algorithm has chosen its output.

\subsubsection{Recording and checking a list}

Let $L_{\le d}$ project onto the span of physical target databases
with at most $d$ entries. The mark, the input and any additional
registers may be in superposition. We work on the full physical
database: a read of an individual permutation image need not preserve
the bucket-orbit subspace.

\begin{lemma}[Growth in one physical read]
\label{terminal:boundary}
For a controlled forward-swap compressed query $\mathsf Q$ and $d<N$,
\begin{equation}
 \norm{P_*\mathsf Q(\Id-P_*)L_{\le d}}
 \le 2\sqrt{2/(N-d)}.
 \label{terminal:boundary-bound}
\end{equation}
In particular, if $v=L_{\le T}v$ and $v^{(s)}$ is obtained by
appending $s$ such queries and register permutations used to store
their responses, then, for $T+s<N$,
\begin{equation}
 \norm{P_*v^{(s)}}\le
 \norm{P_*v}+2s\sqrt{2/(N-T-s)}\,\norm v .
 \label{terminal:read-growth}
\end{equation}
\end{lemma}

\begin{proof}
Fix the mark $b$, the input $x$, and a deletion database $H$ with
$x\notin\dom H$. On $\mathcal F_{x,H}$, write $s_H=N-|H|$.
The target projector is the identity if $H$ already covers $B_b^{+,b}$;
it is zero if $H$ misses at least two of its slots. If $H$ misses
exactly one slot $y$, it is the rank-one projector $Q_e$ onto
$e=\ket{H[x\mapsto y]}$. In this last case
$\ip e{C_xe}=1-1/s_H$. Since $C_x$ is a self-adjoint unitary,
writing $C_x$ in the splitting $Q_e\oplus(\Id-Q_e)$ gives
\[
 \norm{[Q_e,C_x]}=\norm{(\Id-Q_e)C_xe}
 =\sqrt{1-(1-1/s_H)^2}\le\sqrt{2/s_H}.
\]
In the other cases the commutator is zero.
The slot-controlled response swap $U_x$ commutes with $P_*$, so
\[
 [P_*,C_xU_xC_x]=[P_*,C_x]U_xC_x+C_xU_x[P_*,C_x].
\]
Every local space reached from $L_{\le d}$ has $|H|\le d$.
Taking the orthogonal sum over $(b,x,H)$ proves
\eqref{terminal:boundary-bound}; there is no factor for the number
of controls. Each read increases the database size by at most one.
Before read $i$ it is at most $T+i-1$. Split its input at $P_*$,
use the contraction of the already accepted component and apply
\eqref{terminal:boundary-bound} to the other. The register permutations
do not change the target or the norm. Summing the $s$ errors proves
\eqref{terminal:read-growth}.
\end{proof}

For $x,y\in[N]$, let $Q_{x,y}$ be the diagonal projector
testing $D(x)=y$ and let $M_{x,y}=C_x Q_{x,y}C_x$.
A physical swap with a blank response followed by selecting
response $y$ applies exactly $M_{x,y}$ to the database:
between the two occurrences of $C_x$, the swap writes $y$
precisely on the range of $Q_{x,y}$.

\begin{lemma}[Checking retained images]
\label{terminal:verifier}
Suppose a state $w=L_{\le d}w$, $d<N$, retains a reported list of
inputs $x_1,\ldots,x_r$ and first-pass responses
$y_1,\ldots,y_r$. Reject incomplete lists (any $y_i=\bot$), repeated
inputs, and lists whose decoded labels are not all equal. Read the
same inputs again into fresh response registers and accept only if
the second list equals the first. The norm of the accepted component
is at most
\begin{equation}
 \norm{P_* w}+r\sqrt{2/(N-d)}\,\norm w .
 \label{terminal:verifier-bound}
\end{equation}
\end{lemma}

\begin{proof}
On a deletion fiber with $y\notin\im H$, $Q_{x,y}$ and $M_{x,y}$
are the rank-one projections onto $e=\ket{H[x\mapsto y]}$ and
$C_x e$, respectively. Their overlap is $1-1/s_H$. For unit vectors
with overlap of modulus $a$, the difference of the two rank-one
projectors has norm $\sqrt{1-a^2}$: on their span its trace is zero
and the trace of its square is $2(1-a^2)$. If $y\in\im H$, both
projectors vanish. Taking the sum over the fibers gives
\begin{equation}
 \norm{(M_{x,y}-Q_{x,y})L_{\le d}}
 \le\sqrt{2/(N-d)}.
 \label{terminal:pair-verifier}
\end{equation}
For a fixed complete first-pass list set $M_i=M_{x_i,y_i}$ and
$Q_i=Q_{x_i,y_i}$. The selected second-pass component is
$M_r\cdots M_1w$ on that register branch. Compare it to the product
of the literal record tests:
\[
 M_r\cdots M_1-Q_r\cdots Q_1
 =\sum_{i=1}^r M_r\cdots M_{i+1}(M_i-Q_i)Q_{i-1}\cdots Q_1 .
\]
The right factors preserve database size and the left factors are
contractions. Thus
\begin{equation}
 \norm{(M_r\cdots M_1-Q_r\cdots Q_1)L_{\le d}}
 \le r\sqrt{2/(N-d)} .
 \label{terminal:list-verifier}
\end{equation}
No commutativity of the $M_i$ is needed. It remains to bound the
literal term on each nonrejected branch. If two $y_i$ coincide,
the literal product is zero, because a partial injection cannot
assign the same image to two distinct inputs. Otherwise label
injectivity puts the $r$ distinct slots in one bucket. All ordinary
buckets have only $r-1$ slots, so these slots exhaust the exceptional
bucket and the literal projector is dominated by $P_*$. The stored
register branches are orthogonal; the uniform operator bound
\eqref{terminal:list-verifier} therefore applies unchanged to their
direct sum. This proves \eqref{terminal:verifier-bound} without
conditioning or renormalizing the state.
\end{proof}

\subsubsection{Comparison through the reads}

Let $\psi$ be a real state whose permutation slices have completion
degree at most $T$ and $\norm\psi\le1$. Let $v$ be any physical
state supported on $L_{\le T}$ with $\norm v\le1$. Write
$\varepsilon_{\mathrm{ph}}=\norm{v-\widetilde V\psi}$. In
particular the completed phase-query history satisfies
\begin{equation}
 \norm{\xi_T^+-\widetilde V\psi_T}\le C T/N
 \qquad (4T\le N,\ N\ge8)
 \label{terminal:initial-comparison}
\end{equation}
by \eqref{phase:filtered-isometry}. A final workspace-only
operation, as in Lemma~\ref{model:bridge}, leaves this distance
and the $P_*$ amplitude unchanged.

For an integer $d$ with $3d+5\le N$, put
\begin{equation}
 e_N(d)=\frac{60}{\sqrt{N-d}}+\frac{d+1}{N-2d-1}.
 \label{terminal:one-swap-comparison}
\end{equation}
Proposition~\ref{pr:swap} gives a coherent forward-swap comparison with error
at most $e_N(d)\,\norm{\phi}$ on any real state $\phi$ of degree at
most $d$, including a query controlled by previously stored outcomes.
Consequently, if $s$ further swaps are performed on both sides,
$D=T+s$, and $3D+5\le N$, the resulting states obey
\begin{equation}
 \norm{v^{(s)}-\widetilde V\psi^{(s)}}
 \le\varepsilon_{\mathrm{ph}}+s e_N(D).
 \label{terminal:mixed-history-comparison}
\end{equation}
Indeed, before the $i$th read the real degree is at most $T+i-1\le D$.
At that step insert the physical query applied to the encoded real
state. Unitarity bounds the propagated discrepancy by its previous
value; the new local error is at most $e_N(D)$ since the real norm is
at most one. Swapping a response into its unused storage register
commutes with $\widetilde V$. Iterating proves
\eqref{terminal:mixed-history-comparison}. No full-history swap
estimate has been used for the first $T$ phase queries.

\begin{proposition}[Terminal success conversion]
\label{terminal:success-conversion}
Let $p$ be the Born weight for reporting $r$ distinct inputs with
equal labels in $\psi$. For $D=T+2r$ and $3D+5\le N$,
\begin{equation}
 \sqrt p\le \norm{P_*v}
      +6r\sqrt{2/(N-T-r)}
      +\norm{v-\widetilde V\psi}+2r e_N(D).
 \label{terminal:explicit-conversion}
\end{equation}
In particular, for $N\ge32(k+2)$ and $4T\le N$, the corresponding
real and filtered histories satisfy
\begin{equation}
 \sqrt p\le\norm{P_*\xi_T^+}
             +O_k\!\left(T/N+N^{-1/2}\right).
 \label{terminal:success-interface}
\end{equation}
\end{proposition}

\begin{proof}
Store both lists of $r$ swap responses, keeping every slot including
a possible blank. On a fixed real branch with permutation $\pi$,
both passes return the same list $(\pi(x_i))_{i=1}^r$.
Thus the acceptance test of Lemma~\ref{terminal:verifier} selects
exactly the original collision event on that branch. More explicitly,
if $c$ denotes the terminal register configuration (including the
mark) and $u(c,\pi)$ its amplitude, the squared norm accepted in
the real experiment is
\[
 \sum_{c,\pi}|u(c,\pi)|^2
    \mathbf 1_{\{\text{the reported inputs are distinct and
                      have equal labels on }(c,\pi)\}}=p .
\]
The final test acts on the stored registers, not on the permutation
or on the physical database. It therefore commutes with the
permutation-register isometry $\widetilde V$. Applying the test to
both sides of \eqref{terminal:mixed-history-comparison} shows that
$\sqrt p$ is at most the physical acceptance norm plus
$\norm{v-\widetilde V\psi}+2r e_N(D)$.

On the physical side let $w$ denote the state after the first $r$
reads. It has size support at most $T+r$, and
\begin{equation}
 \norm{P_*w}\le\norm{P_*v}+2r\sqrt{2/(N-T-r)}
 \label{terminal:after-reads}
\end{equation}
by Lemma~\ref{terminal:boundary}. Applying
Lemma~\ref{terminal:verifier} with $d=T+r$ adds at most another
$r\sqrt{2/(N-T-r)}$. This even gives coefficient $3r$ in place of
$6r$; we retain the looser bound \eqref{terminal:explicit-conversion}
to match the formal estimate. It is the verifier, not an assumed
commutation with $P_*$, that permits the use of the database event.

Finally $N\ge32(r+1)$ and $T\le N/4$ imply $3D+5\le N$,
$N-D\ge N/2$, and $N-2D-1\ge N/4$. Hence
$e_N(D)\le 60\sqrt{2/N}+4(T+2r+1)/N$. Substitute
\eqref{terminal:initial-comparison} into
\eqref{terminal:explicit-conversion} to get
\eqref{terminal:success-interface}.
\end{proof}

\subsubsection{The query lower bound}

We retain an explicit query--success tradeoff for the exceptional type.
\begin{theorem}[Exceptional collision search]\label{main:theorem}
Fix $k\ge2$, let $n\ge0$, and put $N=kn+k+1$.
For every $q\ge n+1$, sample $f:[N]\to\mathbb Z_q$ uniformly
from exact type $(k^n,k+1)$. If a quantum algorithm uses at most
$T$ function queries and outputs the unique $(k+1)$-collision
with probability $p$, then, for sufficiently large $N$ and
$0\le T\le N/4$,
\begin{equation}
 \sqrt p\le C_k\left(
    N^{-1/2}+\frac TN+
    \frac{T^{\,2-2^{-k}}}{N^{\,3/2-2^{-k}}}\right).
 \label{main:success}
\end{equation}
In particular, for every fixed $p_0\in(0,1)$, success probability
at least $p_0$ requires
\begin{equation}
 T=\Omega_{k,p_0}(N^{\gamma_k}),\qquad
 \gamma_k=\frac{3/2-2^{-k}}{2-2^{-k}}
         =\frac34-\frac1{4(2^{k+1}-1)}.
 \label{main:lower}
\end{equation}
The constants are independent of $q$.
\end{theorem}

\begin{proof}[Proof of Theorem~\ref{main:theorem}]
The exceptional-bucket amplitude bound \eqref{hybrid:amplitude}
and the terminal conversion \eqref{terminal:success-interface}
give, for each fixed injective label table,
\[
 \sqrt p\le C_k\left(N^{-1/2}+\frac TN
                  +\frac{T^{\,2-2^{-k}}}{N^{\,3/2-2^{-k}}}\right).
\]
The bound is uniform in the table. Squaring first and averaging
the conditional probabilities gives the same amplitude bound
for the overall experiment. The exact-type sampling and the
query-model identification in Lemma~\ref{model:bridge} give
\eqref{main:success} for the stated function problem.

Write $a=2-2^{-k}$ and $b=3/2-2^{-k}$, so that
$\gamma_k=b/a<1$. At $T\le cN^{\gamma_k}$ the bound reads
\[
 \sqrt p\le C_k\bigl(N^{-1/2}+cN^{\gamma_k-1}+c^a\bigr).
\]
Choose $c=c(k,p_0)>0$ so that $C_kc^a<\sqrt{p_0}/2$. The first
two terms tend to zero uniformly over the indicated query counts,
and $cN^{\gamma_k}\le N/4$ for large $N$. Thus success probability
$p\ge p_0$ requires \eqref{main:lower}. Any algorithm succeeding
with probability at least $p_0$ on every input of the type also
succeeds with that probability on average under its uniform
distribution. This proves the worst-case lower bound as well.
\end{proof}

\section{Lower bounds for arbitrary bounded types}
\label{g:section}
We now allow arbitrary bounded class sizes and any positive number of
eligible classes. In this section $K$ is the fixed maximum class size,
$r$ is the requested collision order, and $M$ denotes the number of
eligible classes. In particular, $M$ is not a database range size.
The lower bound depends on the intermediate tail counts $B_2,\ldots,B_r$.
Fix integers $K\ge r\ge2$. For an exact type
\[
 \lambda=(1^{n_1},\ldots,K^{n_K}),\qquad
 N=\sum_{d=1}^Kdn_d,
\]
define its tail counts and number of eligible classes by
\begin{equation}
 B_j=\sum_{d=j}^Kn_d\quad(1\le j\le K),\qquad M=B_r\ge1.
 \label{g:tails}
\end{equation}
For $2\le j\le r$ put
\begin{equation}
 F_{j,M}=
 \left(\frac{N^{2^{j-1}}
       \prod_{i=2}^{j-1}B_i^{2^{i-2}}}{M^{2^{j-2}}}
       \right)^{1/(2^j-1)},\qquad
 Q=\max_{2\le j\le r}F_{j,M}.
 \label{g:scale}
\end{equation}
Empty products are one. A class is an equality class of inputs;
a bucket is a set of hidden permutation slots representing that class.
The algorithm receives only the function oracle
$O_f|x,z\rangle=|x,z+f(x)\rangle$, with values in $\mathbb Z_q$.

\begin{theorem}[A lower bound for every bounded exact type]
\label{g:main}
Fix $p_0\in(0,1)$. Uniformly over the types in~\eqref{g:tails},
all $q\ge B_1$, and all sufficiently large $N$, finding one
$r$-collision with success probability at least $p_0$ on a uniformly
random function of type $\lambda$ requires
\[
                       \Omega_{K,r,p_0}(Q)
\]
function queries. The type may be given to the algorithm.
Consequently the same lower bound holds for worst-case bounded-error
search over that exact type.
\end{theorem}

The theorem permits any number $M$ of eligible classes while requesting
one collision. The matching search construction is given in
Theorem~\ref{search:main}.

\begin{corollary}[Tight complexity with a unique largest class]
\label{g:unique}
For fixed $r$, suppose
$\lambda=(1^{n_1},\ldots,(r-1)^{n_{r-1}},r)$, so that
$B_j=1+\sum_{d=j}^{r-1}n_d$ and $M=1$. For every $q\ge B_1$,
the worst-case quantum query complexity of finding the unique
$r$-collision is
\begin{equation}
 \Theta_r\!\left(
 \max_{2\le j\le r}
 \left[N^{2^{j-1}}\prod_{i=2}^{j-1}B_i^{2^{i-2}}
       \right]^{1/(2^j-1)}\right).
 \label{g:unique-tight}
\end{equation}
The lower bound holds on the uniform distribution of the type.
No condition requires an intermediate size to occur even once.
\end{corollary}
\begin{proof}
Substitute $M=1$ into Theorem~\ref{g:main} and combine it with
the upper bound of Theorem~\ref{search:main}.
\end{proof}

For example, for $\lambda=(1^{n_1},2^{n_2},3^{n_3},4)$,
\begin{equation}
 Q=\max\{N^{2/3},\ N^{4/7}B_2^{1/7},\
                 N^{8/15}B_2^{1/15}B_3^{2/15}\}.
 \label{g:four}
\end{equation}
Here $B_2=n_2+n_3+1$ and $B_3=n_3+1$. If $n_2=0$, the
second and third scales still exist: partial records of three-element
classes supply the intermediate progress events.

The proof uses the complete-query return and terminal read estimates
proved above. We first extend the auxiliary database estimates to a
bounded deficit in the number of slots, then compare hidden assignments of capacity multisets and propagate all errors through the intermediate levels.

\subsection{Database conventions and bounded-deficit estimates}
The database basis consists of partial injections
$D:[N]\rightharpoonup[R]$. A range of size $R=N-e$, where
$0\le e=O_K(1)$, is partitioned into buckets of capacities at most $K$.
Auxiliary rectangular databases do not represent a distribution of
total injections into fewer than $N$ slots.
A bucket assignment $d$ records the bucket, rather than the precise
slot, assigned to each recorded input. The normalized orbit vector
$|\bar d\rangle$ is uniform over the physical injections realizing it.
If the capacities are $\kappa_i$, its orbit size is
$\prod_i(\kappa_i)_{o_i(d)}$, where
$o_i(d)=|d^{-1}(i)|$ and $(a)_u$ is the falling factorial.

Let $K_R$ project onto the simultaneous kernel of domain and image
deletion, degree by degree, and let $Z_R=I-K_R$.
Explicitly, in physical degree $h$, their normalized deletion rows are
\[
 (\partial_{\mathrm d}v)_{H,x}
 =\frac{\sum_{y\notin\im H}v_{H[x\mapsto y]}}{\sqrt{R-h+1}},
 \qquad
 (\partial_{\mathrm i}v)_{H,y}
 =\frac{\sum_{x\notin\dom H}v_{H[x\mapsto y]}}{\sqrt{N-h+1}}.
\]
Here $|H|=h-1$. In degree zero $K_R=I$.

For a fixed query input $x$, delete its record to obtain a database
$H$. Write $v_H=|H\rangle$ and let $u_H$ be the uniform normalized
superposition of its extensions at $x$. The decompression operator is
\[
 C_x=I-|v_H-u_H\rangle\langle v_H-u_H|
\]
on this deletion component. It exchanges $v_H$ and $u_H$.
A complete phase-query block is $F=C_xLC_x$, where lookup $L$
applies the phase $\omega_q^{z a(i)}$ to a record in bucket $i$
and the identity to an unrecorded query input, where
$\omega_q=e^{2\pi i/q}$. The injection
$a:[B_1]\hookrightarrow\mathbb Z_q$ gives distinct bucket labels.
The input and phase character may be controlled coherently.
A filtered update is $K_R F K_R$, interleaved with arbitrary algorithm-only
unitaries. These updates are contractions, and their state at time
$t$ has degree at most $t$. No projected state is renormalized.

All capacity assignments to the label indices are hidden and uniformly
random among assignments of the specified capacity multiset. They are retained
in reference registers. For a fixed label table, even allowing algorithm-only unitaries to depend
on that table and the capacity multiset leaves the following estimates valid. They
cannot depend on the hidden capacity assignment or a newly introduced mark.
For fixed $a$, a uniform capacity assignment followed by a uniform
slot permutation yields a uniform exact-type function with image
$a([B_1])$: every function has exactly $\prod_d(d!)^{n_d}$ slot
realizations for its uniquely determined assignment. Averaging over
a uniform injection $a$ gives the uniform type on $\mathbb Z_q$.

For a filtered state $\xi_t=\sum_d|\xi_{t,d}\rangle|\bar d\rangle$,
including its hidden capacity branches, define
\[
 C_j(d)=|\{i:o_i(d)\ge j\}|,\qquad
 \mu_j(t)=\sum_d C_j(d)\|\xi_{t,d}\|^2.
\]
The coefficients are unnormalized algorithm-register vectors.
A superscript will restrict the counter to a specified capacity class.
The basic degree bound gives $\mu_1(t)\le t$.

We need the incoming and outgoing estimates for every bounded range
deficit, rather than only the one-slot deficit used for the exceptional
family. The following argument makes this extension explicit.

\begin{lemma}[Bounded-deficit leakage]
\label{g:bounded-deficit}
Fix a constant $e_0$. For $N-e_0\le R\le N$, sufficiently large $N$,
and $t\le N/4$, complete phase queries satisfy
\begin{equation}
 \|Z_R F K_R\|_{\le t}+\|K_R F Z_R\|_{\le t}
 \le C_{e_0}N^{-1/2}.
 \label{g:leak}
\end{equation}
The norm subscript restricts the physical input degree. The estimate is
uniform in the phases, coherent query controls, and auxiliary registers.
\end{lemma}
\begin{proof}
Lemma~\ref{pk:dgap} applies directly to $[N]\rightharpoonup[R]$.
Its coefficient is $(R-2h+2)/(R-h+1)$ in degree $h$, hence is bounded
below by a positive constant in the stated range.

Let $i_R$ be the basis embedding from range $R$ into range $R+1$,
avoiding the added slot. The zero-sum equations give
\[
 i_R\operatorname{ran}K_{R,h}
   =\operatorname{ran}K_{R+1,h}\cap\operatorname{ran}P_0,
\]
where $P_0$ tests avoidance of that slot. To quantify its angle, let
$P=I-P_0$ and let $v$ be harmonic in the larger space. The image
deletion rows of $Pv$ vanish. A domain row is nonzero only if its
extension uses the added slot, so each contributing database occurs
once. Therefore
\[
 \|\partial Pv\|^2=\frac{\|Pv\|^2}{R-h+2},\qquad
 \|Z_{R+1,h}Pv\|^2\le
       \frac{\|Pv\|^2}{R-2h+3}.
\]
Set $B=K_{R+1,h}PK_{R+1,h}$ on the harmonic space and
$\epsilon^2=(R-2h+3)^{-1}$. The preceding inequality is
$B-B^2\preceq\epsilon^2B$. Thus the nonzero eigenvalues of $B$
are at least $1-\epsilon^2$, and the eigenvalues of
$K_{R+1,h}P_0K_{R+1,h}$ off the displayed intersection are at most
$\epsilon^2$. Comparing the singular values of $K_{R+1,h}P_0$
and its adjoint gives
\[
 \|Z_{R,h}w\|\le(1-\epsilon^2)^{-1/2}
                 \|Z_{R+1,h}i_Rw\|.
\]
At degree zero both invalid components vanish. Iterating over the
at most $e_0$ omitted slots gives a uniformly bounded angle factor.

Embed the range-$R$ database into the square space by the resulting
map $i$, and assign arbitrary unit phases to the added slots. Lookup
intertwines exactly. In a deletion component, put $s=R-|H|$ and
$e=N-R$. Its uniform vectors satisfy
\[
 u_N=a\,iu_R+b\,u_{\rm new},\qquad
 a=\sqrt{\frac{s}{s+e}},\quad b=\sqrt{\frac e{s+e}}.
\]
The compression difference vanishes off the source span of the blank
coordinate and $u_R$. On those two coordinates its Gram matrix is
\[
 \begin{pmatrix}2(1-a)&-b^2\\-b^2&2(1-a)\end{pmatrix},
\]
whose largest eigenvalue is at most $3b^2$. Expanding the two
compression factors of a query gives
\[
 \|(F_Ni-iF_R)L_{\le t}\|
        \le2\sqrt{\frac{3e}{R-t}}.
\]
The embedded rectangular harmonic vector is square harmonic. Apply
Proposition~\ref{pr:leak}, the last display, and the bounded angle
at output degree $t+1$ to obtain the outgoing estimate. Apply the
same estimate to the inverse phase at cutoff $t+1$ and take adjoints
to obtain the incoming estimate; a complete query changes degree by
at most one. For sufficiently large $N$ the square cutoff conditions
hold throughout. Orthogonal sums handle all query controls.
\end{proof}

We next record the moment inequality used both in dense layers and
for auxiliary capacity multisets. It also holds when the counters are restricted
to any specified set of buckets.

\begin{lemma}[Ordinary filtered moment gain]
\label{g:ordinary-gain}
For every $2\le j\le K$, bounded deficit $e_0$, and $t\le N/8$,
\begin{equation}
 \mu_j(t+1)-\mu_j(t)
 \le\frac{C_{K,e_0}}{\sqrt N}
       \bigl(\sqrt{\mu_{j-1}(t)}+
                    \sqrt{\mu_{j-1}(t+1)}\bigr).
 \label{g:ordinary-moment}
\end{equation}
\end{lemma}
\begin{proof}
On a deletion component, write $\widehat C_j=C_j(H)I+B$, where
$B$ tests an extension into a counted bucket of occupancy $j-1$.
There are at most $(K-j+1)C_{j-1}(H)$ available slots of this kind.
For the uniform extension vector $u$, put $a_H=\|Bu\|^2$.
The rank-two calculation of Lemma~\ref{hybrid:local-moment} gives
\[
 \|[B,C_x]\|^2=2a_H-a_H^2,\qquad
 a_H\le\frac{(K-j+1)C_{j-1}(H)}{R-|H|}.
\]
Lookup commutes with $B$. Cauchy--Schwarz over the orthogonal
deletion components therefore bounds the unprojected moment gain by
$C_K\sqrt{\mu_{j-1}(t)/(R-t)}$.

For completeness, the projection correction has the same form for
every range $R$ considered here. For a harmonic degree-$h$ vector $a$,
the domain-row calculation in Lemma~\ref{proj:counter-boundary}
and the rectangular deletion gap give
\[
 \|Z_R\widehat C_j a\|^2
 \le\frac{j(K-j+1)}{R-2h+2}
          \langle a,\widehat C_j\widehat C_{j-1}a\rangle.
\]
This calculation is unchanged for a restricted bucket set: only its
boundary slots enter the row sum. If $v=F\xi_t$, $a=K_Rv$,
$z=Z_Rv$, and $D=t+1$, positivity and the identity
\[
 \langle a,\widehat C_ja\rangle-
 \langle v,\widehat C_jv\rangle
 =-2\operatorname{Re}\langle Z_R\widehat C_ja,z\rangle
       -\langle z,\widehat C_jz\rangle
\]
bound the correction by
\[
 2\|z\|\sqrt{\frac{(K-j+1)D}{R-2D+2}
                      \mu_{j-1}(t+1)}.
\]
Here $C_j\le D/j$ was used. Lemma~\ref{g:bounded-deficit} gives
$\|z\|\le C_{e_0}/\sqrt N$, so this term is at most
$C_{K,e_0}\sqrt{t+1}\sqrt{\mu_{j-1}(t+1)}/N$.
Since $t+1=O(N)$ and $R-t=\Theta(N)$, it is absorbed into
\eqref{g:ordinary-moment}.
\end{proof}

\subsection{A hidden bounded-capacity jump}
\label{g:jump-section}
Consider two capacity multisets related by increasing one capacity-$c$
bucket to $d$, with $1\le c<d\le K$. Write $m$ for the number of
capacity-$c$ buckets in the source and $p$ for the number of
capacity-$d$ buckets in the target. A source capacity assignment and
one of its $m$ candidates correspond bijectively to a target assignment
and one of its $p$ marked $d$-buckets, so
\begin{equation}
 |\mathcal M_-|m=|\mathcal M_+|p.
 \label{g:edge-count}
\end{equation}
In particular, after forgetting the mark the target has the uniform distribution over assignments
of its capacity multiset, and the mark is independent of its execution
conditional on that capacity assignment. More explicitly, identify each
physical branch with its normalized bucket-orbit coordinates. Slot
relabelings intertwine compression, lookup and the harmonic projection,
so the target query matrices depend only on the target capacities and
label table, not on which eligible bucket was designated as the mark.

Write $K_-$ and $K_+$ for the harmonic projections of the source and
target ranges. Their filtered query blocks are
$A_\pm=K_\pm F_\pm K_\pm$.
For each mark $b$, define the capacity map and its normalized direct sum
by
\[
 J_b|\bar d_0\rangle_-=
 \sqrt{\frac{(d)_{o_b(d_0)}}{(c)_{o_b(d_0)}}}
       |\bar d_0\rangle_{+,b},\qquad
 J=\frac1{\sqrt m}\bigoplus_b J_b,\qquad
 S=K_+JK_-.
\]
The assignment symbol $d_0$ here is unrelated to the capacity $d$.
All weights are bounded by a constant depending only on $K$.
On the complement of the source orbit space $J_b$ is zero.
The harmonic projections preserve the orbit spaces.
The map $S$ sends the initial uniform source empty state to the
initial uniformly marked target empty state, by~\eqref{g:edge-count}.

\begin{lemma}[Localized bounded jump]
\label{g:localized-jump}
Suppose $0\le t\le N/8$. Let $\mu_c^-(t)$ be the source moment of full capacity-$c$ candidate
buckets. On the source prequery state, the one-query filtered comparison
error has the form $K_+v_t+r_t$ with
\[
 \Xi_tv_t=v_t,\qquad
 \|v_t\|\le C_K\sqrt{\frac{\mu_c^-(t)}{mN}},
 \qquad \|r_t\|\le\frac{C_K}{\sqrt{mN}}.
\]
Here $\Xi_t$ requires the queried input to be recorded in the marked
target bucket at occupancy $c+1$. For the broader event $P$ of marked
occupancy at least $c+1$, the coarse endpoint satisfies
\[
                  \|PS\xi^-_t\|\le C_K\sqrt{t/(mN)}.
\]
\end{lemma}
\begin{proof}
We calculate the defect on one deletion component, then sum over its
independent query and hidden-parameter controls. In a deletion
component let $a_i$ be the remaining source capacities,
$s=\sum_i a_i$, and $\delta=d-c$. Set
$\rho=\sqrt{(s+\delta)/s}$ and $\alpha=1-\rho^{-1}$.
Write $v$ for the blank coordinate, $|i\rangle$ for a normalized
bucket extension, and define
$r_u=\sqrt{(d)_u/(c)_u}$ for $0\le u\le c$, with
$\widetilde r_b=r_{o_b(H)}$.
Let $u_-$ and $u_+$ be the normalized uniform extension vectors of
the source and target in this component. The exact rank-two identity is
\[
 C_+J_b-J_bC_-=(|v\rangle-|u_+\rangle)\langle p_b|
                  +|q_b\rangle(\langle u_-|-\langle v|).
\]
For $a_b>0$ its vectors are
\[
 p_b=-\widetilde r_b\alpha u_-+
            \frac{\widetilde r_b\delta}{\sqrt{a_b(s+\delta)}}|b\rangle,
 \qquad q_b=\widetilde r_b(\rho-1)u_+.
\]
For $a_b=0$ they are
\[
 p_b=-\widetilde r_b\alpha u_-,\qquad
 q_b=\widetilde r_b\bigl(\alpha u_- -\sqrt{\delta/(s+\delta)}|b\rangle\bigr).
\]
Here $\alpha,\rho-1=O_K(N^{-1})$. Separate the full-candidate term
\[
 G_b^C=-\mathbf1_{\{a_b=0\}}r_c
       \sqrt{\delta/(s+\delta)}|b\rangle
                     (\langle u_-|-\langle v|).
\]
Its range satisfies $\Xi_t$. Its squared direct-sum norm is bounded
by the source full-candidate count divided by $mN$.
After removing this term, the remaining averaged squared defect is
$O_K(N^{-2}+(mN)^{-1})$ times the local source squared norm:
the potentially larger nonfull term has bra $\langle b|$, whose
squared source coordinates sum to at most the source squared norm.

Since $L_+J_b=J_bL_-$, the complete-query leading term can be chosen as
\[
              G_b^F=L_+G_b^C+G_b^CL_-C_-.
\]
It still satisfies $\Xi_tG_b^F=G_b^F$.
The extra term $(C_+-I)L_+G_b^C$ has norm
$O_K(N^{-1/2})\|G_b^C\|$. Orthogonal summation over the query controls,
deletion components and marks gives the asserted raw bounds.

We also require both harmonic-compatibility estimates, including the
coherent reverse sum over marks:
\begin{equation}
 \|Z_+JK_-\|_{\le t}+\|K_+JZ_-\|_{\le t}
                  \le C_K\sqrt{t/(mN)}.
 \label{g:jump-compatibility}
\end{equation}
Let $R_-$ and $R_+=R_-+\delta$ be the two range sizes. On an
orbit assignment $H$ of degree $h-1$, the unnormalized domain row
is $\sum_{i:a_i>0}\sqrt{a_i}\,u_{Hi}$; its squared norm is
divided by $R_\pm-h+1$ in the corresponding normalized row map.
For source-harmonic $u$, the target row of $J_bu$ is, when $a_b>0$,
\[
 \widetilde r_b\left(\sum_{i\ne b}\sqrt{a_i}\,u_{Hi}
       +\frac{a_b+\delta}{\sqrt{a_b}}u_{Hb}\right)
       =\frac{\delta\widetilde r_b}{\sqrt{a_b}}u_{Hb}.
\]
When $a_b=0$ it is zero, since the new target extension has zero
source coefficient. All image rows vanish: the capacity weight is
constant over the inputs summed by each such row. The displayed
coefficient is bounded in terms of $K$; summing over marks and
deletions counts each source database at most $h$ times. Hence
\[
 \|\partial_+Ju\|^2\le
       \frac{C_Kh}{m(R_+-h+1)}\|u\|^2.
\]
The deletion gap proves the forward bound.

For the reverse bound, let $w=\bigoplus_b w^b$ be target harmonic.
When $a_b>0$, the source domain row of $J_b^\dagger w^b$ is
\[
 \widetilde r_b\left(\sum_{i\ne b}\sqrt{a_i}\,w^b_{Hi}
             +\sqrt{a_b+\delta}\,w^b_{Hb}\right)=0.
\]
When $a_b=0$, it equals $-r_c\sqrt\delta\,w^b_{Hb}$.
After the direct-sum adjoint, the normalized row is therefore
\[
 -r_c\sqrt{\frac\delta{m(R_--h+1)}}
               \sum_{b:o_b(H)=c}w^b_{Hb}.
\]
There are at most $h/c$ summands. Cauchy--Schwarz, followed by
summing over rows, counts each contributing target database at most
$c+1$ times, once for each deletion from marked occupancy $c+1$.
Rows with larger marked occupancy are not source-feasible, and image
rows again vanish. Thus
\[
 \|\partial_-J^\dagger w\|^2
       \le\frac{C_Kh}{m(R_--h+1)}\|w\|^2.
\]
The deletion gap and adjoints prove the second bound in
\eqref{g:jump-compatibility}. With hidden capacity assignments, each
marked target branch has one parent; the reverse summation is made
separately in each parent block. Distinct parents and degrees are
orthogonal, which proves the same bounds with all reference registers.
Insert them and~\eqref{g:leak} into the exact identity
\[
 A_+S-SA_-=K_+(F_+J-JF_-)K_-
       -K_+F_+Z_+JK_-+K_+JZ_-F_-K_-.
\]
The filtering remainder is $O_K(\sqrt{t+1}/(N\sqrt m))$.
Together with the raw bounds, $\mu_c^-(t)\le m$, and $t\le N/8$,
this is at most $C_K/\sqrt{mN}$.
Finally $PJ=0$, so the forward compatibility bound implies
$\|PS\xi^-_t\|\le\|Z_+JK_-\xi^-_t\|$ as claimed.
\end{proof}

\begin{lemma}[Squared transfer estimate]
\label{g:energy}
Suppose $p\le m$ and $0\le T\le N/8$. The target moment counting capacity-$d$ buckets
with occupancy at least $c+1$ satisfies
\begin{equation}
 \mu_{c+1}^{(d)}(T)
 \le C_K\sqrt{\frac p{mN}}\sum_{t<T}\sqrt{\mu_c^-(t)}
             +C_K\frac{pT^2}{mN}.
 \label{g:energy-bound}
\end{equation}
The states may have arbitrary algorithm registers and hidden source
capacity assignments.
\end{lemma}
\begin{proof}
Put $e_t=\xi^+_t-S\xi^-_t$ and $E_t=\|e_t\|$.
Then $e_0=0$ and, including the prequery algorithm unitary $U_t$,
\[
 e_{t+1}=a_t+K_+v_t+r_t,\qquad a_t=A_{+,t}U_te_t,
 \qquad \|a_t\|\le E_t.
\]
The mark is uniform among $p$ target $d$-buckets and has no effect
on their unmarked execution. For a fixed queried input and database,
at most one such mark satisfies $\Xi_t$. Therefore
$\|\Xi_t\xi^+_{t+1}\|\le p^{-1/2}$.
Since $a_t$ is harmonic and $v_t$ is localized,
\[
 \langle a_t,K_+v_t\rangle=\langle\Xi_ta_t,v_t\rangle,
 \quad
 \|\Xi_ta_t\|\le p^{-1/2}+\eta_{t+1}+g_t+\rho_t,
\]
where $g_t=\|v_t\|$, $\rho_t=\|r_t\|$ and
$\eta_u=C_K\sqrt{u/(mN)}$. We used
$a_t=\xi^+_{t+1}-S\xi^-_{t+1}-K_+v_t-r_t$ and $\Xi_t\preceq P$.
Consequently
\[
 E_{t+1}^2\le E_t^2+2g_t/\sqrt p+2g_t\eta_{t+1}
                    +3g_t^2+4g_t\rho_t+2E_t\rho_t+\rho_t^2.
\]
Write $A_c^-(t)=\sqrt{\mu_c^-(t)/m}\le1$, so
$g_t\le C_K A_c^-(t)/\sqrt N$.
The terms $g_t^2,g_t\eta_{t+1},g_t\rho_t$ are each at most
$C_K g_t/\sqrt p$, using $p\le m$, $p\le N$ and $t\le N/8$.
For a nonnegative recurrence $E_{t+1}^2\le E_t^2+f_t+
2E_t\rho_t+\rho_t^2$, put $E_* =\max_{t\le T}E_t$,
$F_T=\sum_{t<T}f_t$, and $R_T=\sum_{t<T}\rho_t$.
Summing up to an index attaining $E_*$ gives
$E_*^2\le F_T+2E_*R_T+R_T^2$. Hence
\[
 E_*\le R_T+\sqrt{F_T+2R_T^2}
       \le\sqrt{F_T}+(1+\sqrt2)R_T,
 \qquad E_*^2\le2F_T+12R_T^2.
\]
Thus
\[
 E_T^2\le C_K\left[\frac1{\sqrt{Np}}\sum_{t<T}A_c^-(t)
                                      +\frac{T^2}{mN}\right].
\]
Uniform mark averaging gives $\mu_{c+1}^{(d)}(T)/p=\|P\xi^+_T\|^2$.
Use $\|P\xi^+_T\|\le E_T+\eta_T$, square, and multiply by $p$.
For integer $T\ge1$ the endpoint term is absorbed in $pT^2/(mN)$;
the case $T=0$ is immediate.
\end{proof}

\subsection{A uniform recursion over capacity multisets}
\label{g:recurrence-section}
We now derive one recursion that handles both abundant and missing
intermediate sizes. Its source and target capacity multisets can differ, so the induction
is carried out simultaneously over the required capacity multisets.

\begin{proposition}[Uniform moment recursion over capacity multisets]
\label{g:recurrence}
For each $j\ge2$ with $B_j>0$ and $0\le T\le N/8$, an upper bound
for the target moment is
\begin{equation}
 \mu_j(T)\le
 C_K\sqrt{\frac{B_j}{NB_{j-1}}}
 \sum_{t<T}\bigl(\sqrt{V_{j-1}(t)}+
                      \sqrt{V_{j-1}(t+1)}\bigr)
       +C_K\frac{B_jT^2}{NB_{j-1}}.
 \label{g:moment-recursion}
\end{equation}
Here $V_{j-1}$ is any uniform upper bound on the $(j-1)$st moment
for the target capacity multiset and the auxiliary capacity multisets obtained by reducing
one capacity $d\ge j$ to $j-1$.
Every such auxiliary capacity multiset has exactly the same $B_i$ for
$i\le j-1$.
\end{proposition}
\begin{proof}
If $B_j\ge B_{j-1}/2$, the ordinary filtered recurrence
\eqref{g:ordinary-moment} implies the result, with no need for the
additive term. Otherwise $n_{j-1}=B_{j-1}-B_j>B_{j-1}/2$.
For each $d\ge j$ with $n_d>0$, apply Lemma~\ref{g:energy} to the
source obtained by reducing one $d$-bucket to $j-1$.
Its candidate pool is $m=n_{j-1}+1\asymp B_{j-1}$ and the target
marking pool is $p=n_d\le B_j<m$. The source full-candidate count
is at most its total level-$(j-1)$ count. Summing
\eqref{g:energy-bound} over $d\ge j$, and using
$\sum_{d\ge j}\sqrt{n_d}\le\sqrt{K B_j}$, proves the bound.
Reducing $d$ to $j-1$ leaves every smaller capacity-tail count
unchanged, proving the last assertion.
\end{proof}

Recursive descent has at most $K$ levels and removes at most $K$ slots
at a step. All auxiliary ranges therefore have deficit at most $K^2$
from the starting square range. This is why the bounded-deficit
estimates apply throughout. One may formulate the induction over all
ancestors reachable by these descents; it does not assert that a fixed
range-deficit class is closed under arbitrarily many reductions.

The additive errors can be propagated explicitly without first
restricting $T$ to a proposed lower-bound scale. Put $\alpha_j=2-2^{1-j}$.

\begin{lemma}[Moments with all propagated errors]
\label{g:moment-envelope}
For $j\ge2$ with $B_j>0$, define
\begin{align}
 L_j(T)&=\frac{T^{\alpha_j}}
 {N\prod_{i=2}^{j}B_i^{2^{i-j-1}}},
 \label{g:leading}\\
 E_{l,j}(T)&=\frac{T^2}
 {N B_{l-1}^{2^{l-j}}
      \prod_{i=l+1}^{j}B_i^{2^{i-j-1}}}
                    \quad(2\le l\le j).
 \label{g:errors}
\end{align}
Then, uniformly over the required capacity multisets and $T\le N/8$,
\begin{equation}
 \frac{\mu_j(T)}{B_j}
       \le C_K\left(L_j(T)+\sum_{l=2}^jE_{l,j}(T)\right).
 \label{g:envelope}
\end{equation}
For level one use $\mu_1(T)/B_1\le C_KT/N$.
\end{lemma}
\begin{proof}
The range is $N-O_K(1)$ and capacities are bounded, so
$B_1=\Theta_K(N)$. Set $A_j(T)=\sqrt{\mu_j(T)/B_j}$.
Dividing~\eqref{g:moment-recursion} by $B_j$ gives
\[
 A_j(T)^2\le C_K\left[
 \frac1{\sqrt{NB_j}}\sum_{t<T}
       (A_{j-1}^{\rm src}(t)+A_{j-1}^{\rm src}(t+1))
                       +\frac{T^2}{NB_{j-1}}\right],
\]
where any source is covered by the uniform induction hypothesis and
has unchanged lower tails. Starting with $C_KT/N$, its leading term
is $L_j$, because $\alpha_j=1+\alpha_{j-1}/2$.
An error born at level $l$ is initially $T^2/(NB_{l-1})$.
Taking its square root and summing over time at the next level
preserves the power $T^2$, halves all existing bucket exponents, and
adds exponent $1/2$ on the new $B_j$. This gives exactly~\eqref{g:errors}.
The inequality $\sqrt{u_1+\cdots+u_a}\le\sum_i\sqrt{u_i}$ and
$\sum_{t<T}(t+1)^u\le T^{u+1}$ for $u\ge0$ justify each induction
step, including its same-time term. There are at most $K$ terms and
levels, so the constant depends only on $K$.
\end{proof}

The denominator exponents in each $E_{l,j}$ sum to one. This fact
will control their contribution after any remaining capacity transfer.
It also gives the coarse uniform bound
$\mu_j(T)\le C_K(B_jL_j(T)+T^2/N)$, but keeping the individual
errors is essential beyond the square-root query scale.

\subsection{Converting a union of database events to success}
\label{g:terminal-section}
Let $P_r$ project onto databases having at least one bucket with
$r$ recorded slots. This event is the union over all eligible buckets;
it has no distinguished mark.

\begin{lemma}[Union-event success conversion]
\label{g:terminal}
Let $p_{\rm succ}$ be the real success probability of a $T$-query
function algorithm and $\xi_T$ its square filtered history. For fixed
$K,r$, sufficiently large $N$, and $T\le N/8$,
\begin{equation}
 \sqrt{p_{\rm succ}}\le\|P_r\xi_T\|
       +C_{K,r}\left(\frac TN+\sqrt{\frac{T+1}{N}}\right).
 \label{g:terminal-bound}
\end{equation}
The comparison is uniform over the injective label table and the
hidden capacity assignments. There is no factor $\sqrt M$.
\end{lemma}
\begin{proof}
Corollary~\ref{phase:filtered-comparison} bounds the distance between
the filtered history and the isometrically encoded real history by
$CT/N$. Its complete-phase estimate is uniform over all slot phases,
so it applies to every capacity multiset and injective label table here.
Keep the reported $r$ inputs and use two passes of $r$ forward-swap
reads as in Lemma~\ref{terminal:verifier}. We give the boundary
estimate for the larger event $P_r$ explicitly.

On a physical deletion component with residual database $H$, if $H$
already contains an $r$-collision then $P_r$ is the identity and its
commutator with decompression is zero. Otherwise the event tests
extensions into buckets of occupancy $r-1$. There are at most
$|H|/(r-1)$ such buckets, each with at most $K-r+1$ available slots.
Writing $u$ for the uniform extension vector and $a=\|P_ru\|^2$,
\[
 a\le\frac{(K-r+1)|H|}{(r-1)(N-|H|)},\qquad
 \|[P_r,C_x]\|=\sqrt{a(2-a)}.
\]
The compressed swap is decompression, a lookup diagonal in the
physical database, and recompression. Its boundary norm is therefore
at most $2\|[P_r,C_x]\|=O_K(\sqrt{(T+1)/N})$ on the relevant
degree cutoff. Orthogonal summation over query controls and deletion
components preserves this bound.

After the constant number of reads, the literal verifier accepts only
if the reported distinct inputs are recorded at distinct slots in one
bucket. Injective bucket labels make its event a subevent of $P_r$.
The product comparison~\eqref{terminal:list-verifier}, with $P_r$
dominating its literal term, contributes $O_r(N^{-1/2})$.
Proposition~\ref{pr:swap} and the hybrid sum
\eqref{terminal:mixed-history-comparison} through the constant number
of reads contribute $O_r(N^{-1/2}+(T+1)/N)$. Add the boundary changes and the initial
$O(T/N)$ history error to obtain~\eqref{g:terminal-bound}.
Every estimate holds in direct sums over the hidden parameters and
stored output registers; none conditions on or renormalizes a mark.
\end{proof}

\subsection{Selecting the relevant intermediate level}
\label{g:lower-section}
The maximum in~\eqref{g:scale} specifies the intermediate order that
must remain hard. We now show that this order can always be analyzed
using either the full moment or a capacity layer that is actually
present in the input.

\begin{proof}[Proof of Theorem~\ref{g:main}]

Choose a maximizing index $p$, so $F_{p,M}=Q$, and put
$D=M Q^4/N^2$. If $p\ge3$, comparison with $F_{p-1,M}$ gives
$B_{p-1}\ge D$. For $p=2$, the same statement holds up to a
$K$-dependent constant because $B_1=\Theta_K(N)$ and
$D=Q=N^{2/3}M^{-1/3}$. If $p<r$, comparison with $F_{p+1,M}$ gives
$B_p\le D$.

Suppose first that $M\ge B_{p-1}/(4K)$. Then
$M\le B_p\le B_{p-1}\le4KM$, and the peak inequality
implies $Q=O_K(\sqrt N)$. Since all earlier tails are at least $B_p$,
\[
 B_p E_{l,p}(T)\le T^2/N.
\]
Also
\[
 B_p L_p(T)=\sqrt{B_p/M}\,(T/F_{p,M})^{\alpha_p}.
\]
It follows that
$\mu_p(\varepsilon Q)\le C_K(\varepsilon^{\alpha_p}+\varepsilon^2)$.
Any recorded $r$-collision entails at least one occupied-$p$ bucket,
so Markov's inequality bounds the target database probability by this
quantity.

Otherwise $M<B_{p-1}/(4K)$. Among the capacities
$c\in\{p-1,\ldots,r-1\}$ choose one with
\[
 n_c\ge\frac{B_{p-1}-M}{K}\ge\frac{B_{p-1}}{2K},
 \qquad m=n_c+1.
\]
Thus $m\asymp_K B_c\asymp_K B_{p-1}$, $m>M$, and every tail through
$c$ is at least a constant multiple of $\max\{M,D\}$.
Moreover $F_{c+1,M}\asymp_K Q$. If $c=p-1$ this is immediate.
Otherwise $p<r$, and $B_p\le D$ while $B_i\ge m$ for $p\le i\le c$.
Hence those $B_i$ are all within $K$-dependent constant factors of $D$.
Starting with $F_{p,M}=Q$, iterate
\[
 F_{j+1,M}^{\alpha_{j+1}}
  =N^{1/2}M^{-1/4}B_j^{1/4}
      F_{j,M}^{\alpha_j/2}.
\]
The definition of $D$ proves the claimed comparison at each step.

For each target capacity $d\ge r$ with $n_d>0$, lower one such
bucket to capacity $c$. Its lower tails $B_i$, $i\le c$, are exactly
unchanged. The parent has $m=n_c+1$ candidate buckets, and the final
mark is uniform among the $n_d$ target buckets of capacity $d$.
Apply the localized energy estimate for $c\to d$, and sum over $d$.
Writing $W_{c+1}(T)$ for the expected number of capacity-at-least-$r$
buckets having at least $c+1$ records, this gives
\[
 W_{c+1}(T)\le
 C_K\sqrt{\frac MN}\sum_{t<T}A_c^{\rm src}(t)
           +C_K\frac{MT^2}{mN}.
\]
Lemma~\ref{g:moment-envelope} gives a source moment bound uniform over the different parents and
uses the unchanged tails through $c$. Its leading contribution is
$C_K(T/F_{c+1,M})^{\alpha_{c+1}}$.
Each propagated error has denominator exponents summing to one;
after the last square root their exponents sum to one half. Since
all involved tails are at least a constant multiple of $m$, their
combined contribution is at most
\[
 C_K\frac{T^2}{N}\sqrt{\frac Mm}.
\]
At $T=\varepsilon Q$, this and the final local remainder are
$O_K(\varepsilon^2)$. Indeed, use $m\ge a_K M$ if $Q\le\sqrt N$,
and $m\ge a_KM Q^4/N^2$ if $Q\ge\sqrt N$.
Consequently
\[
 W_{c+1}(\varepsilon Q)
 \le C_K(\varepsilon^{\alpha_{c+1}}+\varepsilon^2).
\]
Every recorded target $r$-collision entails one of these events because
$c+1\le r$. Markov again bounds its database probability.

For sufficiently small $\varepsilon=\varepsilon(K,r,p_0)>0$, the
preceding database probability bounds give
$\|P_r\xi_T\|<\sqrt{p_0}/2$ at $T\le\varepsilon Q$.
Since $B_i\le N$ and $M\ge1$, we have
\[
 Q\le N^{\gamma_K},\qquad
 \gamma_K=1-\frac{2^{K-2}}{2^K-1}<1.
\]
Hence the query budget is at most $N/8$ for large $N$, and the error
in Lemma~\ref{g:terminal} is uniformly $o(1)$.
This rules out real success probability at least $p_0$.
All estimates are uniform for each injective label table. Squaring
before averaging over that table yields the uniform exact-type
lower bound for every $q\ge B_1$. An algorithm with the stated
success guarantee on every input has that guarantee on average,
which also proves the worst-case lower bound.
\end{proof}

\section{Matching search upper bounds}
\label{search:overview}

We now match the type-dependent lower bound by algorithms that use
only the function oracle. Throughout this section, use
$B_j$, $M=B_r$, and $Q$ from~\eqref{g:tails}--\eqref{g:scale}.
The type may be supplied to the algorithm, but the equality classes
and their input positions are hidden. The inverse of the addition
oracle is obtained by conjugating its response register with negation,
so using an inverse call does not require an additional kind of access.

\begin{theorem}[Type-dependent search upper bound]
\label{search:main}
For every fixed $K\ge r\ge2$, every sufficiently large $N$, and every
bounded exact type with $B_r\ge1$, a quantum algorithm finds one
$r$-collision with probability at least $2/3$ on every function of
that type using $O_{K,r}(Q)$ function queries. The implicit constant
is uniform over all feasible alphabets $q\ge B_1$.
\end{theorem}

The constructions on the two sides of $Q=\sqrt N$ use the same
preparation-cost expression. For raw set sizes $a_1,\ldots,a_{r-1}$,
it is
\begin{equation}
 a_1+\sum_{j=2}^{r-1}
 \frac{a_j\sqrt{B_j\prod_{i<j}a_i}}{N^{j/2}},
 \label{search:setup-cost}
\end{equation}
and the target-completion scale is
$N^{r/2}/\sqrt{M\prod_{i<r}a_i}$.
We first choose parameters making these quantities $O_{K,r}(Q)$.
For $Q\le\sqrt N$, sampled tables and successive Grover enumeration
realize the parameters directly. For $Q\ge\sqrt N$, ordered
rolewise restriction and a colored adaptive learning graph suffice.
A coherent decision-to-search conversion avoids logarithmic overhead.

\subsection{Direct search below the square-root scale}
\label{dg:section}

The regime below the square-root scale admits a direct search
algorithm.  It uses classical tables and Grover enumeration of
matching values in independently sampled, disjoint input roles.
In particular, it requires neither a quantum walk on tuple lists nor
a coherent update of those lists.

We first choose the sampling parameters so that every stage has the
desired query cost and a constant expected number of target classes
survives all stages.
\begin{lemma}[Sampling parameters]
\label{dg:parameters}
Suppose $Q\le\sqrt N$.  There are real parameters
$Q\le a_i\le N$ for $1\le i<r$, with $a_1=Q$, such that, on writing
\[
 z_j=\frac{\prod_{i=1}^{j}a_i}{N^j},
\]
one has
\begin{equation}
 \sqrt{a_jB_jz_j}\le Q\quad(2\le j<r),\qquad
 \sqrt{\frac{N}{Mz_{r-1}}}\le Q.
 \label{dg:parameter-costs}
\end{equation}
Moreover $z_1\ge\cdots\ge z_{r-1}$ and
$B_jz_j\ge1$ for $2\le j<r$, while $Mz_{r-1}\ge1$.
\end{lemma}
\begin{proof}
Put $z_1=Q/N$, and for $2\le j<r$ define
\[
 z_j=\min\left\{z_{j-1},
           \frac{Q\sqrt{z_{j-1}}}{\sqrt{NB_j}}\right\},
 \qquad a_j=N z_j/z_{j-1}.
\]
These definitions imply $a_j\le N$ and
\[
 \sqrt{a_jB_jz_j}
 =z_j\sqrt{\frac{NB_j}{z_{j-1}}}\le Q.
\]
They also imply $a_j\ge Q$: the uncapped expression for $a_j$ is
$Q\sqrt{N/(B_jz_{j-1})}\ge Q$, and $N\ge Q$.

Expanding the minima expresses $z_{r-1}$ as the minimum, over
increasing subsequences $2\le i_1<\cdots<i_m\le r-1$, including
the empty subsequence, of
\[
 \frac{Q^{2-2^{-m}}}
 {N\prod_{t=1}^{m}B_{i_t}^{2^{-(m-t+1)}}}.
\]
For a fixed $m$, monotonicity of the $B_i$ makes this expression
smallest at $i_t=t+1$.  The $j=m+2$ term in~\eqref{g:scale} gives
\[
 Q^{4-2^{-m}}
 \ge\frac{N^2}{M}
       \prod_{t=1}^{m}B_{t+1}^{2^{-(m-t+1)}}.
\]
Consequently $z_{r-1}\ge N/(MQ^2)$, which proves the final cost
bound.  Since the $z_j$ are nonincreasing, $B_j\ge M$, and
$Q^2\le N$, all the stated lower bounds on the expected counts
follow.  This argument includes $r=2$, when the subsequence is empty.
\end{proof}

The enumeration subroutine needs a fixed total budget, rather than a
separate high-confidence guarantee for every discovered input.
\begin{lemma}[Fixed-budget enumeration]
\label{search:enumeration}
Let a known set have size $a$, and suppose a query-computable predicate
marks $h\le h_{\max}$ of its positions, where $h_{\max}\ge1$ is known.
For any fixed $\varepsilon>0$, all marked positions can be found with
probability at least $1-\varepsilon$ using a worst-case budget
$O_\varepsilon(\sqrt{a h_{\max}})$ predicate queries.
\end{lemma}
\begin{proof}
If $a=0$, return the empty list. Otherwise repeatedly run verified
randomized Grover search~\cite{BBHT98}, excluding already found
positions, until a single predetermined total budget is exhausted.
Only verified positions are added to the list. When $u>0$ marked
positions remain, the expected time to discover the next one is
$O(\sqrt{a/u})$. Thus the expected time $T_{\rm all}$ at which all
$h$ positions have been found satisfies
\[
 \mathbb E T_{\rm all}
 \le C\sum_{u=1}^h\sqrt{a/u}
 \le 2C\sqrt{ah}\le2C\sqrt{a h_{\max}}.
\]
For $h=0$ the list is already complete. No test certifying that the
remaining set is empty is needed. Markov's inequality gives
$\Pr[T_{\rm all}>D\sqrt{a h_{\max}}]\le2C/D$.
Choose $D\ge2C/\varepsilon$ and stop at that budget, even when the
promise on $h$ fails. This proves the stated guarantee and the
worst-case query bound.
\end{proof}

\begin{theorem}[Search below the square-root scale]
\label{dg:dense}
Fix $K\ge r\ge2$.  For every sufficiently large $N$ and every input
of maximum class size at most $K$ satisfying $Q\le\sqrt N$, an
algorithm finds an $r$-collision with probability at least $2/3$
using $O_{K,r}(Q)$ function queries.  Its guarantee holds
for every such function and is uniform over all feasible alphabets.
\end{theorem}
\begin{proof}
Use the parameters from Lemma~\ref{dg:parameters}.  Independently
assign each input a uniform role in $\{0,1,\ldots,r\}$.  For
$1\le i<r$, retain each input assigned role $i$, independently, with
probability $a_i/N$; call the resulting public input set $A_i$.
Let $A_r$ contain all inputs assigned role $r$.  The sets $A_i$ are
pairwise disjoint.  Their indices are known without querying $f$.

Query all of $A_1$.  At stage $j$, for $2\le j<r$, enumerate the
inputs of $A_j$ whose values occur in the preceding table, storing
their values and one prefix tuple for each value found.  A prefix
tuple has one input from each of $A_1,\ldots,A_j$.  If a value
occurs several times in a role, retain any one representative for
constructing its prefix tuple.  Finally, use Grover search on $A_r$
for an input whose value occurs in the final prefix table.  The
stored prefix and this input give $r$ distinct inputs, because
their roles are distinct.

We analyze the ideal tables obtained when each enumeration is
complete.  Let $H_j$ be the number of marked inputs in $A_j$ at
stage $j$.  For a class to contribute to $H_j$, it must contain
distinct selected inputs in all of the first $j$ roles.  A union
bound over ordered $j$-tuples within each class gives
\begin{equation}
 \mathbb EH_j
 \le\frac{(K)_j}{(r+1)^j}B_jz_j
 \le C_{K,r}B_jz_j
 \qquad(2\le j<r).
 \label{dg:enumeration-mean}
\end{equation}
An unsuccessful earlier enumeration can only reduce the later
marked counts, so this ideal count also bounds the count seen by
the actual algorithm.

For each of the $M$ target classes, fix $r$ distinct inputs for the
analysis.  Let its favorable event mean that these inputs are
assigned, in order, roles $1,\ldots,r$, and that the first $r-1$
are retained in their corresponding sets.  These events are
independent across classes, each with probability
$c_rz_{r-1}$, where $c_r=(r+1)^{-r}$.  If $Y$ counts favorable
classes, then
\[
 \lambda:=\mathbb EY=c_rMz_{r-1}\ge c_r,
 \qquad \mathbb EY^2\le\lambda+\lambda^2.
\]
The second-moment inequality yields
\begin{equation}
 \Pr[Y\ge\lambda/2]
 \ge\frac{\lambda}{4(1+\lambda)}
 \ge p_*:=\frac{c_r}{4(1+c_r)}>0.
 \label{dg:target-event}
\end{equation}
Every favorable class supplies a marked input to the final search
whenever all prefix enumerations are complete.

Choose a constant $C_H=C_H(K,r)$ sufficiently large that
Markov's inequality and~\eqref{dg:enumeration-mean} give
\[
 \Pr[H_j>C_HB_jz_j\text{ for some }2\le j<r]\le p_*/4.
\]
Also, $|A_i|$ is binomial with mean $a_i/(r+1)$.
Since $a_i\ge Q\ge N^{1/3}$, Chernoff bounds imply, for all
sufficiently large $N$,
\[
 \Pr[|A_i|>2a_i\text{ for some }1\le i<r]\le p_*/4.
\]
Here $Q\ge N^{1/3}$ follows already from the $j=2$ term and
$M\le N$.  No independence between these upper-count events and
the target event is required.  Subtracting their failure
probabilities from~\eqref{dg:target-event} shows that
\begin{equation}
 Y\ge\lambda/2,\quad |A_i|\le2a_i,
 \quad H_j\le C_HB_jz_j
 \label{dg:good-event}
\end{equation}
hold jointly with probability at least $p_*/2$.

Abort an attempt if any public size bound $|A_i|\le2a_i$ fails.
For each enumeration, apply Lemma~\ref{search:enumeration} with
$h_{\max}=\lceil C_HB_jz_j\rceil$ and failure probability at most
$p_*/(8r)$. Testing whether a value occurs in the preceding table
uses $O(1)$ function queries; table operations do not add queries.
The resulting query cost is
\[
 O_{K,r}\left(
 \sqrt{|A_j|(C_HB_jz_j+1)}\right)
 =O_{K,r}\left(\sqrt{a_jB_jz_j}\right)
 =O_{K,r}(Q).
\]
The first equality uses $B_jz_j\ge1$.  This budget is fixed even
on inputs to the subroutine with more marked positions; only its
success guarantee is conditional on the count bound.  Conditional
on~\eqref{dg:good-event}, a union bound over the fixed number of
stages leaves enumeration failure probability at most $p_*/8$.

The final randomized Grover search is given a fixed budget
$O_r(\sqrt{N/\lambda})$, which is $O_r(Q)$ by
Lemma~\ref{dg:parameters}.  When there are at least $\lambda/2$
marked inputs, it succeeds with probability at least $2/3$ after
adjusting the constant.  This statement also covers $\lambda/2<1$:
the number of marked inputs is integral, so the favorable event
then guarantees at least one.  Every reported solution is verified.

Combining~\eqref{dg:good-event}, the enumeration guarantee, and the
final search gives a positive success probability depending only
on $K,r$.  A constant number of independent attempts raises it to
$2/3$.  Querying $A_1$ costs at most $2Q$, there are only $r-2$
enumeration stages, and all budgets are fixed.  The resulting
worst-case query bound is therefore $O_{K,r}(Q)$.
\end{proof}

\subsection{A colored learning-graph bound with prescribed prefix populations}
\label{clg:section}

We give an explicit dual-adversary construction for the colored problem.
It adapts the fault correction in Belovs~\cite[Section~6]{Bel12},
with separate input domains for the certificate positions and a fault
budget determined by the maximum multiplicity.  In particular, the
argument below verifies the off-diagonal constraints; the upper bound
does not follow merely by substituting the domain sizes in an existing
complexity formula.

\begin{lemma}[Colored search bound with prescribed prefix populations]
\label{clg:bound}
Let $U_1,\ldots,U_r$ be disjoint sets, with $|U_i|=v_i$, and let
$\mathcal X$ be a finite promise domain of inputs on their union,
with values in a finite alphabet $\Sigma$.
Every equality class of every input has size at most a fixed $K\ge r$.
Let $F(z)=1$ if and only if there is an equal-value tuple in
$U_1\times\cdots\times U_r$.  For $2\le j<r$, suppose every
$z\in\mathcal X$ has at most $C_j$ equal-value tuples in
$U_1\times\cdots\times U_j$.
Put $b=K+1$ and $L_i=b^{i-1}(2^b-1)$.
For positive integers $a_i$ satisfying $L_i a_i\le v_i/4$
for $1\le i<r$, the bounded-error quantum query complexity of $F$ is
\begin{equation}
 O_{K,r}\!\left(
 a_1+\sum_{j=2}^{r-1}
 a_j\sqrt{\frac{C_j\prod_{i<j}a_i}{\prod_{i\le j}v_i}}
 +\sum_{s=1}^{r}
 \sqrt{\frac{\prod_{i\le s}v_i}{\prod_{i<s}a_i}}
 \right).
 \label{clg:main-cost}
\end{equation}
The prefix bounds must hold on both positive and negative inputs.
Uniqueness of the colored $r$-tuple is not required.
\end{lemma}

If the positive promise domain is empty, the conclusion is immediate.
Otherwise, choose a single colored certificate
$(c_1(x),\ldots,c_r(x))$ for every positive input $x$, by an arbitrary
fixed deterministic rule.  Write $\nu_x=x_{c_1(x)}$ and
$Z_x=\{t:x_t=\nu_x\}$ for its entire equality class.
This choice is used only to define the feasible matrices; no
certificate or equality-class information is supplied to the algorithm.

\par\medskip\noindent\textbf{Raw sets and certificate stages.}\enspace
Let $\mathcal R$ consist of families of pairwise disjoint sets
\[
 R_i(d_1,\ldots,d_{i-1},J)\subseteq U_i,
 \qquad
 i<r,\quad d_h\in[b],\quad
 \varnothing\ne J\subseteq[b],
\]
with every set at level $i$ having size $a_i$.
There are $L_i$ sets at that level.
For each family $R$, fix an order of its elements in which all
level-$i$ elements precede all level-$(i+1)$ elements.
The setup indices are the successive queries in this order; an index
retains the full tag $R$, even when only part of the family is loaded.

Call $R$ consistent with a positive input $x$ if
$\bigcup R$ is disjoint from $Z_x$.  Thus consistency excludes the
whole selected equality class, not just the chosen certificate.
Let $\mathcal R_x$ be the consistent families and set
\begin{equation}
 N_{\mathcal R}=|\mathcal R|,\qquad N_{\mathcal R}(x)=|\mathcal R_x|,\qquad
 q_0=N_{\mathcal R}^{-1},\qquad
 \gamma_x=\frac{N_{\mathcal R}}{N_{\mathcal R}(x)},\qquad p_x=N_{\mathcal R}(x)^{-1}.
 \label{clg:normalization}
\end{equation}
In particular, $q_0\gamma_x=p_x$.
The volume condition on the raw sets implies
$1\le\gamma_x=O_{K,r}(1)$; this is also proved in the diagonal
calculation below.

For a nonempty $J\subseteq[b]$, put $\mu(J)=\min J$.
Given $R\in\mathcal R_x$ and nonempty subsets
$J_1,\ldots,J_{s-1}\subseteq[b]$, form a family $S$ by adjoining
$c_i(x)$ to the cell
\[
 R_i\bigl(\mu(J_1),\ldots,\mu(J_{i-1}),J_i\bigr)
 \qquad(1\le i<s).
\]
Denote this family by
$R[J_1\leftarrow c_1(x),\ldots,J_{s-1}\leftarrow c_{s-1}(x)]$.
The certificate-stage indices at stage $s$ are pairs $(S,j)$ with
$j\in U_s\setminus\bigcup S$.
The enlarged cells in $S$ determine $J_1,\ldots,J_{s-1}$.
The index $(S,c_s(x))$ is taken by $x$ and $R$ for every such
branch sequence.  For fixed $x$, each taken index determines $R$
uniquely: remove $c_i(x)$ from its enlarged cell for every $i<s$.
All indices may be included, even if no positive input takes them.

These are indices for the matrices constructed below.  Their use does
not require interpreting the signed correction as an ordinary
nonnegative flow.

\par\medskip\noindent\textbf{Partially revealed assignments.}\enspace
For any loaded family $S$ and input $z$, define an assignment
$\alpha_z^S:\bigcup S\to\Sigma\cup\{\star\}$ recursively by level.
Every element of a level-$1$ cell is revealed, so its assignment is
its input value.  For $i>1$ and
$t\in S_i(d_1,\ldots,d_{i-1},J)$, reveal $z_t$ precisely when it
occurs among the revealed values in
\begin{equation}
 \bigcup_{J'\ni d_{i-1}}
 S_{i-1}(d_1,\ldots,d_{i-2},J').
 \label{clg:parents}
\end{equation}
Otherwise put $\alpha_z^S(t)=\star$.
The symbol $\star$ is not an input value.
Write $x\sim_S y$ when $\alpha_x^S=\alpha_y^S$.

During setup, a newly loaded element can change only its own
assignment: all lower levels are already loaded, and there are no
loaded higher levels.  At certificate stages assignments are
recomputed, so the insertion of a certificate element can uncover
previously hidden raw elements.  The signed construction below
accounts for exactly this effect.

\par\medskip\noindent\textbf{Positive semidefinite matrices.}\enspace
We construct real positive semidefinite matrices $X_j$, indexed by
$\mathcal X$, such that
\begin{equation}
 \sum_{j:x_j\ne y_j}X_j[x,y]=1
 \qquad\bigl(F(x)=1,\ F(y)=0\bigr).
 \label{clg:feasibility}
\end{equation}
By~\eqref{prelim:lg-dual-cost}, these matrices give a query upper bound
within a universal constant of $\max_z\sum_jX_j[z,z]$.

Fix an index $e$ loading input $j$, let $S$ be its source family,
and fix an assignment $\alpha$ on $S$.
Let $\delta_e(z)$ mean that $z$ is positive and takes index $e$
for some consistent raw family.  Every vector defined below is zero
unless $\alpha_z^S=\alpha$; its positive coordinates are also zero
unless $\delta_e(z)$ holds.  These support conditions are understood
in all displayed formulas.  All weights are positive and constant
within the stage and case under consideration.

For a level-$1$ setup index use the vector
\begin{equation}
 u_{e,\alpha}(z)=
 \begin{cases}
  \gamma_z/\sqrt w,&F(z)=1,\ \delta_e(z),\\
  \sqrt w,&F(z)=0,
 \end{cases}
 \qquad
 Y_{e,\alpha}=q_0u_{e,\alpha}u_{e,\alpha}^{*}.
 \label{clg:rank-one-setup}
\end{equation}
For a level-$i$ setup index with $i>1$, let $\chi_e(z)$ indicate
that $z_j$ lies in the revealed parent-value set specified by
\eqref{clg:parents} and $\alpha$.  Define two vectors by
\[
 u_{e,\alpha}(z)=
 \begin{cases}
  \gamma_z/\sqrt{w_1},&F(z)=1,\ \delta_e(z),\ \chi_e(z)=1,\\
  \sqrt{w_1},&F(z)=0,\\
  0,&\text{otherwise},
 \end{cases}
\]
\begin{equation}
 v_{e,\alpha}(z)=
 \begin{cases}
  \gamma_z/\sqrt{w_0},&F(z)=1,\ \delta_e(z),\ \chi_e(z)=0,\\
  \sqrt{w_0},&F(z)=0,\ \chi_e(z)=1,\\
  0,&\text{otherwise},
 \end{cases}
 \qquad
 Y_{e,\alpha}=q_0(uu^{*}+vv^{*}).
 \label{clg:rank-two-setup}
\end{equation}
Consequently, for a positive $x$ taking $e$ and a negative $y$
with $x\sim_Sy$, the off-diagonal contribution is $p_x$ if
at least one of $x_j,y_j$ is revealed after the setup query, and
is zero if both are hidden.

For a stage-$s$ certificate index with branch prefix
$J_1,\ldots,J_{s-1}$, put
\begin{equation}
 \sigma_s=(-1)^{s-1+\sum_{i<s}|J_i|},
 \qquad
 u_{e,\alpha}(z)=
 \begin{cases}
  \gamma_z/\sqrt{w_s},&F(z)=1,\ \delta_e(z),\\
  \sigma_s\sqrt{w_s},&F(z)=0,
 \end{cases}
 \qquad Y_{e,\alpha}=q_0uu^{*}.
 \label{clg:signed-rank-one}
\end{equation}
Its positive-negative entry is $\sigma_s p_x$ when the source
assignments agree, and zero otherwise.
Finally define
\begin{equation}
 X_j=\sum_{e\text{ loading }j}\ \sum_{\alpha}Y_{e,\alpha}.
 \label{clg:matrix-sum}
\end{equation}
Every summand is a Gram matrix of one or two real vectors.
In particular, negative off-diagonal signs in
\eqref{clg:signed-rank-one} do not compromise positive
semidefiniteness.

\par\medskip\noindent\textbf{The off-diagonal identity.}\enspace
Fix a positive $x$, a negative $y$, and $R\in\mathcal R_x$.
We prove that the indices taken for this choice of $x,R$ contribute
exactly $p_x$ to the left side of \eqref{clg:feasibility}.
Let $s$ be the first certificate position with
$x_{c_s(x)}\ne y_{c_s(x)}$.
Such a position exists because otherwise $y$ has the same colored
certificate as $x$.

First, observe two persistence properties.  Before position $s$,
all inserted certificate values equal $\nu_x$ on both inputs.
Because no raw element of $x$ has value $\nu_x$, these insertions
do not change any raw assignment on $x$.  On $y$ they can only
uncover entries with value $\nu_x$.  Hence they cannot repair an
existing disagreement on a raw element.  At position $s$, the
inserted $c_s(x)$ is revealed on $x$, along its chosen branch, with
value $\nu_x$.  Its assignment on $y$ is either hidden or a
different value.  This disagreement persists at every later
certificate stage.  The same argument applies if a disagreement
is first created by uncovering a raw entry with value $\nu_x$.

If $x\not\sim_Ry$, there is a first setup query after which the
assignments disagree.  The queried true values differ, and at
least one is revealed.  Thus this query contributes $p_x$ by
\eqref{clg:rank-one-setup} or \eqref{clg:rank-two-setup}.
Every earlier query either has equal true values or hides both
values, and contributes zero.  Every later setup source disagrees;
the persistence properties show that every certificate source also
disagrees.  This proves the required contribution in this case.

Now suppose $x\sim_Ry$.  All setup contributions vanish.
Certificate stages earlier than $s$ query equal true values, and
stages later than $s$ have disagreeing source assignments.
It remains to sum the stage-$s$ contributions over its branch
prefixes.

For $0\le h<s$ and fixed nonempty $J_1,\ldots,J_h$, let
\[
 S_h=R[J_1\leftarrow c_1(x),\ldots,J_h\leftarrow c_h(x)],
\]
and let $\mathcal C(S_h)$ be the sum of stage-$s$ entries obtained
by all continuations $J_{h+1},\ldots,J_{s-1}$.
We claim
\begin{equation}
 \mathcal C(S_h)=
 \begin{cases}
 (-1)^{h+\sum_{i\le h}|J_i|}p_x,&x\sim_{S_h}y,\\
 0,&x\not\sim_{S_h}y.
 \end{cases}
 \label{clg:block-identity}
\end{equation}
For $h=s-1$, this is exactly \eqref{clg:signed-rank-one}.
We prove the remaining cases by downward induction on $h$.
If the current source disagrees, persistence gives zero for every
continuation.

Suppose instead that the current source agrees.  A raw input
$t$ with $y_t=\nu_x$ will be called a fault.  There are at most
$K$ such inputs, because the multiplicity bound holds for $y$.
For each $d\in[b]$, consider the row group at the next raw level
\begin{equation}
 \bigcup_{\varnothing\ne J\subseteq[b]}
 R_{h+2}\bigl(\mu(J_1),\ldots,\mu(J_h),d,J\bigr).
 \label{clg:fault-row}
\end{equation}
Let $I\subseteq[b]$ be the indices of groups containing no fault.
If $h+2=r$, there is no such raw level, and we set $I=[b]$.
Otherwise the groups are disjoint, so $|I|\ge b-K\ge1$.

Adjoining $c_{h+1}(x)$ in the cell indexed by $J_{h+1}$ can
newly reveal $\nu_x$ only in groups with indices in $J_{h+1}$.
Every fault there was hidden before the insertion, since the
source assignments agree and the corresponding $x$ value is not
$\nu_x$.  Thus the updated assignments agree exactly when
$\varnothing\ne J_{h+1}\subseteq I$.
If there are no faults in these immediate child groups, no raw
value changes its visibility, so there is no subsequent cascade
to higher raw levels.  This also explains the stated boundary
convention when no further raw level exists.

Applying the induction hypothesis to each child gives
\[
 \begin{aligned}
 \mathcal C(S_h)
 &=\sum_{\varnothing\ne J\subseteq I}
   (-1)^{h+1+\sum_{i\le h}|J_i|+|J|}p_x\\
 &=(-1)^{h+\sum_{i\le h}|J_i|}p_x,
 \end{aligned}
\]
because $I$ is nonempty and
$\sum_{\varnothing\ne J\subseteq I}(-1)^{|J|}=-1$.
This proves \eqref{clg:block-identity}.  Taking $h=0$ yields the
desired contribution $p_x$.

Finally, for fixed $x$, no taken index is associated with two
different consistent families $R$.  Summing the per-family
identity over $\mathcal R_x$ therefore yields
$N_{\mathcal R}(x)p_x=1$, proving \eqref{clg:feasibility} exactly.

\par\medskip\noindent\textbf{Cost.}\enspace
The next proposition bounds the diagonal sum of these explicit
matrices.  Combined with \eqref{clg:feasibility} and positive
semidefiniteness, it proves Lemma~\ref{clg:bound} by the dual
adversary characterization.  The positive-dependent factor
$\gamma_x$ is essential to exact normalization; its boundedness
ensures that it changes the diagonal estimate by only a constant
depending on $K,r$.

\subsubsection{Diagonal costs of the colored construction}
\label{sec:colored-lg-cost}

We keep the domain sizes of the roles separate when estimating the
positive-semidefinite matrices of the preceding construction.  In
particular, the counting argument below uses complete colored prefixes,
rather than the number of uncolored pairs or of shorter prefixes.

Fix constants $K\ge r\ge2$.  Let $U_1,\ldots,U_r$ be disjoint public
sets, with $|U_i|=v_i$, and assume that every equality class has size at
most $K$.  For an input $z$, write
\[
 C_j(z)=\bigl|\{(u_1,\ldots,u_j)\in U_1\times\cdots\times U_j:
                 z_{u_1}=\cdots=z_{u_j}\}\bigr|,
 \qquad 2\le j<r.
\]
Let $C_j\ge1$ be specified upper bounds on $C_j(z)$ throughout the
promise, for both positive and negative inputs.  Positive inputs have
at least one colorful target tuple, and negative inputs have none.
For each positive input, fix one such tuple deterministically and use
its equality class as the chosen target class in the construction.

Put $b=K+1$.  At level $i<r$, the fault-tolerant construction has
$L_i=b^{i-1}(2^b-1)=O_{K,r}(1)$ labeled raw-set copies, each of size
$a_i$.  All copies at one level
are disjoint subsets of $U_i$.  In the following diagonal estimates, the $a_i$
are positive integers, and
\begin{equation}
 v_i\ge4K,\qquad L_i a_i\le v_i/4\quad(1\le i<r).
 \label{eq:colored-raw-capacity}
\end{equation}
Copies at different levels are automatically disjoint because their
roles are disjoint.  The constants below may depend on $K,r$ and hence
on the numbers of fault copies, but not on $v_i,a_i,C_j$.

\begin{proposition}[Diagonal costs of the colored construction]
\label{clg:diagonal-cost}
Under the preceding assumptions, the feasible matrices of the colored
construction have maximum diagonal objective at most
\begin{equation}
 O_{K,r}\left(
 a_1+
 \sum_{j=2}^{r-1}
 a_j\sqrt{\frac{C_j\prod_{i<j}a_i}{\prod_{i\le j}v_i}}
 +\sum_{j=1}^{r}
 \sqrt{\frac{\prod_{i\le j}v_i}{\prod_{i<j}a_i}}
 \right).
 \label{clg:cost}
\end{equation}
\end{proposition}

\begin{proof}
\par\medskip\noindent\textbf{A common normalization preserves positivity.}\enspace
Let $\mathcal R$ be the set of all raw configurations.  Its cardinality
and the corresponding reference weight are
\[
 N_{\mathcal R}
   =\prod_{i<r}\frac{(v_i)_{L_i a_i}}{(a_i!)^{L_i}},
 \qquad q_0=N_{\mathcal R}^{-1}.
\]
For a positive input $x$, let $Z_x$ be its chosen target equality
class and put $d_i(x)=|Z_x\cap U_i|$.  We use only configurations
avoiding the whole class $Z_x$, not merely the selected target tuple.
Writing $\mathcal R_x$ for these configurations, we have
\[
 N_{\mathcal R}(x):=|\mathcal R_x|
   =\prod_{i<r}\frac{(v_i-d_i(x))_{L_i a_i}}{(a_i!)^{L_i}},
 \qquad
 \gamma_x=\frac{N_{\mathcal R}}{N_{\mathcal R}(x)}.
\]
The normalization is uniformly bounded.  Indeed,
\begin{equation}
 1\le\gamma_x
   =\prod_{i<r}\frac{(v_i)_{d_i(x)}}
                         {(v_i-L_i a_i)_{d_i(x)}}
   \le2^{\sum_{i<r}d_i(x)}\le2^K.
 \label{eq:colored-normalization-bound}
\end{equation}
Here each denominator factor is at least $v_i/2$, by
\eqref{eq:colored-raw-capacity}, and the target class has at most
$K$ elements.

Use the local Gram factors in
\eqref{clg:rank-one-setup}, \eqref{clg:rank-two-setup}, and
\eqref{clg:signed-rank-one}, with common prefactor $q_0$.
Multiply every positive-input coordinate of every Gram vector by
$\gamma_x$; leave negative-input coordinates unchanged.  Equivalently,
replace each local matrix $X$ by $DXD$, where $D$ is diagonal with
$D_{x,x}=\gamma_x$ on positive inputs and $D_{y,y}=1$ on negative
inputs.  The modified matrices remain positive semidefinite.  A
positive--negative entry associated with one admissible configuration
is now weighted by
\[
 q_0\gamma_x=N_{\mathcal R}(x)^{-1}.
\]
Thus the feasibility calculation, performed separately for each
$R\in\mathcal R_x$, has exactly the required normalization.  On the
diagonal, the positive prefactor is $q_0\gamma_x^2$ and the negative
prefactor is $q_0$.

\par\medskip\noindent\textbf{Counting uncovered raw elements.}\enspace
Let $V_i(R)\subseteq U_i$ be the union of the $L_i$ raw copies at
level $i$, so that $|V_i(R)|=L_i a_i$.  Let $H_j(z,R)$ denote the
number of elements uncovered during setup stage I.$j$, summed over
all copies at that level.  Every such element is the final element
of a chain of equal values containing one raw element from each of
the roles $1,\ldots,j$.  This remains true with the fault-copy
dependencies: those dependencies restrict which previous copies
can supply a match, and cannot create an uncovered element without
a complete colored prefix.  Consequently,
\[
 H_j(z,R)\le
 \sum_{\substack{(u_1,\ldots,u_j)\in U_1\times\cdots\times U_j\\
                  z_{u_1}=\cdots=z_{u_j}}}
       \prod_{i\le j}\mathbf1_{\{u_i\in V_i(R)\}}.
\]
For uniform $R\in\mathcal R$, the unions $V_i(R)$ are independent
across roles and are uniform subsets of the indicated sizes.  Define
\begin{equation}
 A_j=C_j\prod_{i\le j}\frac{a_i}{v_i}.
 \label{eq:colored-setup-matching-scale}
\end{equation}
It follows that, uniformly over the promised inputs,
\begin{equation}
 \mathbb E_{R\in\mathcal R}H_j(z,R)
   \le C_j\prod_{i\le j}\frac{L_i a_i}{v_i}
   =O_{K,r}(A_j).
 \label{eq:colored-uncovered-mean}
\end{equation}
Conditioning on whole-class avoidance changes this upper bound by
at most $\gamma_x\le2^K$.  Hence
$\mathbb E_{R\in\mathcal R_x}H_j(x,R)=O_{K,r}(A_j)$ as well.
All raw setup stages precede certificate insertion, so no inserted
target element contributes an additional shorter-prefix term to
\eqref{eq:colored-uncovered-mean}.

\par\medskip\noindent\textbf{Setup-stage diagonal bounds.}\enspace
For stage I.$j$, with $2\le j<r$, let $w_{0,j},w_{1,j}>0$ be its
two weights.  Equation~\eqref{clg:rank-two-setup} gives the following
diagonal contributions, before normalization: on a negative input
the contribution is $w_{1,j}$ for every loaded element, plus
$w_{0,j}$ when that element is uncovered; on a positive input it
is $1/w_{1,j}$ for an uncovered element and $1/w_{0,j}$ for a
covered element.  Each input satisfies exactly one assignment at a
fixed arc, so summing over assignment labels introduces no factor
depending on the output alphabet.  Summing over configurations and using
\eqref{eq:colored-normalization-bound}--\eqref{eq:colored-uncovered-mean}
therefore gives
\begin{align*}
 D_{0,j}&\le O_{K,r}\bigl(A_jw_{0,j}+a_jw_{1,j}\bigr),\\
 D_{1,j}&\le O_{K,r}\bigl(A_j/w_{1,j}+a_j/w_{0,j}\bigr).
\end{align*}
For example, the positive sum is
$\gamma_x$ times the conditional expectation over
$\mathcal R_x$ of the indicated reciprocal-weight expression;
the bounded factor $\gamma_x$ has already been included above.
Taking
\[
 w_{0,j}=\sqrt{a_j/A_j},\qquad
 w_{1,j}=\sqrt{A_j/a_j}
\]
bounds both diagonal costs by
\begin{equation}
 O_{K,r}\bigl(\sqrt{a_jA_j}\bigr)
  =O_{K,r}\left(
      a_j\sqrt{\frac{C_j\prod_{i<j}a_i}{\prod_{i\le j}v_i}}
            \right).
 \label{eq:colored-setup-cost}
\end{equation}
Stage I.1 has $L_1a_1=O_{K,r}(a_1)$ loads per configuration;
setting its single weight to $1$ gives cost $O_{K,r}(a_1)$.

\par\medskip\noindent\textbf{Certificate-stage diagonal bounds.}\enspace
At stage II.$j$, fix one of the $O_{K,r}(1)$ fault-branch patterns.
For each $i<j$, one specified raw copy at level $i$ has been enlarged
from $a_i$ to $a_i+1$.  The ratio between the number of enlarged
level-$i$ configurations and the number of raw level-$i$
configurations is exactly
\[
 \frac{
   (v_i)_{L_i a_i+1}/[(a_i+1)!(a_i!)^{L_i-1}]
 }{
   (v_i)_{L_i a_i}/(a_i!)^{L_i}
 }
 =\frac{v_i-L_i a_i}{a_i+1}
 \le\frac{v_i}{a_i}.
\]
There are at most $v_j$ choices for the element loaded at stage
II.$j$.  Summing over the constant number of branch patterns, the
total number of its arcs is therefore at most
\[
 O_{K,r}\left(
 N_{\mathcal R}\,v_j\prod_{i<j}\frac{v_i}{a_i}\right).
\]
For a fixed positive input, each $R\in\mathcal R_x$ contributes
only $O_{K,r}(1)$ taken arcs at this stage.  The signs in~\eqref{clg:signed-rank-one} do not affect their
diagonals.  With
stage weight $w_j$, the two diagonal costs consequently satisfy
\[
 D_{0,j}\le O_{K,r}\left(
      w_jv_j\prod_{i<j}\frac{v_i}{a_i}\right),
 \qquad
 D_{1,j}\le O_{K,r}(w_j^{-1}),
\]
where the positive estimate uses
$q_0\gamma_x^2N_{\mathcal R}(x)=\gamma_x=O_K(1)$.
Set
\[
 S_j=\frac{\prod_{i\le j}v_i}{\prod_{i<j}a_i},
 \qquad w_j=S_j^{-1/2}.
\]
The cost of certificate stage II.$j$ is then $O_{K,r}(\sqrt{S_j})$.
This calculation applies also to $j=r$, where there is no raw
level-$r$ set.

Summing the maximum diagonal costs of all stages proves
\eqref{clg:cost}.
\end{proof}

Thus the full colored-prefix bounds $C_j$ control every setup
stage, including all its fault copies.  No additional count of
shorter prefixes or of uncolored collisions is needed in the
diagonal estimate.  The fault-copy constants must depend on the
global multiplicity bound $K$: a negative colorful instance can
have more than $r-1$ occurrences of one value even though it lacks
a colorful $r$-tuple.

The statement was made for integer parameters to expose the exact
configuration counts.  Real parameters at least $1$ can first be
scaled by a sufficiently small constant depending on $K,r$ and
then rounded up.  Provided the role sizes tend to infinity and the
unrounded parameters are at most the corresponding role sizes,
this ensures~\eqref{eq:colored-raw-capacity} and changes every term
in~\eqref{clg:cost} by only a constant factor.

\subsection{Ordered rolewise restriction above the square-root scale}
\label{role:isolation}

We now assume $Q\ge\sqrt N$.
The colored decision bound allows us to retain different fractions
of the input in the different certificate roles. Shrinking these
roles in order preserves the preparation costs and controls every
certificate-loading stage.

\begin{lemma}[Raw parameters in the larger-query regime]
\label{role:parameters}
For a sufficiently large constant $C=C(K,r)$, there are positive
integer parameters $a_i$ and $R=CQ$ such that
\begin{align}
 a_i&\le c'_{K,r}N, &
 \prod_{i<r}a_i&\asymp_{K,r}\frac{N^r}{MR^2},
 \label{role:raw-parameters}\\
 a_1+\sum_{j=2}^{r-1}
 \frac{a_j\sqrt{B_j\prod_{i<j}a_i}}{N^{j/2}}
 &\le O_{K,r}(Q), &
 \sqrt{\frac{N^j}{M\prod_{i<j}a_i}}
 &\le O_{K,r}(Q)\quad(1\le j\le r).
 \label{role:raw-costs}
\end{align}
The constant $c'_{K,r}$ can be made arbitrarily small by increasing
$C$. Every $a_i$ tends to infinity as a fixed positive power of $N$.
\end{lemma}
\begin{proof}
The recurrence in the proof of Lemma~\ref{dg:parameters} does not
use $Q\le\sqrt N$ until its final assertion about expected counts.
It therefore gives real parameters $Q\le\bar a_i\le N$ with
$\bar a_1=Q$, preparation cost $O_{K,r}(Q)$, and
\[
 A:=\prod_{i<r}\bar a_i\ge\frac{N^r}{MQ^2}.
\]
Put
\[
 \theta=\left(\frac{N^r}{MR^2A}\right)^{1/(r-1)},
 \qquad a_i^{\circ}=\theta\bar a_i.
\]
Then $\theta\le C^{-2/(r-1)}$, the product of the $a_i^{\circ}$
is exactly $N^r/(MR^2)$, and every preparation term decreases.
The final certificate term equals $R$. All preceding certificate
terms are no larger, because $a_i^{\circ}\le N$.

For completeness, these real parameters are large enough to round.
The inequalities $B_i\le N$ and $Q\ge\sqrt N$ imply
\begin{equation}
 M\le N^{1-2^{1-r}}.
 \label{role:target-count-hard}
\end{equation}
Indeed, take an index $j$ attaining the maximum in
\eqref{g:scale}, bound its numerator by
$N^{3\cdot2^{j-2}-1}$, and compare with
$N^{(2^j-1)/2}$. For $r\ge3$, using $A\le N^{r-1}$ gives
\[
 a_i^{\circ}\ge
 C^{-2/(r-1)}(N/M)^{1/(r-1)}Q^{(r-3)/(r-1)}.
\]
This grows polynomially by~\eqref{role:target-count-hard}.
For $r=2$, directly $a_1^{\circ}=Q/C^2$.
Take $a_i=\lfloor a_i^{\circ}\rfloor$. Their product changes by
$1+o(1)$, setup costs decrease, and all certificate costs change
by $1+o(1)$. This proves the assertions.
\end{proof}

Fix $\rho_0=1/(2r)$ and then choose a sufficiently small constant
$0<\kappa<\rho_0^r$ depending only on $K,r$. Choose a large constant
$D=D(K,r)$, large enough to accommodate all raw-set copies in
the colored construction, and choose the constant $C$ in
Lemma~\ref{role:parameters} still larger if necessary.
Starting with every role probability equal to $\rho_0$, decrease the
first role probability to $Da_1/N$, then the second to $Da_2/N$,
and so on, stopping as soon as
\begin{equation}
 \prod_{i=1}^r \rho_i=\kappa/M.
 \label{role:prob-product}
\end{equation}
The last role need not be decreased. In fact, the product with
the first $r-1$ roles at their caps is
\[
 \rho_0D^{r-1}\frac{\prod_{i<r}a_i}{N^{r-1}}
 \le\frac{\rho_0D^{r-1}N}{MC^2Q^2}
 \le\frac{\kappa}{M}
\]
for sufficiently large $C$. All caps are below $\rho_0$ by the same
choice. Thus the stopping point exists.

Assign each original input independently to role $i$ with probability
$\rho_i$, and discard it otherwise. The total retained probability is at
most $1/2$. Write $U_i$ for the resulting public role and $v_i=|U_i|$.
The role assignment is internal randomization of the algorithm; it
reveals no equality class.

The next statement supplies both a target and the prefix bounds used
by the colored decision algorithm. The stronger unique-target event
is convenient for the reduction, although the colored algorithm
itself accepts any positive number of targets.
\begin{lemma}[Isolation and prefix populations]
\label{role:isolation-event}
With probability $\Omega_{K,r}(1)$, the roles contain exactly one
colored $r$-tuple, their sizes satisfy
$N\rho_i/2\le v_i\le2N\rho_i$, and simultaneously
\begin{equation}
 C_j(f)\le A_{K,r}B_j\prod_{i\le j}\rho_i
 \qquad(2\le j<r).
 \label{role:prefix-counts}
\end{equation}
Here $C_j(f)$ counts equal-value tuples in $U_1\times\cdots\times U_j$.
\end{lemma}
\begin{proof}
For an original class of size $d\ge r$, the probability that it
meets every role is at most $(d)_r\prod_i\rho_i$. Its probability of
having exactly one element in every role, with all others discarded,
is
\[
 (d)_r\prod_i\rho_i\left(1-\sum_i\rho_i\right)^{d-r}
 =\Theta_{K,r}\left(\prod_i\rho_i\right).
\]
The role assignments in different classes are independent. By
taking $\kappa$ sufficiently small, the expected number of classes
meeting every role is at most $1/2$. For each choice of a unique
surviving class, multiply its exact-one-per-role probability by the
probability that no other class meets every role. The latter is at
least $1/2$ by a union bound and independence. These events are
disjoint for different choices of the surviving class, so their
sum has probability $\Omega_{K,r}(M\prod_i\rho_i)=\Omega_{K,r}(1)$.

Linearity of expectation gives
\[
 \mathbb E C_j(f)=
 \left(\sum_{d\ge j}(d)_j n_d\right)\prod_{i\le j}\rho_i
 \le (K)_jB_j\prod_{i\le j}\rho_i.
\]
Markov's inequality and a union bound, with a sufficiently large
constant $A_{K,r}$, make the probability of violating any prefix
bound smaller than half the isolation probability.
Finally $\rho_i\ge\prod_u\rho_u=\kappa/M$, so
\eqref{role:target-count-hard} implies $N\rho_i\ge\kappa N^{2^{1-r}}$.
Chernoff bounds give all the stated role-size estimates with
probability $1-o(1)$. Intersecting the events proves the lemma.
\end{proof}

Before applying the decision algorithm, we establish a conversion
that exploits the isolated colorful tuple. The role domains remain
fixed, so every virtual input is covered by the same decision circuit.
\begin{lemma}[Coherent search for an isolated colorful target]
\label{search:parity}
Fix disjoint public roles $U_1,\ldots,U_r\subseteq[N]$ and public
bounds $C_j$ on their colored prefix counts. Suppose a bounded-error
decision algorithm uses at most $D_{\rm dec}$ queries on every input
in the promise of Lemma~\ref{clg:bound}, over any prescribed finite
alphabet. On any such input with exactly one colorful $r$-tuple,
its first-role input can be found with constant success probability
using $O(D_{\rm dec})$ original function queries. The algorithm has
a fixed query budget on all inputs, including those outside the
promise.
\end{lemma}
\begin{proof}
Let $x_*$ be the first-role input of the unique tuple. Put
$m=\lceil\log_2 N\rceil$ and encode $x\in[N]$ by the $m$-bit
representation $\operatorname{enc}(x)$ of $x-1$. For
$v\in\{0,1\}^m$, define a virtual input on the unchanged role domains
by
\begin{equation}
 f_v(x)=
 \begin{cases}
  (1,x),&x\in U_1\text{ and }
          v\cdot\operatorname{enc}(x)=0\pmod2,\\
  (0,f(x)),&\text{otherwise}.
 \end{cases}
 \label{search:parity-mask}
\end{equation}
The tagged alphabet is the disjoint union
$(\{0\}\times\mathbb Z_q)\sqcup(\{1\}\times[N])$.
Every dummy value occurs once. Masking therefore creates no new
colored collision, decreases every colored prefix count, and
preserves the maximum class-size bound $K$. Since the original
colorful tuple is unique, the decision answer on $f_v$ is
\begin{equation}
 b(v)=v\cdot s\pmod2,\qquad s=\operatorname{enc}(x_*).
 \label{search:parity-answer}
\end{equation}

Choose one decision circuit $A$ for the full fixed promise consisting
of all inputs on these roles with class sizes at most $K$ and prefix
counts at most the supplied $C_j$. This circuit works for every mask
$v$; its non-query operations do not depend on $v$. Amplify its error
to a fixed constant $\delta<1/100$, using only a constant number of
repetitions, and purify its workspace and measurements. Denote the
resulting unitary circuit with virtual oracle $f_v$ by $A_v$.

A virtual query can be implemented coherently with a constant number
of queries to $f$: compute $f(x)$ into an ancillary register, compute
the public mask and the tagged value in~\eqref{search:parity-mask},
add that value to the virtual response register, and uncompute the
ancilla. The register $v$ is preserved. The inverse circuit has the
same constant-overhead simulation. Thus the controlled family $A_v$
is available without assuming controlled access to an unknown
black-box unitary beyond the stated function-query model.

Let $Z_{\rm ans}$ multiply answer $1$ by $-1$ and leave answer $0$
unchanged, and set $U_v=A_v^\dagger Z_{\rm ans}A_v$. The component
of $A_v|0\rangle$ with an incorrect answer has norm at most
$\sqrt\delta$. Consequently
\begin{equation}
 \bigl\|U_v|0\rangle-(-1)^{b(v)}|0\rangle\bigr\|
 \le2\sqrt\delta.
 \label{search:phase-error}
\end{equation}
Apply this compute--phase--uncompute circuit to a uniform superposition
of $v$. Orthogonality of the preserved control-register sectors gives
\begin{equation}
 \left\|
 2^{-m/2}\sum_v|v\rangle U_v|0\rangle
 -2^{-m/2}\sum_v(-1)^{v\cdot s}|v\rangle|0\rangle
 \right\|^2\le4\delta.
 \label{search:coherent-error}
\end{equation}
In particular, the number $2^m$ of masks introduces no error factor.
The Bernstein--Vazirani Hadamard transform~\cite{BV97} on the control
register maps the second state
exactly to $|s\rangle|0\rangle$. The measured control therefore
returns $s$ with probability at least $1-4\delta$. Decode it as
$x_*$, rejecting an out-of-range index or an index outside $U_1$.
The circuit makes $O(D_{\rm dec})$ function queries. Its budget is
fixed even on an input for which the promise or uniqueness fails.
\end{proof}

We next check the costs of every stage, including the stages before
the last restricted role.
\begin{proposition}[Search above the square-root scale]
\label{role:search}
For $Q\ge\sqrt N$, every function of the prescribed bounded type
admits an $r$-collision search algorithm using
$O_{K,r}(Q)$ function queries and succeeding with
probability at least $2/3$.
\end{proposition}
\begin{proof}
Use Lemma~\ref{role:parameters} and the random role assignment above.
Abort any trial whose public role sizes violate
$N\rho_i/2\le v_i\le2N\rho_i$. The constant $D$ ensures that every
remaining trial satisfies the raw-set capacity conditions of the
colored construction. Supply the bounds
$\lceil A_{K,r}B_j\prod_{i\le j}\rho_i\rceil$ to that algorithm.
The ceiling changes them by only a constant factor: $B_j\ge M$,
and $\prod_{i\le j}\rho_i\ge\prod_{i\le r}\rho_i=\kappa/M$.

On the event of Lemma~\ref{role:isolation-event}, these are valid
prefix bounds. In each setup term of~\eqref{clg:cost}, the role
probabilities cancel between the prefix population and the role
sizes. Thus that term is
\[
 O_{K,r}\left(
 \frac{a_j\sqrt{B_j\prod_{i<j}a_i}}{N^{j/2}}\right)
 =O_{K,r}(Q).
\]
Let $t$ be the last role whose probability was decreased. If
$j\le t$, all earlier probabilities are $\rho_i=Da_i/N$ and hence
\[
 \frac{\prod_{i\le j}v_i}{\prod_{i<j}a_i}=O_{K,r}(N).
\]
Its certificate-stage cost is $O_{K,r}(\sqrt N)=O_{K,r}(Q)$.
If $j>t$, every probability after $j$ is $\rho_0$, so
\eqref{role:prob-product} gives
\[
 \frac{\prod_{i\le j}v_i}{\prod_{i<j}a_i}
 =O_{K,r}\left(\frac{N^j}{M\prod_{i<j}a_i}\right)
 \le O_{K,r}(Q^2)
\]
by~\eqref{role:raw-costs}. Proposition~\ref{clg:diagonal-cost}
therefore gives a bounded-error colored decision algorithm with
$O_{K,r}(Q)$ queries.

On the event of Lemma~\ref{role:isolation-event}, the colorful target
is unique. Apply Lemma~\ref{search:parity} to the fixed colored
promise with the supplied prefix bounds. It finds its first-role
input $x$ with constant success probability in $O_{K,r}(Q)$ function
queries. Query $f(x)$ and use verified randomized Grover search,
excluding previously found inputs, to find $r-1$ further preimages
in the original domain. There are only a fixed number of these
searches, and their total fixed budget can be chosen
$O_{K,r}(\sqrt N)\le O_{K,r}(Q)$ with constant conditional success
probability. Verify all $r$ distinct reported inputs.

All circuits are run with fixed query budgets, even on trials
outside the prefix-count promise, and every final collision is
verified. The unknown prefix-count event need not be checked by
the algorithm. On the event in Lemma~\ref{role:isolation-event},
the search succeeds with a fixed positive probability. A constant
number of trials raises total success to $2/3$. The worst-case query cost is $O_{K,r}(Q)$, with no logarithmic
search-conversion factor.
\end{proof}

\subsection{Combining the two regimes}
\begin{proof}[Proof of Theorem~\ref{search:main}]
The type determines $Q$. If $Q\le\sqrt N$, apply
Theorem~\ref{dg:dense}. Otherwise apply Proposition~\ref{role:search}.
Both are bounded-query search algorithms that succeed on every
input of the stated type, use only function queries, and have
constants independent of the output alphabet. This proves the
claimed bound.
\end{proof}

The colored learning-graph construction adapts the corrected
construction in~\cite{Bel12}; the role-dependent tuple
counts and the ordered restriction are the additional ingredients
used here. The proof does not require a multidimensional quantum
walk that outputs a marked vertex. It also does not establish a
time bound for implementing the adversary-based decision algorithm.
For fixed $K,r$, the large number of correction branches is a
constant in the query model.

Only one collision is requested. An upper bound for outputting
many distinct target classes would require a separate enumeration
or reuse analysis and is not asserted by Theorem~\ref{search:main}.

\begin{proof}[Proof of Theorem~\ref{intro:main}]
The quantity $Q$ in~\eqref{g:scale} is exactly
$\mathcal Q_r(\lambda)$ in~\eqref{intro:profile-scale}.
Theorem~\ref{g:main}, with $p_0=2/3$, gives the lower bound
under the uniform distribution of the type, and hence for worst-case
bounded-error search. Theorem~\ref{search:main} gives an algorithm
with the matching query count on every input of that type.
Both estimates are uniform over all feasible alphabets.
\end{proof}

\section{Consequences and further directions}
\label{scope:final}
The complexity formula separates several sources of difficulty that a
single eligible-bucket count cannot distinguish. An input can contain
many partial collisions while containing few classes in which those
collisions can be extended. The lower bound selects the most restrictive
intermediate order. The upper bound allocates raw sets to balance the
same sequence of intermediate populations.

The exceptional type illustrates why oracle access must remain explicit.
In a public slot representation of $(k^n,k+1)$, independent forward
permutation queries allow the exceptional slots to be searched for
in $O_k(\sqrt N)$ queries. The larger function-query lower bound
uses the restriction to complete function queries. Filtering controls
the canonical return of such a query; it does not establish improved
soundness for arbitrary unfiltered permutation-query algorithms.
Indeed, Proposition~\ref{phase:unfiltered-barrier} shows that the
unfiltered process has a square-root soundness obstruction even
for complete function queries. Its harmonic projection is thus a
substantive change to the compared evolution, not merely an
improvement in the bound for the same simulation.

Further extensions require additional work. First, outputting collisions
in several distinct classes requires tracking depletion and the reuse
of intermediate data. The single-target matching theorem does not
provide that analysis. Second, the learning-graph result is a query
bound. An equally efficient implementation in time for every bounded
type would require suitable data structures and coherent updates.
The three-collision construction in the appendix gives a concrete
alternative update mechanism, but is not an all-order implementation
theorem. Allowing the maximum class size or collision order to grow
with $N$ would also require explicit control of the constants suppressed
throughout this paper.

\newpage
\section*{AI methodology}
\addcontentsline{toc}{section}{AI methodology}
GPT-6 Astra was used in developing the proofs in this work.
The authors proposed the main ideas: combining compressed
permutation techniques with Belovs's transfer map; identifying
the square-root obstruction in the unfiltered CPO simulation,
which motivated the filtered CPO approach; obtaining sharper
upper bounds through more careful control of intermediate
parameters in adaptive learning graphs and nested quantum walks;
and characterizing the lower bound for arbitrary types as the
maximum of several type-dependent bounds. Through iterative
discussions with the authors, GPT-6 Astra developed the detailed
proofs for all of these components.

GPT-6 Astra also assisted with English exposition and the
preparation of references. The introduction and technical overview
combine text written by the authors with model-generated drafts
that the authors edited and revised. The authors take full
responsibility for the final manuscript, including its mathematical
arguments and references.


\clearpage
\appendix

\section{An alternative nested-walk construction for three-collision search}
\label{app:qrw}
The general upper bound in Theorem~\ref{search:main} uses a learning
graph above the square-root scale. This appendix records a different
construction for three-collisions. Its distinct ingredient is an
all-pairs inner-buffer update that remains reversible when eligible
pairs overlap. We give the stationary-distribution, spectral-gap,
and initialization arguments explicitly. This alternative incurs
polylogarithmic overhead from coherent walk operations; it is not
needed for the logarithm-free bound in the main text.

\subsection{Problem, conclusion, and scope}
Let $f:[N]\to\mathbb Z_q$ be any function with exact type
$\lambda=(1^{n_1},\ldots,K^{n_K})$, where $K\ge3$ is fixed. Its
equality classes are the nonempty sets $f^{-1}(y)$. Set
\[
 B=\sum_{d=2}^{K}\binom d2 n_d,\qquad
 M=\sum_{d=3}^{K}n_d\ge1.
\]
Thus $B$ is the number of unordered equal pairs, and $M$ is the number
of classes that contain a three-collision. In particular,
$3M\le B\le (K-1)N/2$ and $B=\Theta_K(B_2)$ for
$B_2=\sum_{d\ge2}n_d$. The algorithm knows the type, and has only
ordinary function-query access. It is not given the members of any
equality class or access to a hidden permutation.

\begin{theorem}\label{ts:main}
For fixed $K$, a three-collision can be found with probability at least
$2/3$, for every such function, using
\[
 \widetilde O_K(Q)\quad\text{function queries},\qquad
 Q=\max\left\{\frac{N^{2/3}}{M^{1/3}},
                  \frac{N^{4/7}B^{1/7}}{M^{2/7}}\right\}.
\]
The bound is uniform over every feasible alphabet. The tilde suppresses
polylogarithmic factors in $N$ arising from coherent search and
approximate reflections.
\end{theorem}

The theorem concerns search, not only a decision promise. Its quantum-walk
compilation uses the finding theorem for nested updates of
Childs, Jeffery, Kothari, and Magniez~\cite[Definition~3.2 and
Theorem~3.3]{CJKM13}. The modifications below are additional arguments;
the type-dependent theorem is not a theorem quoted from that paper.
Theorem~\ref{g:main} gives the matching type-dependent lower bound.
The construction in this appendix is specific to three-collision search;
it does not assert a higher-order nested-walk extension.

\subsection{Parameters and the two search regimes}\label{ts:parameters-section}
Write $Q_2=N^{2/3}M^{-1/3}$ and
$Q_3=N^{4/7}B^{1/7}M^{-2/7}$. Temporarily ignore constant factors and
integer rounding, and put
\begin{equation}
 s_0=Q,\qquad
 L_0=\max\left\{1,\min\left\{
       \frac{Q^{3/2}\sqrt B}{N},\frac{BQ}{N}\right\}\right\}.
 \label{ts:parameters}
\end{equation}
The following elementary identities explain both the desired costs and
the normalization bounds used later.
\begin{lemma}\label{ts:parameter-facts}
Up to constants depending only on $K$, these parameters satisfy
\begin{align*}
 s_0+\frac{NL_0}{\sqrt{Bs_0}}
 +\frac{N}{\sqrt{Ms_0}}+\sqrt{\frac{NB}{ML_0}}&=O_K(Q),\\
 \min(M,L_0)\frac{s_0}{N}&=O_K(1),\\
 \mu_0\frac{L_0}{B}&=O_K(1),\qquad
 \frac{M s_0^3}{N^3}=o(\min(1,\mu_0)),\quad
 \mu_0=\frac{Bs_0^2}{N^2}.
\end{align*}
Also $Q=O_K(N^{5/7})$, and $\mu_0/L_0=o(1)$ whenever $L_0\to\infty$.
\end{lemma}
\begin{proof}
The condition $Q_3\ge Q_2$ is equivalent to $B^3M\ge N^2$ and to
$Q_3\le B$. In this case the first expression in the inner minimum
in~\eqref{ts:parameters} is smaller. Its square is $Q^3B/N^2\ge B/M$,
so it is at least one. Balancing this choice of $L_0$ gives the first
inequality. Moreover,
\[
 \left(\frac{MQ}{N}\right)
 \left(\frac{L_0Q}{N}\right)=\frac BN,
 \qquad \frac{\mu_0L_0}{B}=\frac{B}{MN}.
\]
If $Q_2\ge Q_3$, then $B\le Q$ and the second expression in the
minimum applies. When it is at least one the same two identities hold,
using $Q^3=N^2/M$; the setup term is $\sqrt{BQ}\le Q$.
When it is below one, set $L_0=1$. Then
$N/\sqrt{BQ}\le Q$ since $B\ge M$, and
$\sqrt{NB/M}\le N/\sqrt{MQ}\le Q$ since $BQ<N$.
The remaining normalization assertions follow from the displayed
identities and $B=O_K(N)$. In the $Q_2$ branch,
$Ms_0^3/N^3=1/N$; in the $Q_3$ branch it is
$N^{-9/7}B^{3/7}M^{1/7}=O_K(N^{-5/7})$.
When $\mu_0<1$, its ratio to $\mu_0$ is
$Ms_0/(BN)=O_K(s_0/N)=o(1)$.
Finally, $\mu_0/L_0$ is $Q/N$ or $\sqrt{BQ}/N$, respectively,
unless $L_0=1$.
\end{proof}

In the walk construction take $s=\lfloor a_Ks_0\rfloor$ and
$L=\max\{1,\lfloor b_KL_0\rfloor\}$ for sufficiently small positive
constants. They can be chosen in order so that the setup has many more
available partner inputs than $L$, and
$m_{\max}L/B$ is below the constant required by the spectral-gap
lemma. Every cost in Lemma~\ref{ts:parameter-facts} changes only by a
$K$-dependent factor. Cases where $L_0$ is bounded by a sufficiently
large $K$-dependent constant use $L=1$ and the separate initialization
below; this also changes the costs only by a constant.

For $Q\le\sqrt N$, Theorem~\ref{dg:dense} with $r=3$ already
provides a direct search algorithm: its tail-count scale is no larger
than $Q$, because $B_2\le B$. Since also $B=O_K(B_2)$, both scales
agree up to a $K$-dependent constant. The nested walk addresses
$Q>\sqrt N$.

\subsection{A symmetric update for lists from distinct equality classes}
\label{ts:coherent-updates}

This subsection gives an exact implementation of the outer update.  The
stationary distribution is allowed to depend on the number of admissible
inner buffers.  This removes the need to assume that all collision pairs
in the input are disjoint.

Fix a function $f:[N]\to\mathbb Z_q$ whose equality classes have sizes
at most a fixed constant $K$.  Let $M$ be the number of classes of size
at least three.  Fix disjoint sets $A_1,A_2\subseteq[N]$ and a fixed set
$\mathcal Z\subseteq\mathbb Z_q$ of masked values.  Define the eligible
pair pool
\[
 \mathscr P=\bigl\{\{x,y\}:x\in A_1,\ y\in A_2,
              \ f(x)=f(y)\notin\mathcal Z\bigr\},
 \qquad B=|\mathscr P|.
\]
For the one-pair initialization branch, $\mathscr P$ may instead be
the full pool of equal pairs. The proofs below apply verbatim: the
cross-pair restriction is then simply omitted.
The mask changes only which pairs are eligible.  In particular, inputs
whose values belong to $\mathcal Z$ are not excluded from an inner
buffer merely because of the mask.  All queried values are stored as
their actual values under $f$.

Write $w_b$ for the number of eligible pairs in equality class $b$.
Then $0\le w_b\le W_K:=\binom K2$.  An eligible class of size two has
exactly one eligible pair.  Let $\Omega_L$ be the family of sets $S$ of
$L$ eligible pairs, with at most one pair from each equality class.
For any such pair set $W$, let $\operatorname{end}(W)$ be the union of its endpoints and
let $F(W)$ be the set of its function values.  Thus $|\operatorname{end}(S)|=2L$ for
$S\in\Omega_L$.

Fix an integer buffer size $s$ and a nonempty set of permitted batch
sizes
\[
 \mathcal J\subseteq
 \{1,\ldots,\min\{L,\lfloor s/2\rfloor\}\}.
\]
For an input set $R$, put
\[
 J(R)=\{e\in\mathscr P:e\subseteq R\},\qquad X(R)=|J(R)|.
\]
For $S\in\Omega_L$, a buffer is \emph{good} if
\[
 \begin{split}
 \mathrm{Good}_S=\bigl\{R\subseteq[N]\setminus \operatorname{end}(S):\;&|R|=s,
       \ f(R)\cap F(S)=\varnothing,\\
     &\text{each equality class meets $R$ in at most two inputs},
       \quad X(R)\in\mathcal J\bigr\}.
 \end{split}
\]
Let $Z_S=|\mathrm{Good}_S|$.  The occupancy condition implies that
$J(R)$ consists of disjoint pairs from distinct equality classes.
Their classes are also disjoint from those represented by $S$.

For a pair set $W$ with distinct values and an integer $u\ge0$, define
the residual-buffer family
\[
 \begin{split}
 \mathcal F_{W,u}=\bigl\{C\subseteq[N]\setminus \operatorname{end}(W):\;&|C|=u,
      \ f(C)\cap F(W)=\varnothing,\\
     &\text{each equality class meets $C$ in at most two inputs},
       \quad J(C)=\varnothing\bigr\}.
 \end{split}
\]
Write $G_W(u)=|\mathcal F_{W,u}|$.
Here $J(C)=\varnothing$ forbids eligible cross-pairs only.  It does
not forbid, for example, a pair whose value is masked.

\begin{lemma}[Exact local diffusion and symmetric garbage]
\label{ts:update}
Suppose $Z_S>0$ for every $S\in\Omega_L$.  There is a Markov chain
$P$ on $\Omega_L$ with the following transition probabilities.  If
$S\cup S'$ represents distinct equality classes and
$m=|S\setminus S'|=|S'\setminus S|\in\mathcal J$, then
\begin{equation}
 P(S,S')=
 \frac{G_{S\cup S'}(s-2m)}{Z_S\binom Lm};
 \label{ts:transition}
\end{equation}
otherwise $P(S,S')=0$.  Its stationary distribution is
\begin{equation}
 \pi(S)=\frac{Z_S}{\sum_{U\in\Omega_L}Z_U}.
 \label{ts:stationary}
\end{equation}
Given the queried values in $S$ and a coherent uniform good buffer,
the local diffusion associated with $P$ uses no further function
queries.  Its edge garbage is invariant under exchanging $S$ and
$S'$, so the corresponding garbage swap also uses no further queries.
\end{lemma}

\begin{proof}
For $R\in\mathrm{Good}_S$, let $m=X(R)$ and extract \emph{all}
the pairs $J=J(R)$.  Set $C=R\setminus \operatorname{end}(J)$.  Every class represented
by $J$ has exactly two inputs in $R$, so $C$ contains no further input
from any of these classes.  It also avoids the classes of $S$.
Consequently,
\[
 C\in\mathcal F_{S\cup J,s-2m}.
\]
Choose $I\subseteq S$, $|I|=m$, uniformly, and put
$S'=(S\setminus I)\cup J$.  Then $S\cup J=S\cup S'$.

Conversely, fix $S'$ occurring in~\eqref{ts:transition} and
$C\in\mathcal F_{S\cup S',s-2m}$.  The sets
\[
 I=S\setminus S',\qquad J=S'\setminus S,\qquad R=C\cup \operatorname{end}(J)
\]
recover the original choices uniquely.  The buffer $C$ avoids every
class represented by $S\cup S'$.  Adding $\operatorname{end}(J)$ therefore creates
exactly the $m$ eligible pairs $J$, with occupancy exactly two in
their classes, and creates no other eligible pair.  Thus
$R\in\mathrm{Good}_S$.  This proves a bijection between the choices
$(R,I)$ and the resulting pairs $(S',C)$.

Include the queried function values in the notation for each basis
state.  Starting with the normalized uniform good-buffer state, the
query-free extraction and uniform choice of $I$ give
\[
 \begin{split}
 &\frac1{\sqrt{Z_S}}
    \sum_{R\in\mathrm{Good}_S}|S,R\rangle\\
 &\quad\longmapsto
 \sum_{S'}
 \sqrt{\frac{G_{S\cup S'}(s-2m)}{Z_S\binom Lm}}
 |S,S'\rangle\,|\psi_{S\cup S',m}\rangle,
 \end{split}
\]
where terms with zero residual count are omitted and
\[
 |\psi_{W,m}\rangle=
 \frac1{\sqrt{G_W(s-2m)}}
 \sum_{C\in\mathcal F_{W,s-2m}}|C,f|_C\rangle.
\]
The bijection proves that the coefficients in
\eqref{ts:transition} sum to one.  It also makes the extraction
reversible: from $S,S',C$ one recovers both $R$ and the chosen $I$.
All values needed for these operations were already stored.

Detailed balance follows from the exact identity
\[
 \pi(S)P(S,S')
 =\frac{G_{S\cup S'}(s-2m)}
        {\binom Lm\sum_{U\in\Omega_L}Z_U}
 =\pi(S')P(S',S).
\]
Finally, the normalized garbage state depends only on $S\cup S'$
and $m$.  Swapping the two outer registers leaves this state
unchanged.  This proves the asserted zero-query garbage swap.
\end{proof}

One may replace $P$ by $(I+P)/2$ to make the chain lazy.  This does
not change its stationary distribution and changes its gap and
implementation costs by at most constant factors.

\par\medskip\noindent\textbf{Uniform comparison of residual-buffer counts.}\enspace
The next elementary switching bound controls the dependence of
$G_W(u)$ on the selected equality classes.

\begin{lemma}[Deleting a few positions from a residual buffer space]
\label{ts:buffer-switching}
Let a domain of $n$ inputs be partitioned into classes of size at
most $K$.  Consider its size-$u$ subsets satisfying an occupancy
cap of two and containing no pair from a fixed eligible cross-pair
pool.  If $n>Ku$, the probability that a uniform such subset
contains a specified input is at most
\[
 \frac{u}{n-Ku}.
\]
In particular, suppose $u=o(N)$, $n=\Theta(N)$, and $D=o(N)$.
Deleting any $D$ positions changes the number of admissible subsets
by a multiplicative factor between
$\exp(-O_K(Du/N))$ and $1$, provided the remaining domain still has
size $\Theta(N)$.
\end{lemma}

\begin{proof}
Fix an input $x$.  For every admissible set $C$ containing $x$,
replace $x$ by an input $y$ from a class not represented in $C$.
At least $n-Ku$ choices of $y$ are available.  The replacement is
admissible: removal preserves both constraints, and the new input
lies in an otherwise empty class.  For any resulting set not
containing $x$, at most $u$ choices of its element $y$ can invert
this operation.  If $a$ and $b$ count admissible sets containing
and not containing $x$, respectively, then
\[
 (n-Ku)a\le ub.
\]
This implies the displayed marginal bound.  Apply that bound
successively while deleting the prescribed positions.  The ratio
of the final count to the original count is at least
\[
 \prod_{j=0}^{D-1}
 \left(1-\frac{u}{n-j-Ku}\right).
\]
For sufficiently large $N$, the denominators have order $N$ and
the subtracted fractions are $o(1)$.  Taking logarithms gives
the claimed lower bound.  Deletion cannot increase the count.
\end{proof}

\begin{lemma}[Residual counts are comparable]
\label{ts:buffer-count}
Assume $s=o(N)$, $L=o(N)$, and
\begin{equation}
 \frac{\min\{M,L\}\,s}{N}\le A
 \label{ts:hidden-capacity-condition}
\end{equation}
for a fixed constant $A$.  Fix $m\in\mathcal J$ and put $u=s-2m$.
For any two pair sets $W,W'$ of size $L+m$ with distinct equality
classes,
\[
 G_W(u)=\Theta_{K,A}\bigl(G_{W'}(u)\bigr).
\]
The constants are uniform over the fixed cross-pair pool and its
fixed value mask.
\end{lemma}

\begin{proof}
All eligible classes of size two have the same residual-buffer
constraint: both of their inputs cannot be selected together.
In particular, their counting polynomial is $1+2z$, independently
of their actual input positions and output labels.  Thus, after
relabeling whole classes of this kind, the residual domains for
$W$ and $W'$ can be aligned on all but
\[
 D=O_K\bigl(\min\{M,L+m\}\bigr)
\]
positions.  To see this explicitly, keep the exceptional classes
of size at least three that remain in both domains.  Delete the
exceptional classes that remain in only one domain, and match the
numbers of remaining eligible size-two classes by deleting the
excess from the other domain.  The number of exceptional selected
classes is at most $\min\{M,L+m\}$, and each has at most $K$
inputs.  All other classes, including masked classes, are
unchanged.  This yields isomorphic common residual domains after
deleting at most $O_K(\min\{M,L+m\})$ positions from each.

The full residual domains have size $N-O_K(L+m)=N-o(N)$.
Lemma~\ref{ts:buffer-switching} compares each count to that on the
common domain.  Since $m\le L$ and $u\le s$,
\[
 \frac{Du}{N}
 =O_K\!\left(\frac{\min\{M,L\}\,s}{N}\right)=O_K(A).
\]
Both comparisons therefore lose at most a constant factor.
\end{proof}

\par\medskip\noindent\textbf{Preparing and reflecting about good inner buffers.}\enspace
The inner state is obtained from an ordinary Johnson walk; none
of the validity constraints has to be built into its transition
rule.  This distinction is useful because the ordinary walk has
a known uniform stationary state and gap of order $1/s$.

\begin{lemma}[Density of good buffers]
\label{ts:inner-density}
Assume $K$ and $A$ are fixed, $s\ge2$, $s=o(N)$, $L=o(N)$,
\eqref{ts:hidden-capacity-condition}, and $B\ge2W_KL$.  Put
\[
 \mu=\frac{Bs^2}{N^2},\qquad
 \nu=\frac{Ms^3}{N^3},
\]
and suppose $\nu=o(\min\{1,\mu\})$.
There are constants $\mu_0,c_-,c_+>0$, depending only on $K$, with
the following property.  Choose the batch window to be $\{1\}$
when $\mu<\mu_0$, and otherwise to be
\[
 \mathcal J=\{m\in\mathbb Z:
       \max\{1,\lceil c_-\mu\rceil\}\le m
       \le\lfloor c_+\mu\rfloor\}.
\]
Assume this window is contained in
$\{1,\ldots,\min\{L,\lfloor s/2\rfloor\}\}$.
Then, uniformly over $S\in\Omega_L$,
\begin{equation}
 \frac{Z_S}{\binom{N-2L}{s}}
   =\Theta_{K,A}\bigl(\min\{1,\mu\}\bigr).
 \label{ts:good-density}
\end{equation}
In particular, the stationary distribution in
\eqref{ts:stationary} is pointwise comparable to the uniform
distribution on $\Omega_L$.
\end{lemma}

\begin{proof}
A raw buffer is a uniform $s$-subset of $[N]\setminus \operatorname{end}(S)$.
Beyond the $2L$ known endpoints, the classes represented by $S$
contain at most
\[
 D_S\le(K-2)\min\{M,L\}
\]
further inputs.  The probability of avoiding these inputs is
\[
 \frac{\binom{N-2L-D_S}{s}}{\binom{N-2L}{s}}
 =\Omega_{K,A}(1),
\]
by the product formula for the ratio, $s,L=o(N)$, and
\eqref{ts:hidden-capacity-condition}.  Condition on this avoidance.
The buffer is now uniform on a domain $U_S$ of size
$n_S=N-O_K(L)\sim N$ obtained by deleting the complete classes
represented by $S$.

Let $B_S$ be the number of eligible pairs remaining on $U_S$.
The assumption $B\ge2W_KL$ gives $B/2\le B_S\le B$.
For the eligible-pair count $X=X(R)$,
\[
 \mathbb E X=B_S\frac{(s)_2}{(n_S)_2}
             =\Theta_K(\mu).
\]
Pairs on disjoint input sets contribute
$O_K(\mu^2)$ to $\mathbb E[X(X-1)]$.  Each eligible pair
overlaps only $O_K(1)$ other eligible pairs; an overlapping pair
of pairs requires three sampled inputs.  Consequently,
\begin{equation}
 \mathbb E[X(X-1)]
       =O_K\!\left(\mu^2+\mu\frac{s}{N}\right),
 \qquad
 \mathbb E X^2=O_K(\mu+\mu^2).
 \label{ts:pair-second-moment}
\end{equation}
All these estimates are uniform in $S$.

For a sufficiently small fixed $\mu_0$, if $\mu<\mu_0$ then
\[
 \Pr[X=1]\ge \mathbb E X-\mathbb E[X(X-1)]
                =\Omega_K(\mu).
\]
Here the first inequality follows pointwise from
$\mathbf1_{\{x=1\}}\ge x-x(x-1)$ for nonnegative integers $x$.
If $\mu\ge\mu_0$, the second-moment inequality gives a constant
lower bound on $\Pr[X\ge c_-\mu]$ for a suitable $c_->0$.
Markov's inequality makes $\Pr[X>c_+\mu]$ smaller than half
that lower bound by choosing $c_+$ sufficiently large.  The
constant can also be increased so that the integer window is
nonempty throughout $\mu\ge\mu_0$.  Thus
\[
 \Pr[X\in\mathcal J]=\Omega_K(\min\{1,\mu\}).
\]

There are at most $\binom K3 M$ triples within equality classes.
A union bound therefore gives
\[
 \Pr[\text{some class has occupancy at least three in $R$}]
       =O_K(\nu)=o(\min\{1,\mu\}).
\]
Removing these buffers preserves the preceding lower bound.
Multiplication by the probability of avoiding the extra inputs
proves the lower bound in~\eqref{ts:good-density}.  The upper
bound follows from
$\Pr[X\ge1]\le\min\{1,\mathbb E X\}$, also without conditioning
on the avoidance event.  Finally, all $Z_S$ are comparable to
the same quantity in~\eqref{ts:good-density}, which proves the
claim about $\pi$.
\end{proof}

On the raw Johnson graph $J(N-2L,s)$, setup costs $s$ function
queries, an update costs $O(1)$ queries, and the good-buffer
predicate costs no further queries once the buffer values are
stored.  Its gap is $\Theta(1/s)$ for $s=o(N)$.
Thus standard coherent quantum-walk search and its inverse
prepare, and reflect about, the normalized good-buffer state
with the inner search cost
\begin{equation}
 \widetilde O_{K,A}\!\left(
       \sqrt{\frac{s}{\min\{1,\mu\}}}\right)
 =\widetilde O_{K,A}\!\left(
       \max\left\{\sqrt s,\frac{N}{\sqrt{Bs}}\right\}\right).
 \label{ts:inner-reflection-cost}
\end{equation}
The one-time initial preparation also includes the $s$ setup
queries.  The logarithmic factors allow the coherent operations
to be implemented to the inverse-polynomial accuracy required
by the enclosing walk.  Lemma~\ref{ts:update} then supplies the
local diffusion and garbage swap exactly, without additional
function queries.

\subsection{The outer walk on pairs from distinct classes}
\label{ts:outer-chain}

We first analyze the outer chain independently of its quantum
implementation.  Let $E=\bigsqcup_{b=1}^{H}E_b$, where $E_b$ is the set
of pair choices in class $b$, and write
\[
 1\le w_b:=|E_b|\le W,\qquad B=\sum_{b=1}^{H}w_b.
\]
For bounded multiplicities one may take $W=\binom K2$.
For $1\le L\le H$, put
\[
 \Omega_L=\{S\subseteq E:|S|=L,\ |S\cap E_b|\le1\text{ for every }b\}.
\]
Let $\operatorname{cl}(S)$ denote the classes represented by $S$, and
let $\pi_0$ be uniform on $\Omega_L$.  Under $\pi_0$, a class set $A$
has probability proportional to $w(A):=\prod_{b\in A}w_b$;
conditional on $A$, the item in each selected class is uniform.
We use $e_j$ for an elementary symmetric polynomial, with $e_0=1$
and $e_j=0$ outside its possible range, and set
\[
 Z_j=e_j(w_1,\ldots,w_H),\qquad
 Z_j(R)=e_j((w_b)_{b\notin\operatorname{cl}(R)}),\qquad
 z_j(S)=Z_j(S).
\]

For a reversible Markov kernel $P$ with stationary measure $\pi$, write
\[
 \mathcal E_{\pi,P}(f)=\frac12\sum_{S,T}\pi(S)P(S,T)(f(S)-f(T))^2.
\]
The Poincar\'e spectral gap is the largest $\gamma$ for which
$\mathcal E_{\pi,P}(f)\ge\gamma\operatorname{Var}_{\pi}f$ for all real
$f$.  When the kernel is not already lazy, we explicitly pass to
$(I+P)/2$ before using an absolute spectral gap for quantum search.

\subsubsection{Weighted subsets and the marked fraction}

The following coupling will also be used to compare two batch updates.
\begin{lemma}[Nested weighted subsets]
\label{ts:weighted-containment}
For positive weights on a finite set, let $\nu_j(A)=w(A)/e_j(w)$ on
its $j$-subsets.  For each feasible $j\ge1$, there is a coupling of
$A_{j-1}\sim\nu_{j-1}$ and $A_j\sim\nu_j$ with
$A_{j-1}\subseteq A_j$.
\end{lemma}
\begin{proof}
The elementary symmetric polynomials satisfy
$e_i^2\ge e_{i-1}e_{i+1}$.  For completeness, this follows by adding
the weights one at a time.  If the old coefficient sequence is
$(a_i)$, adjoining a weight $w$ gives $b_i=a_i+wa_{i-1}$, and
\[
 \begin{split}
 b_i^2-b_{i-1}b_{i+1}
 ={}&(a_i^2-a_{i-1}a_{i+1})
 +w(a_i a_{i-1}-a_{i-2}a_{i+1})\\
 &+w^2(a_{i-1}^2-a_{i-2}a_i)\ge0.
 \end{split}
\]
The middle term is nonnegative because the ratios of consecutive
positive terms of a log-concave sequence are nonincreasing; the
boundary cases use the zero coefficients outside its support.

We now induct on the number of elements.  If the last element has
weight $w$, its inclusion probability under $\nu_j$ is
\[
 p_j=\frac{w e_{j-1}}{e_j+w e_{j-1}},
\]
where these polynomials concern the remaining elements.
Log-concavity implies $p_{j-1}\le p_j$.  Couple the two last-element
indicators monotonically.  If both are absent, apply induction to
sizes $j-1,j$ on the remaining set.  If both are present, apply it to
sizes $j-2,j-1$.  If the indicator changes from absent to present,
choose the same $(j-1)$-subset of the remaining elements.  Each case
has the required conditional marginals and preserves containment;
zero-probability boundary cases can be omitted.
\end{proof}

Suppose that $M$ of the $H$ classes are eligible, and mark an outer
state if it represents at least one eligible class.  This event has
the required probability even though different classes have different
numbers of pair choices.
\begin{lemma}[Marked fraction]
\label{ts:marked}
For every $1\le L\le H$, the marked fraction under $\pi_0$ satisfies
\[
 \frac1{W+1}\min\left\{1,\frac{ML}{H}\right\}
 \le\varepsilon_0
 \le W\min\left\{1,\frac{ML}{H}\right\}.
\]
In particular, it is
$\Theta_W(\min\{1,ML/B\})$.
\end{lemma}
\begin{proof}
Let $I_b$ indicate inclusion of class $b$.  Its probability is
\[
 p_b=\frac{w_b e_{L-1}(w_{-b})}
 {e_L(w_{-b})+w_b e_{L-1}(w_{-b})}.
\]
The identity
\[
 L e_L(w_{-b})=
 \sum_{\substack{A\subseteq[H]\setminus\{b\}\\|A|=L-1}}
 w(A)\sum_{c\notin A\cup\{b\}}w_c
\]
shows that
\[
 \frac{H-L}{L}
 \le\frac{e_L(w_{-b})}{e_{L-1}(w_{-b})}
 \le\frac{W(H-L)}L.
\]
Consequently $L/(WH)\le p_b\le WL/H$.

The indicators are pairwise negatively correlated.  Indeed, for
distinct $b,c$ put $a_i=e_i(w_{-\{b,c\}})$.  Clearing denominators
reduces $\mathbb E(I_bI_c)\le p_bp_c$ to
$a_{L-2}a_L\le a_{L-1}^2$, which was established in the preceding
proof.  Thus, if $X$ counts eligible classes represented in the state,
\[
 \frac{ML}{WH}\le\mathbb EX\le\frac{WML}{H},\qquad
 \mathbb EX^2\le\mathbb EX+(\mathbb EX)^2.
\]
Cauchy--Schwarz and the union bound give
\[
 \frac{\mathbb EX}{1+\mathbb EX}
 \le\Pr(X>0)\le\min\{1,\mathbb EX\}.
\]
The claimed constants follow.  Finally, $H\le B\le WH$.
\end{proof}

\subsubsection{Batch heat-bath updates}

We next obtain a gap bound for the chain that will serve as a
comparison chain.  For $1\le m\le L$, the batch heat-bath kernel
$P_m^{\mathrm{hb}}$ removes a uniform $m$-subset of the current state
and then chooses a uniform completion in $\Omega_L$ of the retained
set.  It is reversible for $\pi_0$: its transition probability is a
sum, over retained sets $R\subseteq S\cap T$ of size $L-m$, of
$1/(\binom Lm Z_m(R))$, which is symmetric in $S,T$.

\begin{lemma}[Batch heat-bath gap]
\label{ts:heatbath-gap}
If $L\le B/(3W)$, then every nonconstant eigenvalue of
$P_m^{\mathrm{hb}}$ has absolute value at most $1-m/(2L)$.
For $L=1$, the chain is independent uniform resampling and has gap
one, without the displayed restriction.
\end{lemma}
\begin{proof}
Use the exchange metric $d(S,T)=L-|S\cap T|$.  It is the graph
distance for valid single-item exchanges: first correct the chosen
item in classes common to the two states, and then replace classes
present only on one side.  We give a coupling for adjacent states
$S=R_0\cup\{u\}$ and $T=R_0\cup\{v\}$.

Couple the removal sets by identifying $u$ with $v$ and leaving all
common items fixed.  With probability $m/L$, the differing items
are removed; identical completions then coalesce.  Otherwise, the
retained states still differ by $u,v$.  If these items have the same
class, their completion spaces coincide, and identical completions
preserve distance one.

Suppose that their classes are different.  Let $C$ be the classes
available to both completions.  The first completion can additionally
use the class of $v$, and the second the class of $u$.  Their
class weights are denoted by $w_v,w_u$ in this paragraph.  Their
probabilities of using these extra classes are
\[
 p_v=\frac{w_v e_{m-1}(C)}{e_m(C)+w_v e_{m-1}(C)},\qquad
 p_u=\frac{w_u e_{m-1}(C)}{e_m(C)+w_u e_{m-1}(C)}.
\]
The elementary-symmetric-polynomial identity used above gives
\[
 \frac{e_m(C)}{e_{m-1}(C)}
 \ge\frac{\sum_{b\in C}w_b-(m-1)W}{m}.
\]
At most $L-m+1$ classes are excluded in defining $C$, so
\[
 \max\{p_u,p_v\}\le\frac{mW}{B-WL}.
\]

Couple the two extra-class indicators maximally.  If neither extra
class is used, choose the same weighted $m$-subset of $C$.  The final
distance is one.  If both are used, choose the same weighted
$(m-1)$-subset of $C$; the final distance is at most two.  If exactly
one is used, apply Lemma~\ref{ts:weighted-containment} to couple the
common-class subsets of sizes $m-1$ and $m$ by containment.  Couple
pair choices identically in shared classes.  The final distance is
again at most two.  Therefore
\[
 \begin{split}
 \mathbb E d(S',T')
 &\le\left(1-\frac mL\right)
       \left(1+\frac{mW}{B-WL}\right)\\
 &\le1-\frac mL+\frac{mW}{B-WL}
 \le1-\frac{m}{2L}.
 \end{split}
\]
The last inequality uses $3WL\le B$.
Composing the adjacent-state couplings along shortest paths gives
the same contraction for arbitrary states.  Equivalently, the
Lipschitz seminorm of $P_m^{\mathrm{hb}}f$ contracts by this factor.
Applying that assertion to any nonconstant eigenfunction proves the
eigenvalue bound.  If $L=1$, removing the sole item leaves the empty
state and the subsequent completion is an independent uniform
sample, proving the final assertion.
\end{proof}

\subsubsection{Strict exchanges}

The actual update inserts pairs from classes absent from the entire
old list.  We therefore remove the heat-bath transitions that reuse
a deleted class.  Write $S\sim_m T$ if $|S\cap T|=L-m$ and every
class represented by $T\setminus S$ is absent from $S$.  This
condition is symmetric.  The removed and inserted sets are uniquely
determined on such an edge.

\begin{lemma}[Strict-exchange comparison]
\label{ts:strict-gap}
Suppose $1\le m\le L$ and
\begin{equation}
 \frac{mL}{B}\le\frac1{100W}.
 \label{ts:strict-density}
\end{equation}
The strict-exchange kernel
\[
 P_m^{\mathrm{str}}(S,T)
 =\frac{\mathbf1_{\{S\sim_mT\}}}{\binom Lm\,z_m(S)}
\]
is reversible for $\pi_m^{\mathrm{str}}(S)\propto z_m(S)$, and
\begin{equation}
 \frac{49}{50}\le\frac{z_m(S)}{Z_m}\le1.
 \label{ts:completion-ratio}
\end{equation}
Its Poincar\'e gap is at least $m/(4L)$, and the absolute gap of its
lazy version is at least $m/(8L)$.
More precisely, the symmetric substochastic kernel
\[
 K_m(S,T)=\frac{\mathbf1_{\{S\sim_mT\}}}{\binom Lm\,Z_m}
\]
satisfies
\begin{equation}
 \frac12\sum_{S,T}\pi_0(S)K_m(S,T)(f(S)-f(T))^2
 \ge\frac{m}{4L}\operatorname{Var}_{\pi_0}f.
 \label{ts:strict-raw-dirichlet}
\end{equation}
\end{lemma}
\begin{proof}
Under the weighted $m$-subset distribution on all classes, the
inclusion probability of a fixed class is at most
\[
 \frac{mW}{B-W(m-1)}\le\frac{2mW}{B}.
\]
This follows directly by bounding the average remaining weight in
the elementary-symmetric-polynomial identity.  A union bound over
the $L$ classes of $S$ proves~\eqref{ts:completion-ratio}.

Condition~\eqref{ts:strict-density} implies $L\le B/(3W)$.
In a heat-bath step, call reuse of any removed class bad.  Conditional
on the retained set, the inclusion probability of one removed class
is at most
\[
 \frac{mW}{B-W(L-1)}\le\frac{2mW}{B}.
\]
The total bad-step probability is consequently at most
$\alpha=2Wm^2/B$.  The bad-edge kernel is symmetric.
Centering $f$ at its $\pi_0$-mean and using
$(f(S)-f(T))^2\le2f(S)^2+2f(T)^2$ bounds its Dirichlet form by
$2\alpha\operatorname{Var}_{\pi_0}f$.
Lemma~\ref{ts:heatbath-gap} therefore gives
\[
 \mathcal E_{\mathrm{good}}(f)
 \ge\left(\frac{m}{2L}-\frac{4Wm^2}{B}\right)
       \operatorname{Var}_{\pi_0}f,
\]
where the good kernel retains only strict exchanges.

On a strict edge the retained set is necessarily $S\cap T$, so the
good heat-bath transition is $1/(\binom Lm Z_m(S\cap T))$.
The preceding union bound, applied to its $L-m$ classes, gives
$Z_m(S\cap T)/Z_m\ge49/50$.  Consequently
\[
 \begin{split}
 \mathcal E_{\pi_0,K_m}(f)
 &\ge\frac{49}{50}\mathcal E_{\mathrm{good}}(f)\\
 &\ge\frac{49}{50}\left(\frac12-\frac1{25}\right)
           \frac mL\operatorname{Var}_{\pi_0}f
 \ge\frac{m}{4L}\operatorname{Var}_{\pi_0}f.
 \end{split}
\]
Here the notation for the Dirichlet form also applies to the
substochastic kernel; adding its missing mass on the diagonal does
not change that form.

The row sum of $K_m$ is $d_m(S)=z_m(S)/Z_m$.  Put
$\bar d_m=\mathbb E_{\pi_0}d_m$.  Conditioning on an accepted step
gives $P_m^{\mathrm{str}}$, with stationary density
$d_m/\bar d_m$ relative to $\pi_0$.  Thus
\[
 \mathcal E_{\pi_m^{\mathrm{str}},P_m^{\mathrm{str}}}(f)
 =\frac{\mathcal E_{\pi_0,K_m}(f)}{\bar d_m},\qquad
 \operatorname{Var}_{\pi_m^{\mathrm{str}}}f
 \le\frac{\operatorname{Var}_{\pi_0}f}{\bar d_m}.
\]
The second inequality follows by choosing the $\pi_0$-mean as the
centering constant and using $d_m\le1$.  These inequalities prove
the Poincar\'e gap.  Lazification makes all eigenvalues nonnegative
and divides the gap by two.
\end{proof}

\subsubsection{The exact buffer-weighted chain}

We finally allow the batch size and the residual-buffer count to
depend on the transition.  This comparison does not assume that
buffer counts agree across outer states.

\begin{theorem}[Buffer-weighted variable-batch gap]
\label{ts:comparison}
\label{ts:gap}
Let $\mathcal I\subseteq\{1,\ldots,L\}$ be nonempty, with endpoints
$m_{\min}$ and $m_{\max}$, and assume
\[
 \frac{m_{\max}L}{B}\le\frac1{100W}.
\]
For every $m\in\mathcal I$, suppose that a positive symmetric
buffer count on strict exchange edges satisfies
\begin{equation}
 g_m\le G_{S\cup T}(s-2m)\le C_0g_m
 \qquad(S\sim_mT),
 \label{ts:buffer-comparison-assumption}
\end{equation}
where $g_m>0$ and $C_0\ge1$ is independent of $m,S,T$.
Define conductances and degrees by
\[
 \begin{split}
 c(S,T)&=\sum_{m\in\mathcal I}
 \frac{\mathbf1_{\{S\sim_mT\}}}{\binom Lm}
 G_{S\cup T}(s-2m),\\
 A(S)&=\sum_Tc(S,T),\qquad P(S,T)=\frac{c(S,T)}{A(S)}.
 \end{split}
\]
Then $P$ is reversible for $\pi(S)\propto A(S)$, and
\begin{equation}
 \frac{49}{50C_0}\le\frac{\pi(S)}{\pi_0(S)}
 \le\frac{50C_0}{49}.
 \label{ts:stationary-comparison}
\end{equation}
The absolute spectral gap of the lazy chain $(I+P)/2$ is at least
\begin{equation}
 \frac{49}{400C_0}\frac{m_{\min}}L.
 \label{ts:weighted-lazy-gap}
\end{equation}
Its marked fraction is
$\Theta_{W,C_0}(\min\{1,ML/B\})$.
\end{theorem}
\begin{proof}
Symmetry of the conductances proves reversibility.  For fixed $m$,
each inserted set $J$ can be paired with exactly $\binom Lm$ removed
sets, and $S\cup T=S\cup J$ does not depend on the removed set.
In particular, the degree contribution is exactly
\begin{equation}
 A_m(S)=
 \sum_{\substack{J\in\Omega_m\\
          \operatorname{cl}(J)\cap\operatorname{cl}(S)=\varnothing}}
 G_{S\cup J}(s-2m).
 \label{ts:weighted-degree}
\end{equation}

Compare first with the baseline conductances
\[
 c^0(S,T)=\sum_{m\in\mathcal I}
       \frac{g_m}{\binom Lm}\mathbf1_{\{S\sim_mT\}}.
\]
Their degrees are $A^0(S)=\sum_mg_mz_m(S)$.  With
$Z_*=\sum_mg_mZ_m$, Lemma~\ref{ts:strict-gap} gives
\begin{equation}
 \frac{49}{50}Z_*\le A^0(S)\le Z_*.
 \label{ts:baseline-degree}
\end{equation}
Let $P^0$ and $\pi^0$ be the associated normalized kernel and
stationary measure, and put $D^0=\sum_SA^0(S)$.
Equation~\eqref{ts:strict-raw-dirichlet} implies
\[
 \begin{split}
 \mathcal E_{\pi^0,P^0}(f)
 &=\frac{|\Omega_L|}{D^0}
       \sum_mg_mZ_m\mathcal E_{\pi_0,K_m}(f)\\
 &\ge\frac{m_{\min}}{4L}\operatorname{Var}_{\pi_0}f,
 \end{split}
\]
because $D^0\le|\Omega_L|Z_*$.  Also
$\pi^0/\pi_0\le50/49$ by~\eqref{ts:baseline-degree}, so
\[
 \operatorname{gap}(P^0)
 \ge\frac{49}{200}\frac{m_{\min}}L.
\]

Assumption~\eqref{ts:buffer-comparison-assumption} gives
$c^0\le c\le C_0c^0$.  Writing $D=\sum_SA(S)$, direct comparison
of Dirichlet forms and stationary variances yields
\[
 \mathcal E_{\pi,P}(f)
 \ge\frac{D^0}{D}\mathcal E_{\pi^0,P^0}(f),\qquad
 \operatorname{Var}_{\pi}f
 \le C_0\frac{D^0}{D}\operatorname{Var}_{\pi^0}f.
\]
For the variance inequality, center at the $\pi^0$-mean and use
$A(S)\le C_0A^0(S)$.  The Poincar\'e gap of $P$ is therefore at
least $49m_{\min}/(200C_0L)$.  Lazification proves
\eqref{ts:weighted-lazy-gap}.

Finally, $49Z_*/50\le A(S)\le C_0Z_*$ proves the density bounds
\eqref{ts:stationary-comparison}.  Applying those bounds to the
marked event and using Lemma~\ref{ts:marked} proves its stated
probability.  Lazification changes neither the stationary measure
nor the marked event.
\end{proof}

The theorem supplies the required outer parameters whenever the
exact quantum update realizes these conductances and the buffer
comparison~\eqref{ts:buffer-comparison-assumption} holds.  It does not
by itself construct the conditional completion states or their
coherent update.  Parameter regimes with only boundedly many pair
classes can be handled separately instead of imposing a strict
exchange when no new class is available.

\subsection{Preparing the outer state without a uniqueness promise}
\label{ts:setup}
\label{ts:initialization}

The update construction uses a stationary distribution proportional to
the number of admissible inner buffers.  We do not assume that the
initial list produced by a sequence of searches has this distribution.
Instead, we first obtain a state that is uniform on the unmarked lists,
and then project onto the stationary state of the implemented walk.
The one preliminary test of whether a list is marked costs
$\widetilde O(\sqrt N)$; this is affordable in the present regime $Q>\sqrt N$.

Write $B=\sum_d\binom d2 n_d$ and $M=\sum_{d\ge3}n_d$.  Throughout this
subsection the maximum class size is the fixed constant $K$, and
\[
 Q=\max\{N^{2/3}M^{-1/3},\ N^{4/7}B^{1/7}M^{-2/7}\}.
\]
The choices of $s=\Theta_K(Q)$ and $L$ in the parameter calculation
satisfy $L\le c_K Bs/N$ outside the one-pair initialization regime.
We first treat list sizes exceeding a sufficiently large fixed
$K$-dependent cutoff; the estimates below approach probability one
when $L\to\infty$, and exceed a fixed constant close to one uniformly
above that cutoff. Smaller list sizes are handled separately below.

Independently place each input into one of two parts
$\widetilde A$ and $C$, with equal probabilities.  Their sizes are
$\Theta(N)$ except with exponentially small probability.  Prepare a
uniform queried $s$-subset $I$ of $\widetilde A$.  A candidate partner
is an input $y\in C$ whose value occurs exactly once in $I$.  If a value
has already been selected, exclude that value from later searches.
Thus repeated values in the seed are ignored; the seed is not required
to be collision-free.

For a seed $I$, let
\[
 H(I)=\bigl|\{y\in C:
       |\{x\in I:f(x)=f(y)\}|=1\}\bigr|.
\]
Use coherent quantum searches to select $L$ partners of distinct
values, keeping each selected input in superposition.  Each partner
has a uniquely determined anchor in $I$.  The selection can be
implemented with error $N^{-c}$ per stage, for an arbitrary fixed
$c>0$, whenever $H(I)\ge h_0+KL$, where $h_0=c'_K Bs/N$.
The cost per stage is $\widetilde O_K(N/\sqrt{Bs})$.
Validity flags may be measured, but the partner registers are not
measured.  The algorithm declares failure on an invalid stage.
All search operations are symmetric among inputs satisfying their
current search predicate.  Consequently, their successful amplitudes
are equal on those inputs, including any phase depending on the
number of candidates.  This observation will also account for seeds
on which the prescribed high-success guarantee does not apply.

Let $S$ be the resulting list of $L$ pairs and let
$F=I\setminus\operatorname{anchors}(S)$. Sort the list reversibly,
retaining its ordering register, and measure $F$ together with its
queried values. On an unmarked outcome the ordering register is a
known uniform state and can then be erased without queries. Define
\[
 A=\widetilde A\setminus F.
\]
The pair universe for the ensuing walk consists of equal-value pairs
across $A\times C$ whose values do not occur in $F$.  A fixed mask
implements this restriction: on the remaining inputs use the virtual
function
\begin{equation}
 g_F(x)=
 \begin{cases}
 (0,f(x)),& f(x)\notin f(F),\\
 (1,x),& f(x)\in f(F).
 \end{cases}
 \label{ts:fixed-mask}
\end{equation}
The two tags denote disjoint alphabets.  A query to $g_F$ uses a
constant number of queries to $f$.  In particular, inputs whose values
are masked remain available to the inner buffer as singletons.  We do
\emph{not} require the buffer to avoid every class touched by $F$.
The pair universe is restricted to $A\times C$, but the inner
buffer is sampled from the full original domain, including the
masked positions, apart from the known endpoints of its current
outer list.  Its ambient size is therefore always $N-2L$.
The final third-input search also uses the original domain of $f$.

\begin{lemma}[Uniformity after an unmarked outcome]
\label{ts:unmarked-flat}
Fix the measured set $F$ and the classical transcript of the
partner-search procedure.  On the component in which every selected
pair belongs to a class of size exactly two in the original input,
the amplitudes of all valid outer lists are equal.  Thus an ideal
unmarked projection produces the uniform superposition of unmarked
lists in the modified pair universe.
\end{lemma}
\begin{proof}
For an unmarked list $S$, every selected anchor belongs to a true
two-element class, whose other element is its selected partner in
$C$.  This class has no member in $F$.  Conversely, every unmarked
list in the modified pair universe determines the unique seed
$I=F\cup\operatorname{anchors}(S)$.

Adding these $L$ anchors to $F$ contributes exactly one candidate
partner per anchor and does not change the candidate status of any
other value.  Hence
\[
 H(I)=H(F)+L.
\]
After $j$ selected partners, the candidate count is exactly
$H(F)+L-j$.  In an ideal uniform search, the amplitude of any ordered
list is therefore proportional to
\[
 \bigl((H(F)+L)_L\bigr)^{-1/2}.
\]
For an implemented symmetric search, an additional successful
amplitude, phase, or transcript factor can depend on these candidate
counts, but not on the particular unmarked list.  The initial seed
amplitudes are equal as well.  All orders of a fixed unmarked list
have the same amplitude, so their order register can be erased by a
query-free change of basis.  Measuring $F$ leaves equal amplitudes on
all unmarked lists, as claimed.  No assertion about the amplitudes
of marked lists is needed.
\end{proof}

The test used to realize the unmarked projection is a verified
Grover search for an input outside the stored endpoints whose value
belongs to the list.  It costs $\widetilde O(\sqrt N)$, never returns an
incorrect witness, and can have an arbitrarily small fixed failure
probability on every marked list.  On an unmarked list its failure
workspace is independent of that list.  If the marked component has
constant weight, this preliminary test already succeeds with
constant probability.  Otherwise its failure branch has constant
weight and, by Lemma~\ref{ts:unmarked-flat}, is arbitrarily close to
the uniform unmarked state.  This argument does not condition a
small absolute error on an event of vanishing probability.

\begin{lemma}[Retention of the input populations]
\label{ts:setup-retention}
With probability bounded below by a positive constant depending only
on $K$, the modified pair universe contains $\Theta_K(B)$ pairs and
has $\Theta_K(M)$ classes that extend to a three-collision in the
original input.  For $L\to\infty$, the partner-selection procedure
succeeds with probability $1-o(1)$ on the seeds needed for this
conclusion.
\end{lemma}
\begin{proof}
The random two-coloring is independent across original equality
classes.  A class of size at least two crosses the two parts with
probability at least $1/2$, and its number of cross pairs is at most
the constant $\binom K2$.  The mean number of cross pairs is $B/2$,
and its variance is $O_K(B)$.  Thus the pair count is $\Theta_K(B)$
with probability $1-o(1)$ when $B\to\infty$.  The number of eligible
classes crossing the parts has mean at least $M/2$ and second moment
$O_K(M+M^2)$.  It is therefore at least a fixed positive multiple of
$M$ with constant probability, including when $M$ is bounded.

For a uniform seed of size $s$ in a part of size $\Theta(N)$, the
probability of touching any fixed class is $O_K(s/N)$.  Therefore the
expected number of target classes touched by $I$ is $O_K(Ms/N)$,
and the expected number of candidate pairs belonging to a class
touched by $I$ is $O_K(Bs/N)$.  Since $s=o(N)$ and $F\subseteq I$,
Markov's inequality shows that masking $f(F)$ loses $o(M)$ target
classes and $o(B)$ pairs in probability.  For bounded $M$ this says,
more strongly, that no target class is lost with probability
$1-o(1)$.

For completeness, put $Y=\sum_{x\in I}\deg_C(x)$, where
$\deg_C(x)$ is the number of equal-value inputs in $C$.
On partitions with $\Theta_K(B)$ cross pairs,
\[
 \mathbb E Y=\Theta_K(Bs/N),\qquad \operatorname{Var} Y\le(K-1)\mathbb E Y.
\]
The variance bound follows from the negative correlation of
fixed-size sampling indicators and $0\le\deg_C(x)\le K-1$.
If a class has $a$ inputs in $\widetilde A$, $b$ in $C$, and seed
occupancy $X$, its contribution to $Y-H(I)$ is
$bX\mathbf1_{\{X\ge2\}}\le bX(X-1)$. Summing the hypergeometric
second factorial moments therefore gives
\[
 \mathbb E(Y-H(I))=O_K(Bs^2/N^2)=O_K(s/N)\mathbb E Y.
\]
Chebyshev's inequality for $Y$ and Markov's inequality for
$Y-H(I)$ imply $H(I)\ge\mathbb E Y/2$ except with probability
$O_K(1/\mathbb E Y+s/N)$. Since
$Bs/N\ge L/c_K\to\infty$, the partner count is at least
$c'_K Bs/N+KL$ with probability $1-o(1)$ after choosing the constant
in the constraint on $L$ sufficiently small.  Removing one selected
value removes at most $K$ candidates, so the same lower bound needed
by each subsequent search holds.

The partner searches are controlled by the seed.  Averaging over
their outcomes leaves the diagonal seed distribution unchanged.
Moreover, the later map is injective in the relevant information:
$I=F\cup\operatorname{anchors}(S)$.  Thus a bad-retention event for
$F$ cannot acquire more unconditional probability than the
corresponding bad event for $I$.  This justifies combining the
population estimates with the measurements and validity flags in
the actual preparation procedure.
\end{proof}

\begin{lemma}[Projection onto the implemented stationary state]
\label{ts:stationary-setup}
Suppose the implemented outer walk has stationary distribution
$\pi$ pointwise comparable, within $K$-dependent constants, to the
uniform distribution on its outer lists, and its stationary marked
mass is at most a fixed constant smaller than one.  From the uniform
unmarked state one can prepare the implemented stationary walk state
with constant success probability.  In the parameter range used
here the additional query cost is $\widetilde O_K(Q)$.
\end{lemma}
\begin{proof}
Let $\rho_S$ be the good-buffer probability for an outer list $S$.
The inner analysis gives, uniformly in $S$,
\[
 \rho_S=\Theta_K(\min\{1,\mu\}),\qquad
 \mu=Bs^2/N^2.
\]
First prepare the uniform queried inner buffer, at cost $O(s)$.
If $\mu$ is small, use an ordinary Grover amplification with one
common safe iteration count and then measure the good-buffer flag.
The iteration count is chosen using uniform upper and lower bounds
on $\rho_S$, so it does not overshoot: successful amplitudes are
positive and bounded below by a fixed constant on every $S$.
If $\rho_S$ is already bounded below, merely measure this flag.
The reflection about the uniform inner state is implemented by the
ordinary Johnson walk.  The cost is
\[
 \widetilde O_K\left(s+
       \frac{\sqrt s}{\sqrt{\min\{1,\mu\}}}\right).
\]
This use of a positive successful amplitude is deliberate; arbitrary
uncontrolled $S$-dependent phases would not justify a stationary
overlap bound.

Consequently, after a successful flag the outer coefficients on the
unmarked lists are positive and comparable to uniform coefficients,
and each list carries its normalized good-buffer state.  Apply the
query-free local diffusion to put this state in the row-star space
of the implemented walk.  Pointwise comparison of $\pi$ with
uniform measure, together with the constant unmarked mass, gives
constant overlap with the stationary walk state.

For the lazy reversible walk, phase estimation projects onto this
state at cost $\widetilde O(U/\sqrt\delta)$.  Within the row-star space,
irreducibility makes the stationary state the only zero-phase
component; zero-phase states orthogonal to both star spaces have
zero initial overlap.  The inner reflection, as in the nested-update
construction of~\cite{CJKM13}, and the outer phase
estimation are implemented to inverse-polynomial precision, which
adds only logarithmic factors.

The update and gap bounds established for the buffer construction
are
\[
 U=\widetilde O_K\left(
       \frac{\sqrt s}{\sqrt{\min\{1,\mu\}}}\right),
 \qquad
 \delta=\Omega_K\left(\frac{\max\{1,\mu\}}L\right).
\]
They give
\[
 \frac U{\sqrt\delta}
 =\widetilde O_K\left(\frac{N\sqrt L}{\sqrt{Bs}}\right)
 \le \widetilde O_K(Q).
\]
The last inequality follows from the parameter bounds, or from
$ML/B=O_K(1)$ and $N/\sqrt{Ms}=O_K(Q)$.
The other displayed initialization costs are also $\widetilde O_K(Q)$.
Thus a constant number of attempts suffices in expectation; a
constant-factor cutoff gives a bounded-query procedure with constant
success probability.
\end{proof}

\par\medskip\noindent\textbf{Bounded list sizes.}\enspace
\label{ts:bounded-list}
If the unrounded optimal list size remains bounded, use $L=1$ and
the full, uncolored pool of equal pairs.  At these parameters
$\mu=Bs^2/N^2=o(1)$ for $s=\Theta_K(Q)$.
Prepare a uniform queried $s$-subset and search for a subset
containing exactly one equal pair.  After coherently extracting
that pair, the joint state differs by
\[
 O_K\left(\sqrt{Bs^2/N^2+s/N}\right)=o(1)
\]
from a uniform pair together with an unrestricted uniform residual
buffer avoiding its two endpoints.  Indeed, under the latter ideal
state another disjoint pair occurs with probability
$O_K(Bs^2/N^2)$, and another pair overlapping the selected pair
occurs with probability $O_K(s/N)$.  Projection onto the complement
of these events gives exactly the extracted state, by the bijection
between an exactly-one-pair subset and its pair and residual subset.
Unquery and erase the residual buffer, and subsequently initialize
the inner data as in Lemma~\ref{ts:stationary-setup}.  The query cost
is
\[
 \widetilde O_K\left(s+\frac{N}{\sqrt{Bs}}\right)=\widetilde O_K(Q),
\]
because $B\ge M$ and $Q^3\ge N^2/M$.

If $M/B$ is at least a sufficiently small fixed constant, measuring
this almost uniform pair and running the third-input search already
solves the problem in $\widetilde O_K(Q)$ queries.  Otherwise the constant
can be chosen so that $B$ is large enough for the strict one-pair
exchange estimate, and the stationary marked mass is bounded away
from one.  The preceding stationary projection and search apply.
This also treats the finite-candidate cases without presuming that
a nontrivial strict-exchange graph exists.

Combining the partner-selection cost, the preliminary third-input
test, and Lemma~\ref{ts:stationary-setup}, the outer initialization
in the regime $Q>\sqrt N$ has total cost
\[
 \widetilde O_K\left(s+\frac{NL}{\sqrt{Bs}}+\sqrt N+Q\right)
 =\widetilde O_K(Q).
\]
All constructions use only function queries and the values already
stored by the algorithm.

\subsection{Completion of the three-collision search bound}
\label{ts:completion}

We now combine the exact update with its implementation costs. In this
subsection $B$ again denotes the original total pair count. A cross-pair
pool constructed by initialization has size $B'=\Theta_K(B)$ and
contains pairs from $M'=\Theta_K(M)$ target classes on the successful
constant-probability preparation event. The estimates in the preceding
sections are uniform over every such pool and its fixed mask.

\begin{proof}[Proof of Theorem~\ref{ts:main}]
Theorem~\ref{dg:dense} covers $Q\le\sqrt N$.
Suppose $Q>\sqrt N$. Take the constant-scaled parameters from
Subsection~\ref{ts:parameters-section} and put $\mu=Bs^2/N^2$. The constants defining the batch
window can be chosen uniformly for $B'=\Theta_K(B)$.

First dispose of a harmless bounded-ratio case. If $M/B$ is bounded
below by a sufficiently small positive constant depending only on $K$,
then $B=\Theta_K(M)$ and $Q>\sqrt N$ implies $B=O_K(\sqrt N)$.
Prepare a nearly uniform equal pair by the sparse-buffer construction
of Subsection~\ref{ts:initialization}, and measure it. The probability
that its class has size at least three is $\Omega_K(M/B)=\Omega_K(1)$.
A Grover search for a third input costs $O(\sqrt N)=O(Q)$, while
pair preparation costs at most $\widetilde O_K(Q)$. Constant repetition
proves the claim. This case includes bounded $B$.

In the remaining case $M/B$ is sufficiently small. For unbounded
$L_0$, choose $s=a_KQ$ and $L=b_KL_0$ with suitable fixed constants.
For bounded $L_0$, use $L=1$ instead. Lemma~\ref{ts:parameter-facts}
implies, uniformly in these regimes,
\[
 s,L=o(N),\quad
 \frac{\min(M,L)s}{N}=O_K(1),\quad
 \frac{Ms^3}{N^3}=o(\min(1,\mu)).
\]
The scaled constants ensure $B'\ge2W_KL$ and
$m_{\max}L/B'\le1/(100W_K)$. If the unscaled $L_0$ is not bounded,
$\mu/L=o(1)$, so the window in Lemma~\ref{ts:inner-density} lies
within $\{1,\ldots,\min(L,\lfloor s/2\rfloor)\}$ for large $N$.
If $L_0$ is bounded, then $\mu=o(1)$ and the window is simply
$\{1\}$. Its smallest and largest batch sizes have order
$\max(1,\mu)$, with constants depending only on $K$.

Lemma~\ref{ts:update} realizes the exact local diffusion and garbage
swap, using no new queries. Lemma~\ref{ts:buffer-count} verifies the
normalizer hypothesis of Theorem~\ref{ts:comparison}. The lazy outer
walk therefore satisfies
\[
 \delta=\Omega_K\!\left(\frac{\max(1,\mu)}{L}\right),
 \qquad
 \varepsilon=\Theta_K\!\left(\frac{ML}{B}\right),
 \qquad \pi\asymp_K\pi_0.
\]
Here $ML/B$ is at most a sufficiently small fixed constant by the
choice of $b_K$ (and by the bounded-ratio case just removed when
$L=1$). In particular, the unmarked stationary mass is bounded below.
All inner marked-state reflections have cost
\[
 U=\widetilde O_K\!\left(\sqrt{\frac{s}{\min(1,\mu)}}\right).
\]
The checking procedure searches the original domain for an input
whose function value is represented by the pair list and whose index
is outside its stored endpoints. Because the list uses distinct
classes, this is exactly the condition that some stored pair belongs
to a target class. It costs $C=\widetilde O(\sqrt N)$; a successful check
also supplies a verifiable third input.

Subsection~\ref{ts:initialization} prepares the actual stationary walk
state, with constant success probability, within
\[
 \widetilde O_K\!\left(s+\frac{NL}{\sqrt{Bs}}+\sqrt N
                       +\frac{U}{\sqrt\delta}\right).
\]
The first three terms are $\widetilde O_K(Q)$, and the fourth is dominated
by the eventual search cost. Apply the finding theorem for nested
updates~\cite[Theorem~3.3]{CJKM13}. Since
\[
 \min(1,\mu)\max(1,\mu)=\mu,
\]
the update contribution becomes
\[
 \frac{U}{\sqrt{\varepsilon\delta}}
 =\widetilde O_K\!\left(\sqrt{\frac{Bs}{M\mu}}\right)
 =\widetilde O_K\!\left(\frac{N}{\sqrt{Ms}}\right),
\]
and the checking contribution is
\[
 \frac C{\sqrt\varepsilon}
 =\widetilde O_K\!\left(\sqrt{\frac{NB}{ML}}\right).
\]
Lemma~\ref{ts:parameter-facts} bounds both by $\widetilde O_K(Q)$.
The initialization, search, and reported third-input verification
have a fixed positive total success probability. Independent
repetition raises it to $2/3$. All subroutines have fixed query budgets. Any expected-cost
preparation is cut off at a sufficiently large constant multiple of
its expectation; Markov's inequality preserves a fixed positive
success probability. They use
only queries to $f$ and classical information obtained from those
queries. None of the bounds depends on $q$ beyond realizability.
\end{proof}

\par\medskip\noindent\textbf{What the argument establishes.}\enspace
The new step is the choice to extract every eligible pair in a good
inner buffer. The resulting symmetric residual state determines a
reversible outer chain with a generally nonuniform stationary
distribution. Counting and initialization are adapted to that
distribution rather than identifying it with a Johnson stationary
state. This proves a three-collision search upper bound with the
two type-dependent scales in Theorem~\ref{ts:main}. Extending
the construction to higher collision orders requires additional
levels of coherent data and their own update analysis.

\end{document}